\documentclass[useAMS,usenatbib]{mn2e}
\usepackage{times}
\usepackage{verbatim}
\usepackage{amsmath}
\usepackage{graphicx}
\usepackage{color}
\usepackage{rotating}
\usepackage{lscape}

\newcommand{\er}{$\log R_{e}$}
\newcommand{\vd}{$\log \sigma_0$}
\newcommand{\sbmag}{$\langle \mu_{e} \rangle$}
\newcommand{\sblum}{$\log \langle I_{e} \rangle$}
\newcommand{\kms}{km s$^{-1}$}
\newcommand{\Ng}{$N_{\rm{g}}$}
\newcommand{\VR}{V$\backslash$R}

\title[6dFGS: The Fundamental Plane]{The 6dF Galaxy Survey: 
  The Near-Infrared Fundamental Plane of Early-Type Galaxies} 
\author[C. Magoulas et al.]{Christina Magoulas$^{1,2}$, Christopher M. Springob$^2$, Matthew Colless$^2$, D.~Heath Jones$^{2,3}$,\newauthor
Lachlan A. Campbell$^{4}$, John R. Lucey$^{5}$, Jeremy Mould$^{6}$, Tom Jarrett$^{7}$, Alex Merson$^{5}$, \newauthor
Sarah Brough$^{2}$ \\ 
$^1$School of Physics, University of Melbourne, Parkville, VIC 3010, Australia\\
$^2$Australian Astronomical Observatory, PO Box 296, Epping, NSW 1710, Australia\\
$^3$School of Physics, Monash University, Clayton, VIC 3800, Australia\\
$^4$Department of Physics \& Astronomy, University of Western Kentucky, Bowling Green, KY 42102-3576, USA\\
$^5$Department of Physics, University of Durham, Durham DH1 3LE, UK\\
$^6$Centre for Astrophysics and Supercomputing, Swinburne University, Hawthorn, VIC 3122, Australia\\
$^7$Spitzer Science Center, California Institute of Technology, Pasadena, CA 91125, USA
}

\date{Draft version of \today}

\begin{document}

\maketitle

\begin{abstract}
  We determine the near-infrared Fundamental Plane for $\sim$10$^4$
  early-type galaxies in the 6dF Galaxy Survey (6dFGS). We fit the
  distribution of central velocity dispersion, near-infrared surface
  brightness and half-light radius with a three-dimensional Gaussian
  model using a maximum likelihood method. The model provides an
  excellent empirical fit to the observed Fundamental Plane distribution
  and the method proves robust and unbiased. Tests using simulations
  show that it gives superior results to regression techniques in the
  presence of significant and correlated uncertainties in all three
  parameters, censoring of the data by various selection effects, and
  outliers in the data sample. For the 6dFGS $J$ band sample we find a
  Fundamental Plane with
  $R_{e}$\,$\propto$\,$\sigma_0^{1.52\pm0.03}I_{e}^{-0.89\pm0.01}$,
  similar to previous near-infrared determinations and consistent with
  the $H$ and $K$ band Fundamental Planes once allowance is made for
  differences in mean colour. The overall scatter in $R_e$ about the
  Fundamental Plane is $\sigma_r$\,=\,29\%, and is the quadrature sum of
  an 18\% scatter due to observational errors and a 23\% intrinsic
  scatter. Because of the Gaussian
  distribution of galaxies in Fundamental Plane space, $\sigma_r$ is
  {\em not} the distance error, which we find to be $\sigma_d$\,=\,23\%.
  Using group richness and local density as measures of environment, and
  morphologies based on visual classifications, we find that the
  Fundamental Plane slopes do not vary with environment or morphology.
  However, for fixed velocity dispersion and surface brightness, field
  galaxies are on average 5\% larger than galaxies in groups or
  higher-density environments, and the bulges of early-type spirals are
  on average 10\% larger than ellipticals and lenticulars. The residuals
  about the Fundamental Plane show significant trends with environment,
  morphology and stellar population. The strongest trend is with age,
  and we speculate that age is the most important systematic
  source of offsets from the FP, and may drive the other trends through
  its correlations with environment, morphology and metallicity. These
  results will inform our use of the near-infrared Fundamental Plane in
  deriving relative distances and peculiar velocities for 6dFGS
  galaxies.
\end{abstract}


\section{Introduction}

Empirical correlations between observable galaxy parameters guide our
understanding of the physical mechanisms that regulate the formation and
evolution of galaxies. One of the first early-type galaxy scaling
relations was recognised by \citet{Faber:1976}, and connects galaxy
luminosity, $L$, and stellar velocity dispersion, $\sigma$. The
Faber-Jackson (FJ) relation has the form of a power law, $L \propto
\sigma^{\gamma}$, where $\gamma$ is usually observed to be in the range
3 to 5. A similar relation between galaxy luminosity and effective
radius, $R_{e}$, was derived around the same time \citep{Kormendy:1977}.
The Kormendy relation also has power-law form, $L \propto
R_{e}^{\epsilon}$, with $\epsilon$ usually found to be in the range $-1$
to $-2$. Both relations show a wide range of slopes depending on the
properties of the sample under consideration (e.g.\ absolute magnitude
and morphological type) and substantial intrinsic scatter, in the range
0.2--0.5\,dex
\citep[e.g.][]{Desroches:2007,Nigoche-Netro:2008,Nigoche-Netro:2010}.

However, subsequent examination of the three-dimensional (3D)
logarithmic space of size, surface brightness and velocity dispersion
revealed that early-type galaxies populate a more tightly correlated
two-dimensional (2D) plane with significantly lower intrinsic scatter
\citep{Dressler:1987,Djorgovski:1987}.  This Fundamental Plane (FP) has
the power-law form $R_{e} \propto \sigma_0^{a} \langle I_{e}
\rangle^{b}$, where $R_{e}$ is the effective radius, $\langle I_{e}
\rangle$ is the mean surface brightness enclosed within the effective
radius, and $\sigma_0$ is the central stellar velocity dispersion.

Since the original formulation of the FP relation, the size and quality
of early-type galaxy samples have been steadily improved
\citep[e.g.][]{Bernardi:2003b,DOnofrio:2008,LaBarbera:2008,Hyde:2009,Gargiulo:2009,LaBarbera:2010a,Graves:2010b}
in an effort to explain important properties such as the FP's observed
orientation (or \emph{tilt}) and its intrinsic scatter (or
\emph{thickness}).

The \emph{tilt} of the FP is the difference between the observed
coefficients of the plane, $a$ (for \vd) and $b$ (for \sblum), and the
values $a=2$ and $b=-1$ that would follow if galaxies were homologous
virialised systems with constant mass-to-light ratio. The physical
origin of this tilt is usually interpreted as being due to some
combination of systematic deviations either from dynamical homology
(i.e.\ differences in density profile or orbital structure) or from a
fixed mass-to-light ratio ($M/L$). Both effects clearly contribute in
some degree, but neither one by itself appears to explain the entirety
of the FP tilt, leaving its origin an open and much-debated question
\citep[see,
e.g.,][]{Ciotti:1996,Busarello:1997,Graham:1997,Trujillo:2004,DOnofrio:2006,Cappellari:2006}.

The other notable property of the Fundamental Plane is its remarkably
small intrinsic scatter or \emph{thickness}, which has enabled its use
as a distance indicator for early-type galaxies. The intrinsic scatter
in the distance-dependent quantity, $R_{e}$, is measured to be as small
as 10--15\%, although the effective precision of the distance estimator,
including observational errors, is typically 20--30\% (see discussion in
\S\ref{subsec:litcomp} and Table~\ref{tab:litcomp}).

Several authors
\citep{Scodeggio:1998,Bernardi:2003b,Hyde:2009,LaBarbera:2010b} have
detected a weak steepening of the slope in \vd\ (i.e.\ a decrease in
$a$) in redder passbands. This wavelength variation has also been
observed in near-infrared (NIR) FP samples
\citep[e.g.][]{Pahre:1998a,Jun:2008}, suggesting a variation of stellar
content (and $M/L$) along the FP. In contrast, the slope in \sblum\
(i.e.\ $b$) is found to be largely independent of wavelength.

The Fundamental Plane relation is often claimed to be `universal', in
the sense that the coefficients are similar for galaxies across
environments ranging from the low-density field to high-density clusters
\citep[e.g.][]{Jorgensen:1996,Pahre:1998b,Colless:2001,Reda:2005}.
However there are also suggestions in the literature that there are
mild, but statistically significant, environmental variations
\citep[e.g.][]{Lucey:1991a,deCarvalho:1992,Bernardi:2003b,DOnofrio:2008,LaBarbera:2010c}.
Any variation in the FP between field and cluster galaxies, or for
galaxies in clusters of different richness, would be interesting from
the point of view of the formation of early-type galaxies, but would
complicate the use of the FP as a distance indicator.

The structural similarity of elliptical (E) galaxies and the bulges of
lenticular (S0) and early-type spiral galaxies suggests that the latter
classes of object may also populate the FP \citep{Dressler:1987}, and
\citet{Jorgensen:1996} found that the FPs for E and S0 galaxies were
consistent. In contrast, galaxies with both bulge {\em and} disk
components have been observed to be offset from ellipticals on the FP
\citep{Bender:1992,Saglia:1993}. It is therefore important to examine
whether there are morphological variations in the observed FP, and (if
so) whether these are due to intrinsic differences between E's and the
bulges of S0's and early-type Sp's or to observational contamination of
the bulge parameters by the disk for the latter classes of galaxy. If
such morphological variation exists, for either reason, it would result
at some level in offsets and increased scatter of the FP, and increase
the systematic and random errors (respectively) in the estimated
distances and peculiar velocities.

More recent studies \citep{Graves:2009,LaBarbera:2010b} have focused on
the trends in FP space of stellar population parameters such as age and
metallicity. A separate paper in this series \citep{Springob:2012}
explores the variations of age and metallicity within the 6dFGS FP
sample, and looks for variations of the FP for galaxies with different
stellar populations.

One difficulty in comparing the results from different studies of the FP
is that physical variations can be mimicked by biases resulting from the
interaction of the fitting method with the sample selection criteria or
the complicated error dependencies in the data. The regression methods
typically used to fit the FP broadly fall in the category of linear
least squares, and minimise the residuals of one of the FP variables or
the residuals orthogonal to the plane. The type of least-squares
regression chosen is often determined by the focus of the study (e.g.\
regression on \er\ to estimate distances or regression on \sblum\ for a
stellar population study), though it is well-known that different
regression methods do not necessarily converge on a unique (or even
consistent) best fit, particularly if selection effects or correlated
measurement errors are not fully accounted for \citep{Hogg:2010}. This
tendency to use different regression techniques interchangeably has made
it challenging to compare the results of different FP studies, and in
some cases has led to conclusions that are either incorrect or
misleading.

There is also the additional question of whether the traditional FP
model of a 2D plane with Gaussian scatter is statistically robust or
truly representative of the distribution of galaxies in FP space.
\citet{Saglia:2001} have shown that a 3D Gaussian model provides a more
accurate (and therefore less biased) representation of the galaxy
distribution, at least for the large, bright, early-type galaxies in
most FP samples.

Given these considerations, we have developed a robust maximum
likelihood algorithm for fitting the galaxy distribution in FP space
with a 3D Gaussian model. Through simulations we compare this approach
to the usual least-squares regressions of a plane with Gaussian scatter,
and show that it is superior in virtually all respects: more versatile
in dealing with complex sample selection criteria and correlated
measurement errors, more robust against outliers and blunders in the
data, and providing unbiased and precise estimates of the FP parameters
and their uncertainties.

We apply this method to a sample of $\sim$10$^4$ early-type galaxies
drawn from the 6-degree Field Galaxy Survey. The 6dFGS is a combined
redshift and peculiar velocity survey of galaxies covering the entire
southern sky at $|b|>10^\circ$ \citep{Jones:2004,Jones:2005,Jones:2009}.
The FP sample consists of the brightest (highest S/N) ellipticals,
lenticulars and early-type spiral bulges in the 6dFGS volume out to
$cz=16,500$\,\kms. This sample will ultimately form the basis of the
6dFGS peculiar velocity survey (6dFGSv), with the broad aims of mapping the
density and velocity fields in the nearby Universe and providing tighter
constraints on a range of cosmological parameters \citep{Colless:2005}.

The paper is organised as follows. Section\,\ref{sec:mlfit} outlines the
general 3D Gaussian model and maximum likelihood algorithm that can be
used to fit any FP sample. Section\,\ref{sec:fpdata} describes the FP
sample data from the 6dFGS to which we apply our model. We establish the
validity of our methodology and determine the errors on the fits from
Monte Carlo simulations using mock samples described in
Section\,\ref{sec:genmocks}. The overall FP fit results are given in
Section\,\ref{sec:thefp}; variations of the FP with environment are
addressed in Section\,\ref{sec:fpenviro} and dependencies on galaxy
morphology in Section\,\ref{sec:fpmorph}. Various aspects of our
results are discussed in Section\,\ref{sec:discuss}, including: the
validity of modelling the FP as a 3D Gaussian; the interpretation of the
scatter about the FP and the proper estimation of distance errors; the
physical insights offered by studying the FP in $\kappa$-space; and the
significance of the trends of the residuals about the FP with
environment, morphology and stellar population. Throughout we assume a
flat $\Lambda$CDM cosmology with $\Omega_{m}=0.3$,
$\Omega_{\Lambda}=0.7$ and $H_0=100\,h$\,\kms\,Mpc$^{-1}$; this is only
used for converting between angular and physical scales, and in fact the
specific cosmology chosen makes little difference for this low-redshift
sample.

\section{Maximum Likelihood Gaussian Fit}
\label{sec:mlfit}

\subsection{Motivation} 

The Fundamental Plane relation is defined as 
\begin{equation}
\label{eq:fp}
\log R_e = a \log{\sigma_0} + b \log \langle I_e \rangle + c
\end{equation}
where the coefficients $a$ and $b$ are the {\it slopes} of the plane and
the constant $c$ is the {\it offset} of the plane. In this study we
employ units of $h^{-1}$\,kpc for effective radius $R_e$, \kms\ for
central velocity dispersion $\sigma_0$, and L$_{\sun}$\,pc$^{-2}$ for
mean surface brightness $\langle I_e \rangle$. We prefer to use \sblum\
rather than \sbmag\ (which is in units of mag\,arcsec$^{-2}$), so that
all our FP parameters are unscaled logarithmic quantities; this means
that the relative errors and scatter are directly comparable in all
axes. Throughout the rest of this paper we adopt an abbreviated notation
for the FP parameters: $r \equiv \log R_e$, $s \equiv \log \sigma_0$ and
$ i \equiv \log \langle I_e \rangle$. Hence we write the FP relation as
\begin{equation}
\label{eq:fpdef}
r = a s + b i + c ~.
\end{equation}

Traditional methods for deriving the coefficients of equation
\ref{eq:fp} have preferred using a form of linear regression that
involves minimising residuals in the direction of one of the FP axes
\citep{Dressler:1987}, or orthogonal to the plane itself
\citep{Jorgensen:1996}, or both \citep{Hyde:2009,LaBarbera:2010b}.
Least-squares is used for its simplicity and relatively fast numerical
implementation. However, such regression techniques can be biased by the
choice of variable they minimise, the unacknowledged properties of the
model they assume, the selection effects they fail to model, and the
(possibly correlated) uncertainties they do not include in the fit.
Simple regressions are thus likely to result in unreliable and biased
fits to the FP.

Specifically, we identify the dominant sources of bias in FP samples as
arising in general from: (i)~the model for the FP distribution and its
intrinsic scatter; (ii)~selection effects, in the form of both hard and
soft censoring of the sample; and (iii)~the measurement errors on all
three FP variables, which are often correlated.

(i)~\emph{FP distribution model:} As discussed above, a 3D Gaussian is a
simple and convenient model that empirically is found to be a better
match to the (censored) observed FP distribution of early-type galaxies
than the standard model of a 2D plane-surface with Gaussian scatter in
one direction (see \S\ref{subsec:mockalg}). The standard model
effectively assumes that galaxies uniformly populate the whole plane,
whereas the 3D Gaussian naturally accounts not only for the scatter
about the plane but also the distribution within the plane, at least for
the bright galaxies included in the 6dFGS sample and most others.

(ii)~\emph{Selection effects:} Censoring of the intrinsic FP
distribution is always present for observed FP samples, in both obvious
and not-so-obvious ways. If the fitting technique is to avoid biased
results due to censoring, it must account for all the selection effects.
These include both hard selection limits in FP variables (e.g.\ in
velocity dispersion due to the limiting instrumental resolution) or soft
(i.e.\ graduated) selection limits in any other observable or
combination of observables (e.g.\ the joint selection on size and
surface brightness due to the flux limit of a sample). Using maximum
likelihood fitting it is straightforward to incorporate these limits
(see \S\ref{subsec:selcutfit}); by comparison, for linear regressions it
is significantly more difficult to account for selection effects more
complex than a hard limit in one variable.

(iii)~\emph{Measurement errors:} The modelling of measurement errors in
a FP sample is complicated by the fact that galaxies have different
errors in all three of their FP parameters, and some of these errors are
significantly correlated (notably those in $r$ and $i$). Standard
least-squares regression only accounts for uncorrelated measurement
errors (and in naive applications, only measurement errors in one
parameter). However, a maximum likelihood approach can account exactly
for differing measurements errors and their correlations in a
straightforward way.

\subsection{Least-Squares Regression Bias}
\label{subsec:leastsq}

As discussed above, a maximum likelihood method is clearly to be
preferred in principle. However it does not necessarily follow in
practice that the limitations of the linear regression approach result
in significant biases when fitting the FP. We therefore illustrate the
consequences of using linear regressions to fit mock samples simulated
by drawing galaxies from a 3D Gaussian intrinsic FP and applying
realistic measurement errors and selection effects. The process of
creating these mock samples is outlined in Section \ref{subsec:mockalg}.

\begin{figure*}
\begin{minipage}{170mm}
\centering
\includegraphics[width=1.0\textwidth]{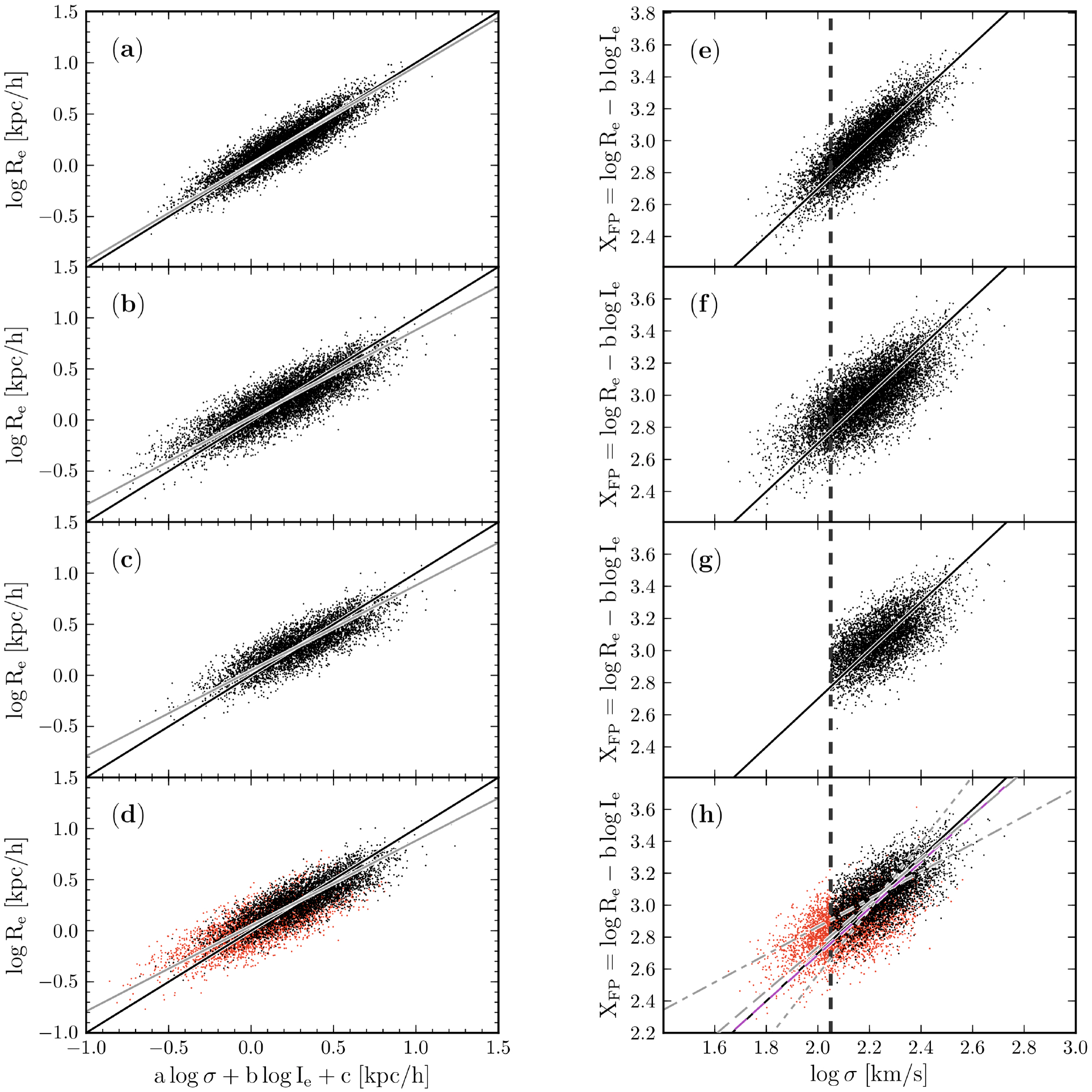} 
\end{minipage}
\caption{{\em Panels (a)--(d):} Comparison of the observed effective
  radius against predicted effective radius (calculated from equation
  \ref{eq:fpdef}) for mock samples all with the same underlying FP ($r =
  1.52s - 0.89i - 0.33$) and intrinsic scatter, but subject to differing
  levels of measurement errors and sample censoring: (a)~no measurement
  errors or censoring (\Ng\ $= 8901$); (b)~measurement errors but no 
  censoring (\Ng $= 8901$); (c)~both measurement errors and censoring
  (\Ng $= 5139$); (d)~as for (c) but with the censored data points shown
  in red (\Ng\ $= 8901$). Note that the sample is skewed from the
  one-to-one line (in black) by the measurement errors and the censoring
  of the sample, as indicated by the best-fit orthogonal regression
  lines for each sample (in grey). {\em Panels (e)--(h):} For the same
  mock samples as in (a)--(d), the correlation between the
  distance-dependent quantity, $X_\mathrm{FP} \equiv r - bi$, and the
  distance-independent quantity, $s \equiv$~\vd. The vertical dashed
  black line indicates the hard cut in \vd\ ($s\geq 2.05$) that is
  applied, along with other selection cuts, in censoring the mock
  samples in panels (g) and (h). In each panel the solid black line
  indicates the intrinsic FP that the mock samples were generated from;
  panel~(h) also shows as grey lines the standard least-squares
  regressions (in 2D) minimising with respect to $X_{\mathrm{FP}}$
  (dot-dash) and $s$ (dotted), and the orthogonal regression (dashed);
  the solid magenta line shows the maximum likelihood fit to a 3D
  Gaussian.}
\label{fig:xfpsig}
\end{figure*}

Three different types of mock samples were fit with each of the
commonly-used linear least-squares regressions (i.e.\ by minimising
residuals in the distance-dependent quantity, $X_\mathrm{FP} \equiv r -
bi$, or the distance-independent quantity, \vd $\equiv s$, or the
residuals orthogonal to the regression line) and also by a maximum
likelihood fit of a 3D Gaussian. In the lefthand panels of
Figure~\ref{fig:xfpsig} we compare the fits to these mocks using the
observed effective radius versus predicted effective radius (calculated
from equation \ref{eq:fpdef}). The simplest mock sample, panel~(a) and
panel~(e), is just the intrinsic distribution with no observational
errors or selection effects applied to it; consequently it is the
tightest sample and the best fit has almost no method-dependent bias.

However, when simulated observational error scatter is added to the mock
FP parameters, panel~(b), the sample is significantly skewed away from
the one-to-one line as a result of the systematic variation in the
observational errors with velocity dispersion, size and surface
brightness, as well as the correlation between the observational errors
in size and surface brightness. The skewing effect is exacerbated when
censoring is also present in the mock sample; panel~(c) shows the
situation where the censored data is absent, while panel~(d) is the same
but with the censored data shown in red (though still not included in
the fits). This censoring is the result of observational selection
effects operating on both velocity dispersion (due to the instrumental
spectral resolution limit) and jointly on size and surface brightness
(due to the sample apparent magnitude limit). The consequences of
this skewing of the sample distribution are illustrated in panels
(a)--(d) by the discrepancy between the 1-to-1 relation (black line) and
the best-fit orthogonal regression (grey line). The overall effect, shown
in panels (c) and (d), is that the best-fit slope is found to be 0.84
rather the true value of unity.

This biasing is also seen in the frequently used 2D projection of the FP
showing the distance-dependent photometric parameter, $X_{\mathrm{FP}}
\equiv r - bi$, and distance-independent spectroscopic parameter, $s
\equiv$~\vd. The righthand panels in Figure~\ref{fig:xfpsig} show this
projection for precisely the same mock FP samples as those in the
corresponding lefthand panels. The most obvious selection effect on the
mock sample in the righthand bottom panel is the velocity dispersion
limit, which censors the red points to the left of the vertical dashed
line at $s=2.05$ (i.e.\ \vd\,=\,112\,\kms). The red points to the right
of this line are those eliminated by the joint selection effect on $r$
and $i$ due to the apparent magnitude limit of the sample, which tends
to censor galaxies with smaller sizes and fainter surface brightnesses,
but in a way that depends on redshift.

These simulations show that the combined effect from all the selection
criteria and measurement errors skews the best fit when not accounted
for correctly (as is the case for least-squares fitting), most
noticeably for the regressions on $X_{\mathrm{FP}}$ and $s$. The
orthogonal fit (dashed grey line) fits the data well in this
projection, but this is a consequence of fixing the value of $b$, a
priori, to approximately the correct value. In this case, $b$ has been
fixed to the canonical value of $b=-0.75$; because this differs from
the input value of $b=-0.88$ for the mock sample, the fit deviates
from the input plane (particularly at the low-$\sigma$ end).
Additionally, Figure~\ref{fig:xfpsig} illustrates why the maximum
likelihood best fit does not appear, by eye, to be a good fit to the
observed data---the observational errors and the selection effects
systematically skew the observed sample away from the underlying
intrinsic distribution.

The conclusion from this exercise is that, for samples with realistic
observational errors and censoring, the input FP is best recovered with
the maximum likelihood method. Regressions on $X_{\mathrm{FP}}$ or $s$
lead to highly biased results, while the 2D orthogonal regression
gives a reasonable fit, at least for this particular combination of
observables, only if $b$ is fixed a priori close to the true value.
However, as shown below, regressions on $r$, $s$, $i$ and the orthogonal
residuals \emph{all} show significant biases when fitting the FP
parameters in 3D, and only the maximum likelihood method accurately
recovers the FP.

\begin{figure*}
\centering
\begin{minipage}{170mm}
\includegraphics[width=1.0\textwidth]{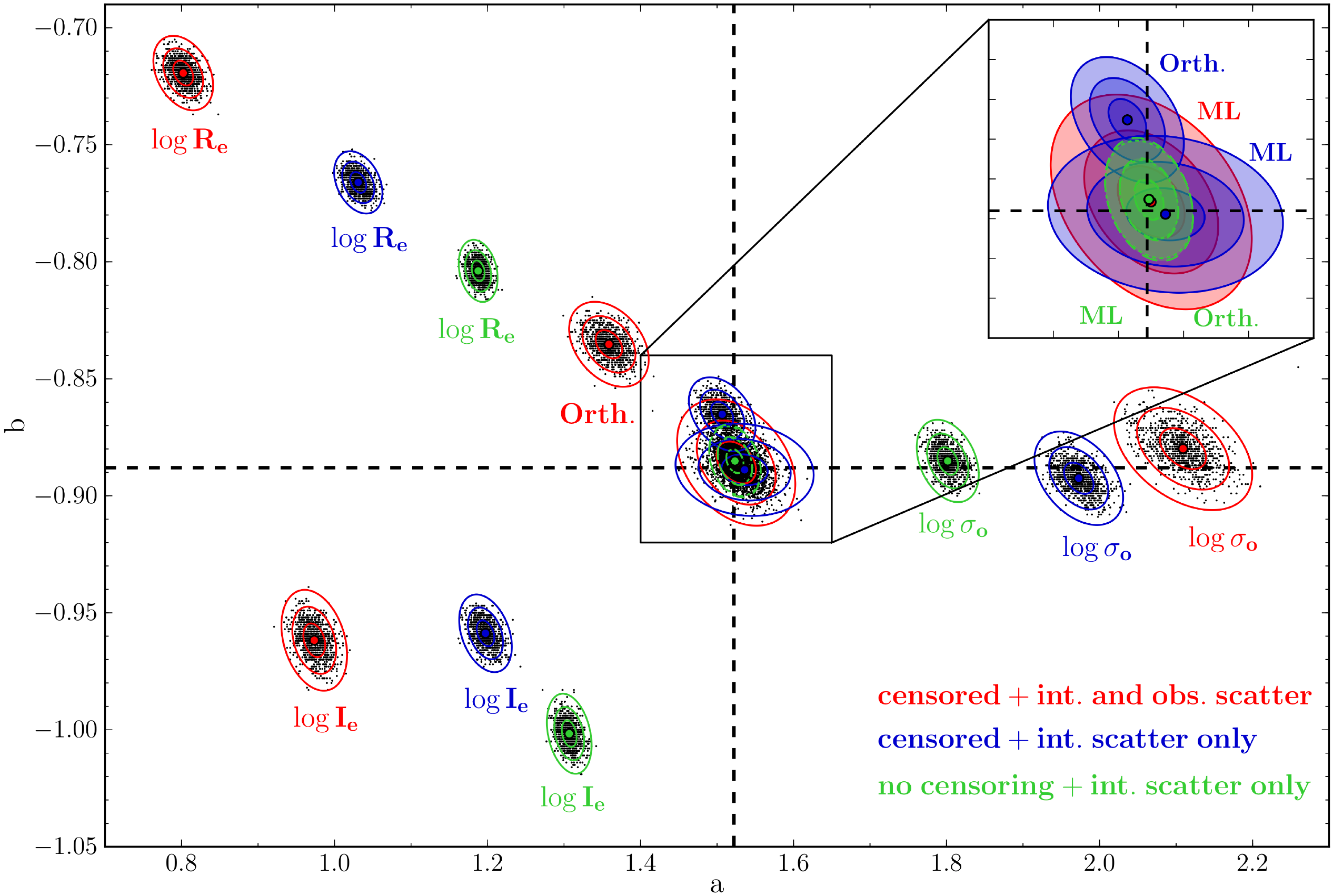}  
\caption{The best-fit values for the FP slopes, $a$ and $b$, for each of
  1000 mock FP samples (black dots) fit with least-squares regressions
  (in 3D) minimising the residuals in each of the three FP variables
  (i.e.\ $r, s, i$) and orthogonal to the plane; also fit with the
  maximum likelihood 3D Gaussian. The labels on each cluster of black
  points indicate the fitting method used; the colours indicate whether
  intrinsic scatter, observational errors and selection effects
  (censoring) are included in the mock samples, as follows: green
  indicates the mocks only include the intrinsic scatter of the FP; blue
  indicates the mocks include intrinsic scatter and censoring; red
  indicates the mocks include intrinsic scatter, observational scatter
  and censoring. The mean values of the fitted slopes (coloured dots)
  and the $1\sigma$, $2\sigma$ and $3\sigma$ contours (coloured
  ellipses) are over-plotted in the colour corresponding to the type of
  mock sample. The dashed lines indicate the input FP coefficients
  ($a=1.52$ and $b = -0.89$) from which all the mock samples were
  drawn.}
\label{fig:fiterror}
\end{minipage}
\end{figure*}

To illustrate the differences resulting from different fitting methods
in 3D and the impact of various problems with the real data, we fit
simulated samples with progressively more realistic properties (just as
in Figure~\ref{fig:xfpsig}). Figure \ref{fig:fiterror} shows the fitted
FP slope values ($a$ and $b$) for 1000 mock samples of various types
(each sample containing 8901 galaxies) using least-squares regression in
3D on each of the FP variables (i.e.\ $r, s, i$) and orthogonal to the
plane, as well as our 3D Gaussian model fitted using a maximum
likelihood method. In green are the results of fits to mocks just
including the intrinsic scatter of the FP; in blue are the fits to mocks
with both intrinsic scatter and sample censoring due to the selection
criteria; and in red are fully realistic mocks including all the
effects of intrinsic scatter, selection criteria and observational
errors.

The linear regressions on individual FP parameters give biased estimates
of $a$ and $b$ even for the `ideal' case (green), and become
progressively more strongly biased as censoring and observational errors
are included (blue and red). The \vd\ slope, $a$, is biased high, even
for the `ideal' case, when an FP sample is fit by minimising the \vd\
residuals as compared to the other fitting techniques. This is
consistent with previous studies \citep{Jorgensen:1996,LaBarbera:2010b}
and is a result of the dominant selection limit in \vd. The sense of the
trends in both $a$ and $b$ for all regression methods agree with those
found by \citet{Saglia:2001}, as shown in their Figure~6.

Figure \ref{fig:fiterror} also indicates that orthogonal regression (in
3D) is the least biased of the regression methods; however, in the most
realistic simulations (red), it nonetheless returns slopes that are
biased by many times the nominal precision of the fits (given by the
$1\sigma$ contour). The maximum likelihood fitting method clearly
out-performs all the regression methods, recovering the FP slopes
without significant bias for \emph{all} types of mock samples (see the
inset, which expands the region centred on the input values of the FP
slopes).

As might be expected, for all fitting methods the error contours on the
fitted slopes become larger when censoring and observational errors are
applied to the mock samples. Not so obviously, the error contours for
the most realistic mocks (red) are largest for the maximum likelihood
fit and the regression on $s$; the apparently greater precision of the
$r$, $i$ and orthogonal regressions are obtained at the expense of very
substantial biases in the fitted slopes. These regression fits thus give
a false sense of precision while at the same time introducing biases
that are many times larger than the nominal errors on the fitted slopes.

\subsection{3D Gaussian Likelihood Function}
\label{subsec:likelihood}

The Fundamental Plane is modelled as a three-dimensional Gaussian in a
similar fashion to the approach adopted by the EFAR survey
\citep{Saglia:2001,Colless:2001} and subsequently by
\citet{Bernardi:2003b}. This choice of model is justified by the good
empirical match it provides to the distribution of galaxies in FP space,
at least for samples limited by their selection criteria to larger,
brighter galaxies.

In one dimension the Gaussian probability distribution for a given
galaxy, $n$, is
\begin{equation}
P(x_n) = \frac{1}{\sqrt{2 \pi \sigma^2}} \exp{-\frac{(x_n-\bar{x})^2}{2 \sigma^2}}
\end{equation}
for a variable, $x_n$, with mean $\bar{x}$ and standard deviation
$\sigma$. Generalising this to three dimensions, the probability density
distribution, $P(\bf{ x_n})$, for a given galaxy, $n$, occupying the
position $\mathbf{x_n} = (r-\bar{r},s-\bar{s},i-\bar{i})$ in FP space
with respect to the mean values $\bar{r}$, $\bar{s}$ and $\bar{i}$ is
\begin{equation}
\label{eq:trigauss}
P({\bf x_n}) = \frac{ \exp[-\frac{1}{2}{\bf x^{T}_n}
               ({\bf \Sigma}+{\bf E_n})^{-1}{\bf x_n}] }
               { (2\pi)^{\frac{3}{2}} |{\bf \Sigma}+{\bf E_n}|^{\frac{1}{2}}f_n } 
\end{equation}
where $f_n$ is the normalisation factor accounting for the fact that,
due to selection effects, the galaxies do not fully sample the entire
Gaussian distribution. The total 3D scatter in FP space is given by the
addition of the FP variance matrix, $\mathbf{\Sigma}$ (specifying the
intrinsic scatter of the FP distribution in 3D) and the observational
error matrix $\mathbf{E_n}$ (specifying the observational errors in $r$,
$s$ and $i$ and their correlations; this is constructed in
\S\ref{subsec:errors}).

\begin{figure}
\centering
\includegraphics[width=0.5\textwidth]{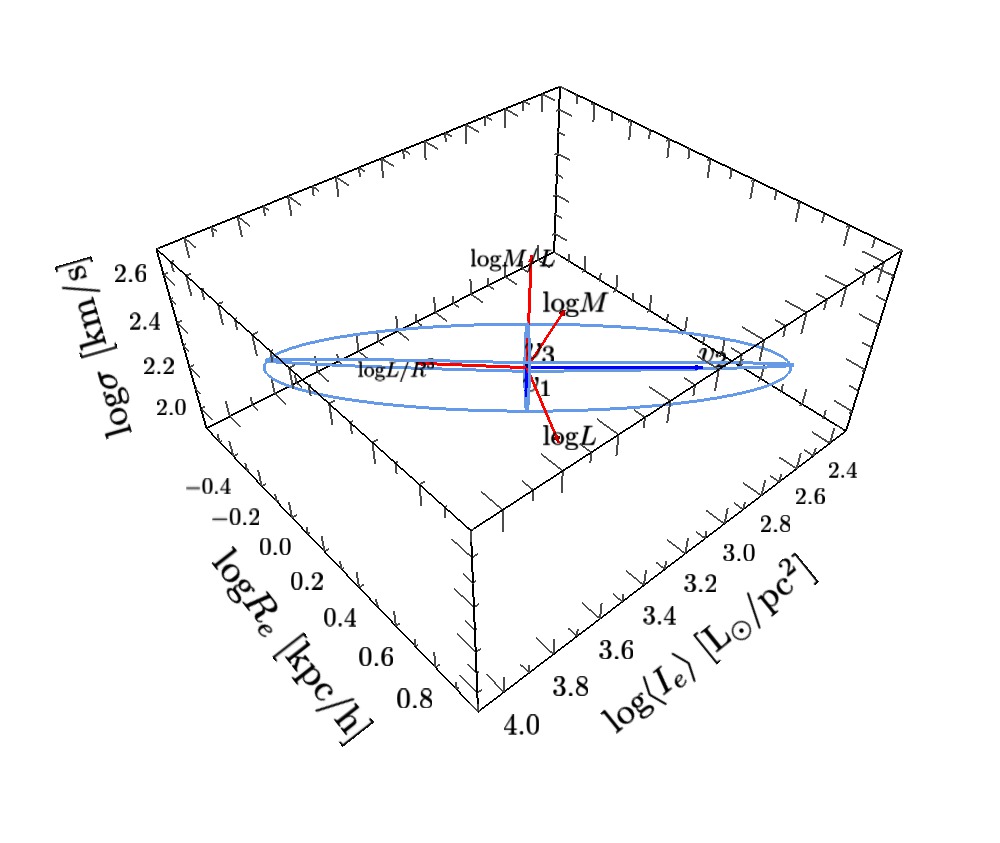} 
\caption{An interactive 3D schematic of FP space, showing the vectors
  ${\bf v_1, v_2, v_3}$ (in dark blue) that define the axes (see
  equation~\ref{eq:vectors}) of our Gaussian model ($3\sigma$ Gaussian
  ellipsoid in cyan) as they are oriented with respect to the three
  observational parameter axes ${\bf r, s, i}$. We also show (in red)
  the vectors corresponding to the physical quantities $\log M$, $\log
  L$, $\log M/L$ and $\log L/R^3$ as defined in
  \S\ref{subsec:likelihood}. We note that the angle between the vectors
  $\log M/L$ and ${\bf -v_1}$ and also $\log L/R^3$ and ${\bf -v_1}$ are
  both within 5$\degr$ of each other. (Readers using Acrobat Reader v8.0
  or higher can enable interactive 3D viewing of this schematic by mouse
  clicking on the version of this figure found in the ancillary files; see Appendix~\ref{sec:usage3D} for more
  detailed usage instructions.)}
\label{fig:3dvectors}
\end{figure}

The Fundamental Plane space can be described either in terms of the
observational parameters or in terms of the unit vectors showing the
principal axes of the 3D Gaussian characterising the galaxy distribution
(hereafter, $\mathbf{v}$-space). The Fundamental Plane itself is defined
by its normal vector, which is the eigenvector of the intrinsic FP
variance matrix $\mathbf{\Sigma}$ with the smallest eigenvalue. A
representation of the $\mathbf{v}$-space axes (${\bf v_1, v_2, v_3}$)
with respect to the axes of the observational parameters ($r$, $s$, $i$)
is shown in Figure~\ref{fig:3dvectors} as a 3D interactive visualisation
that can be accessed by viewing the version of this paper found in the ancillary files with Adobe Reader Version 8.0
or higher. All the interactive 3D figures in this paper were created
with custom C-code and the S2PLOT graphics library \citep{Barnes:2006},
using the approach described in \citep{Barnes:2008}.

The resulting vectors that define the axes of the Gaussian are
\begin{align}
\label{eq:vectors}
\mathbf{\hat{v}_1} &= (1/\sqrt{1 + a^2 + b^2}) \cdot \mathbf{v_1}~,  \notag \\
\mathbf{\hat{v}_2} &= (b/\sqrt{1 + b^2})\cdot \mathbf{v_2}~, \\
\mathbf{\hat{v}_3} &= (ab/\sqrt{(1 + b^2)(1 + a^2 + b^2)}) \cdot \mathbf{v_3}~, \notag
\end{align}
where
\begin{align}
\label{eq:vectors2}
\mathbf{v_1} &= \mathbf{\hat{r}} - a \mathbf{\hat{s}} - b \mathbf{\hat{i}}~,  \notag \\
\mathbf{v_2} &= \mathbf{\hat{r}} + \mathbf{\hat{i}}/b~, \\
\mathbf{v_3} &= -\mathbf{\hat{r}}/b - (1 + b^2)\mathbf{\hat{s}}/(ab) + \mathbf{\hat{i}}~, \notag
\end{align}
in terms of the FP slopes $a$ and $b$. These are the same axes defined
by \citet{Colless:2001} for the EFAR FP study, with the exception that
the value of $b$ quoted in this study is the coefficient of \sblum\
(with units of L$_{\sun}$\,pc$^{-2}$) rather than the coefficient of
$\langle \mu_{e} \rangle$ (with units of mag\,arcsec$^{-2}$) used in the
EFAR study, so that $b_{\mathrm{6dF}} = -2.5\,b_{\mathrm{EFAR}}$.

The direction of the short axis ($\mathbf{\hat{v}_1}$), which runs
\emph{through} (i.e.\ normal to) the plane, is fully determined by the
fitted slopes $a$ and $b$. The long axis ($\mathbf{\hat{v}_2}$), which
runs \emph{along} the plane, is fixed by being orthogonal to
$\mathbf{\hat{v}_1}$ and having no \vd\ component. Although this is
fixed by fiat, in fact (as we show in Section \ref{subsec:scomp}) this
is very close to the longest natural axis of the 3D Gaussian if no
constraints are placed on its direction. The advantage of this
definition of $\mathbf{\hat{v}_2}$ lies in its physical interpretation
as the direction within the FP that has no dynamical component,
connecting only the photometric parameters $r$ and $i$ . The third,
intermediate axis ($\mathbf{\hat{v}_3}$), which runs \emph{across} the
plane, is orthogonal to both $\mathbf{\hat{v}_1}$ and
$\mathbf{\hat{v}_2}$.

Figure~\ref{fig:3dvectors} also shows the relation between the
$\mathbf{v}$-space axes and the physical quantities of dynamical mass
($M$), luminosity ($L$), mass-to-light ratio ($M/L$) and luminosity
density ($L/R^3$). The logarithm of these quantities can be expressed as
a function of the FP parameters, under the assumption of homology, as $m
= r + 2s$ and $l = 2r + i$, where $m \equiv \log{M}$ and $l \equiv
\log(L)$. The logarithm of mass-to-light ratio is then simply $m - l =
-r + 2s - i$ and the logarithm of luminosity density is $l - 3r = -r +
i$. Therefore, in the case of the virial plane, where $a=2$ and $b=-1$,
the principal axes are aligned with these quantities:
$\mathbf{m}-\mathbf{l} = -\mathbf{v_1}$ and $\mathbf{l} - 3\mathbf{r} =
-\mathbf{v_2}$. Even for the actual \emph{tilted} FP we find the angle
between these vectors is small (our observed FP has $\mathbf{v_1}$
offset $5.0\degr$ from $\mathbf{m}-\mathbf{l}$ and $\mathbf{v_2}$ offset
$3.6\degr$ from $\mathbf{l}-3\mathbf{r}$).

The likelihood function, $\mathcal{L}$, is evaluated from the product of
the probability density function (equation~\ref{eq:trigauss}) for each
galaxy, $n$, using
\begin{equation} 
\label{eq:like}
\mathcal{L} = \prod_{n=1}^{N_{g}} P(\mathbf{x_n})^{1/S_n}~.
\end{equation}
The probability density function is weighted by the fraction of the
survey volume in which the galaxy could have been observed, which is
inversely proportional to the selection probability, $S_n$, depending on
the magnitude and redshift selection criteria imposed on the FP sample
(see \S\ref{subsec:selcutfit}). The probability is normalised over the
region of the FP space allowed by the selection criteria, so that 
$\int P(\mathbf{x})~\mathrm{d}^3\mathbf{x} = 1$.

For convenience, the log-likelihood value ($\ln\mathcal{L}$) is used, so
the product in equation~\ref{eq:like} can be reduced to a summation, and
then evaluated for our particular P($\mathbf{x_n}$):
\begin{equation}
\label{eq:likelihood}
\begin{split}
\ln\mathcal{L} = &-\sum_{n=1}^{N_{g}} S_n^{-1}[\frac{3}{2} \ln(2\pi)+\ln(f_n) \\
&+\frac{1}{2}\ln(|\mathbf{\Sigma}+\mathbf{E}_n|) 
  +\frac{1}{2}\mathbf{x^T}_n(\mathbf{\Sigma}+\mathbf{E}_n)^{-1}\mathbf{x}_n]~.
\end{split}
\end{equation}
The leading factor in the summation is the weight of the $n^\mathrm{th}$
galaxy, given by the inverse of its selection probability. Within the
square brackets, the first three terms are the normalisation of the
probability, and the final term is half the $\chi^2$.

\subsection{Likelihood Function Optimisation}
\label{subsec:likelihoodopt}

The log-likelihood of equation~\ref{eq:likelihood} is maximised to
simultaneously fit for the eight FP parameters that define the 3D
Gaussian model discussed in the preceding section. The parameters that
are derived from the fit are: the slopes of the plane ($a$ and $b$,
which define the directions of the 3D Gaussian's axes through equation
\ref{eq:vectors}); the centre of the 3D Gaussian in FP space ($\bar{r},
\bar{s}, \bar{i}$), which can be used to calculate the offset of the FP
($c=\bar{r}-a\bar{s}-b\bar{i}$); and the dispersion of the Gaussian in
each of the three axes ($\sigma_1, \sigma_2$, $\sigma_3$). The set of
parameters $\{ a, b, \bar{r}, \bar{s}, \bar{i}, \sigma_1, \sigma_2,
\sigma_3 \}$ that maximise the log-likelihood of the 3D Gaussian are
therefore those that define the best-fit model to the FP data. Note that
the offset of the FP, $c$, is defined in terms of these parameters as
$c=\bar{r}-a\bar{s}-b\bar{i}$.

The log-likelihood function is maximised with a non-derivative
multi-dimensional optimisation algorithm called BOBYQA \citep[Bound
Optimisation BY Quadratic Approximation;][]{Powell:2006}. BOBYQA is
found to be more robust and efficient than more generic optimisation
algorithms such as the Nelder-Mead simplex algorithm
\citep{Nelder:1965}. It performs well under the particular demands of FP
fitting, namely high dimensionality (simultaneous optimisation of eight
parameters) and large sample size ($\sim$10$^4$ galaxies). The
parameters in the BOBYQA algorithm that can be tuned to suit the
particular function being optimised are the initial and final
tolerances, $\rho_{beg}$ and $\rho_{end}$, and the number of
interpolation points between each iteration, $N_{int}$. After
considerable experimentation, values of these parameters that were found
to be efficient and to give the required accuracy were $\rho_{beg} =
10^{-1}$, $\rho_{end} = 10^{-5}$ and $N_{int} = 30$. The BOBYQA
algorithm with these parameters was used for all the fitting presented
in this work.
 
\subsection{Selection Criteria and Fitting}
\label{subsec:selcutfit}

Fundamental Plane studies must employ some form of model to analyse
censored or truncated data resulting from observational selection
effects. If these models fail to account for statistical effects that
are due to selection, they run the risk of biasing the fitting method
being used to recover the FP. We now describe the dominant selection
limits---both hard and graduated---that pertain to FP data and how a
maximum likelihood fitting method can account for this censoring in a
straightforward and transparent manner.

A central velocity dispersion lower limit is typical of FP surveys,
which are unable to measure dispersions accurately below the
instrumental resolution of the spectrograph. Because this limit is
applied to just one of the FP parameters (i.e.\ $s$), the appropriate 3D
Gaussian normalisation is calculated by integrating over the volume of
the distribution that remains after the velocity dispersion cut, as
outlined in Appendix~\ref{sec:likenorm}. In this way the likelihood is
appropriately normalised and the maximum likelihood method correctly
accounts for the truncation of the FP in velocity dispersion by this
hard selection limit.

Most FP samples are drawn from flux-limited surveys, excluding galaxies
fainter than some apparent magnitude limit. This selection effect can be
accounted for by weighting the individual likelihood of each galaxy by
the inverse of its selection probability $S$; this is analogous to a
$1/V_{max}$ weighting \citep{Schmidt:1968}.

For the case of a FP survey with explicit redshift limits, the selection
probability is proportional to the fraction of the survey volume between
these limits over which a particular galaxy can be observed given the
survey's apparent magnitude limit. This is a function of the limiting
distance $D^{lim}_n$ (in $h^{-1}$\,Mpc) out to which the galaxy $n$,
with an absolute magnitude $M_n$, can be observed given the survey
magnitude limit $m^{lim}$ in a given passband, and can be calculated as
\begin{equation}
\label{eq:dlim}
D^{lim}_n = 10^{0.2(m^{lim} - M_n - 25)}~.
\end{equation}
If the redshift $cz^{lim}_n$ corresponding to this luminosity distance
is larger (smaller) than the high (low) redshift limit of the survey,
$cz_{max}$ ($cz_{min}$), then a galaxy with that absolute magnitude will
definitely have been observed (or not) and the selection probability is
$S = 1~(0)$. However, if $cz^{lim}_n$ is between the minimum and maximum
survey redshifts, then the selection probability is given by the
fractional co-moving volume in which it \emph{could} be observed given
the apparent magnitude limit. Therefore the selection probability
function is
\begin{equation}
\label{eq:sprob}
S_n =\begin{cases}
1 & cz^{lim}_n \geq cz_{max} \\
\frac{V(cz^{lim}_n)-V(cz_{min})}{V(cz_{max})-V(cz_{min})} & cz_{min}<cz^{lim}_n<cz_{max} \\
0 & cz^{lim}_n \leq cz_{min}
\end{cases}
\end{equation}
where $V(cz)$ is the co-moving volume of the survey out to redshift
$cz$. This definition of $S_n$ is similar to the selection probability
of the EFAR survey, although their selection probability function was
based on a size parameter rather than absolute magnitude
\citep{Saglia:2001}.

In addition to these selection effects, a FP sample may contain spurious
outliers whose significance is best characterised by a $\chi^2$ value.
The $\chi^2$ for a particular galaxy $n$ can be calculated as
\begin{equation}
\label{eq:chisq}
\chi^2_n =
\mathbf{x_n^T}(\mathbf{\Sigma}+\mathbf{E}_n)^{-1}\mathbf{x}_n ~.
\end{equation}
Note that this is twice the exponent of the Gaussian probability
distribution of equation~\ref{eq:trigauss} and appears in the final term
of equation~\ref{eq:likelihood}. Thus $\chi^2$ measures the departure of
a galaxy in FP-space from a given 3D Gaussian model, and outliers can be
identified and removed based on their extreme (and extremely unlikely)
values of $\chi^2$. The refined sample, excluding these high-$\chi^2$
outliers, can then be re-fitted to achieve an improved fit that is not
biased by outliers.

\section{6dFGS Fundamental Plane Data and Sample}
\label{sec:fpdata}

\subsection{Fundamental Plane data}
\label{subsec:fpdata}

The 6-degree Field Galaxy Survey (6dFGS) provides a comprehensive census
of galaxies and measured redshifts in the Southern hemisphere out to a
depth of $z\sim0.15$ \citep{Jones:2004,Jones:2005}. Primary targets were
selected from the $K$ band photometry of the Two Micron All Sky Survey
(2MASS) Extended Source Catalog \citep{Jarrett:2000}, with secondary
samples selected to approximately equivalent limits in the 2MASS $J$ and
$H$ bands and the SuperCOSMOS \citep{Hambly:2001} $r_{\mathrm{F}}$ and
$b_{\mathrm{J}}$ bands. The total apparent magnitude limits of the 6dFGS
are $(K, H, J, r_{\mathrm{F}}, b_{\mathrm{J}}) \leq (12.65, 12.95,
13.75, 15.60, 16.75)$. The survey extends across the entire southern sky
and, because of its near-infrared selection, reaches down to 10 degrees
from the Galactic Plane in $J$, $H$ and $K$ (for $b_{\mathrm{J}}$ and
$r_{\mathrm{F}}$ the survey reaches down to 20 degrees from the Galactic
Plane). It is the largest combined redshift \emph{and} peculiar velocity
survey by a factor of two, with the additional advantage of homogeneous
sampling of the galaxy population over a large volume of the local
universe.

We initially select galaxies suitable for the 6dFGS peculiar velocity
subsample (6dFGSv) from the parent redshift survey sample (6dFGSz) by
selecting galaxies with reliable redshifts (i.e.\ redshift quality
$Q=3-5$) and redshifts less than 16500\,\kms\ (i.e.\ $z < 0.055$). The
redshift limit is imposed because at higher redshifts the key spectral
features used to measure \vd\ are shifted out of the wavelength range
for which sufficiently high resolution spectra are available
\citep{Campbell:2009}. These criteria select $\sim$43,000 of the
$\sim$125,000 galaxies in the 6dFGS redshift survey.

The spectra of these galaxies is classified by matching the observed 
spectrum, via cross-correlation, to template galaxy spectra. The sample
 only includes galaxies with spectra that, within the 6dF fibre region, are a better
match to early-type spectral templates (E/S0 galaxies) than to late-type
templates (Sbc or later). The sample can therefore be characterised in
spectral terms as galaxies that, within the 6dF fibre region, have
dominant old stellar populations with little or no ongoing
star-formation. Morphologically the sample galaxies are either
ellipticals and lenticulars or early-type spirals with the bulge filling
the 6dF fibre.

These $\sim$20,000 early spectral type galaxies had their central
velocity dispersions measured using the Fourier cross-correlation
technique \citep{Campbell:2009}. These velocities were then corrected
for the effect of the fibre aperture size to a uniform \er/8 aperture
following the formula of \citet{Jorgensen:1995}. The sample of galaxies
with early-type spectra, sufficiently high signal-to-noise ratio for
reliable velocity dispersion measurements ($S/N > 5$\,\AA$^{-1}$), and
velocity dispersions greater than the instrumental resolution limit ($s
\geq 2.05$, i.e.\ $\sigma_0 \geq 112$\,\kms) contains 11,561 galaxies.

The FP photometric parameters (R$_{e}$ and \sbmag) for our sample were
derived from 2MASS. The relatively large 2MASS point-spread function
(PSF), with FWHM\,$\approx$\,3.2$''$, required a procedure to derive
PSF-corrected parameters. For each galaxy we analysed the pixel data
provided by the 2MASS Extended Source Image Server as follows. We
adopted the apparent magnitude ($m$) measured by 2MASS from the `fit
extrapolation' method (i.e.\ {\tt j\_m\_ext}, {\tt h\_m\_ext}, {\tt
  k\_m\_ext}) and determined the circular apparent half-light radius
(r$_{\rm APP}$) of the target galaxy on the 2MASS image. A model 2D
Gaussian PSF image was derived from stars on the parent 2MASS data
`tile'. GALFIT \citep{Peng:2002} was run with the galaxy image and model
PSF image as inputs to find the best-fit 2D S\'{e}rsic model. The
half-light radius was determined for the S\'{e}rsic model before and
after convolution with the PSF (r$_{\rm MODEL}$ and r$_{\rm SMODEL}$).
The difference r$_{\rm SMODEL}$$-$r$_{\rm MODEL}$ is subtracted from
r$_{\rm APP}$ to derive the PSF-corrected half-light radius (i.e.\ the
effective radius $R_{e}$). The effective radius was observed in angular
units of arcseconds, $R_{e}^{\theta}$, and converted to physical units
of $h^{-1}$\,kpc, $R_{e}$, using the galaxy's angular diameter distance,
$D_{A}(z)$.

The 2MASS data for the $J$, $H$ and $K$ bands were analysed
independently. As the 2MASS PSF is well-determined and we only use the
S\'{e}rsic model to provide the PSF-correction, this procedure is very
robust. The effective surface brightness (\sbmag) is derived via
$\langle \mu_{e} \rangle = m + 2.5\log(2 \pi R_e^2)$. Additionally, each
surface brightness was corrected for the effects of surface brightness
dimming and Galactic extinction, and also K-corrected for the effect of
redshift on the broadband magnitudes \citep{Campbell:2009}.

\begin{figure}
\centering
\includegraphics[width=0.45\textwidth]{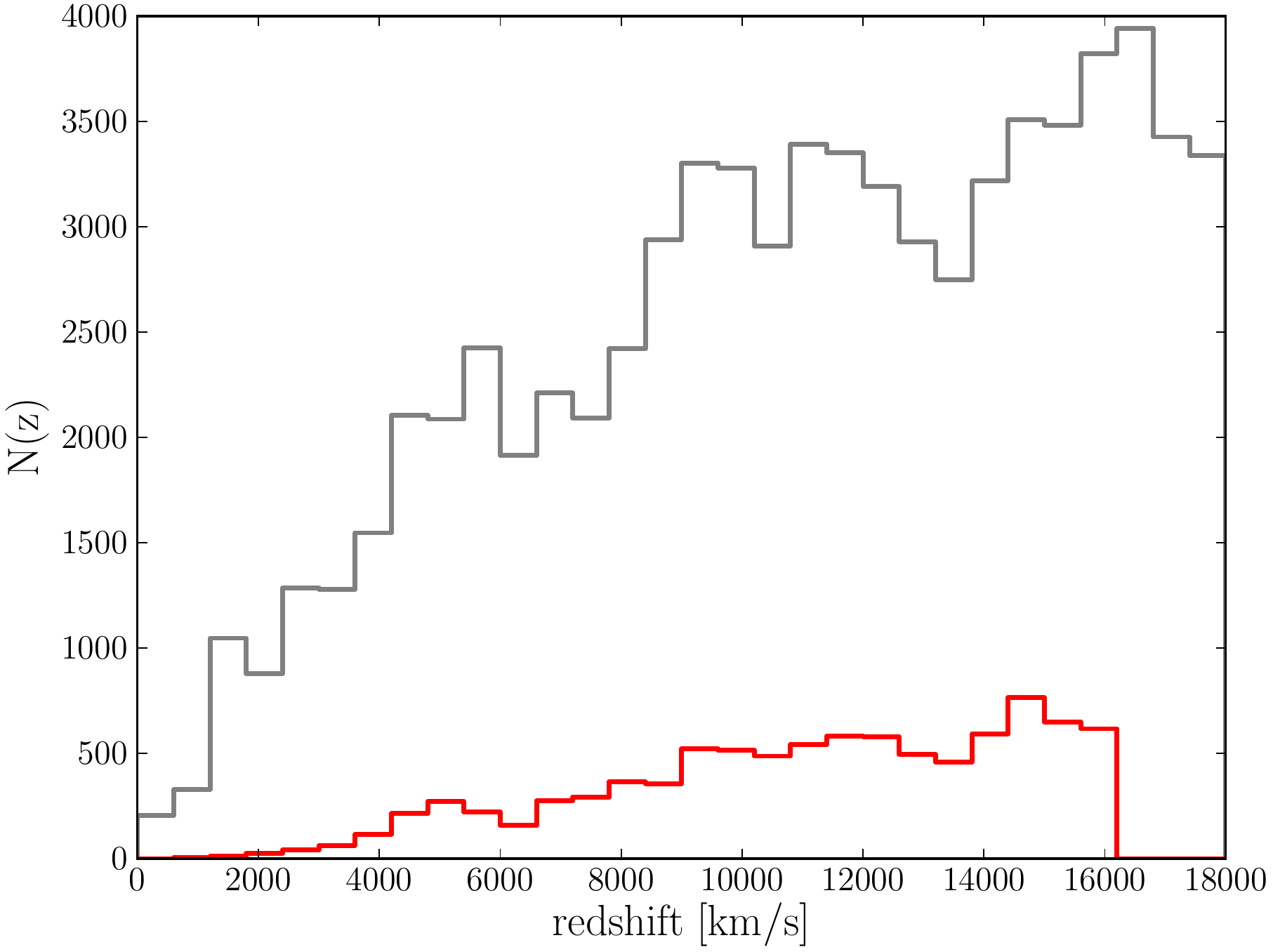} 
\caption{Redshift distribution of the 6dFGSv FP sample (red;
  $N_{g}$=8901) with maximum redshift $cz_{\mathrm{max}}$=16120\,\kms\
  compared to the full 6dFGS redshift sample (grey; $N_{g}$=124646).}
\label{fig:redshift}
\end{figure}

It is most natural to have all FP parameters in logarithmic units, so
surface brightness values were converted from magnitude units (i.e.\
\sbmag\ in mag\,arcsec$^{-2}$) to log-luminosity units (i.e.\ \sblum\ in
L$_{\sun}$\,pc$^{-2}$) using
\begin{equation}
\log \langle I_{e} \rangle = 
0.4 M_{\sun} - 0.4 \langle \mu_{e} \rangle + 8.629 ~,
\end{equation}
where the absolute magnitude of the Sun, $M_{\sun}$, depends on the
passband. For the $J$ band, $M_{\sun}$=3.67; for the $H$ band,
$M_{\sun}$=3.33; and for the $K$ band, $M_{\sun}$=3.29.\footnote{The
  values for the absolute magnitude of the Sun quoted here are from
  http://mips.as.arizona.edu/$\sim$cnaw/sun.html.} The value used for the
magnitude of the Sun does not impact the fit, however, as it is simply a
constant offset that is applied to the surface brightness.

Finally, the FP sample with both spectroscopic measurements from 6dFGS
and photometric measurements from 2MASS in three $J$, $H$ and $K$ bands
contains 11287 early-type galaxies in total. In Figure~\ref{fig:redshift}
we show the 6dFGSv redshift distribution (red), which is truncated at
$cz_{max} = 16120$\,\kms\ (a limit we apply as described in
Section~\ref{subsec:selcuts}). This maximum redshift is approximately at
the median redshift of the full 6dFGS redshift sample (grey); the 6dFGSv
galaxies are sampled across this entire redshift range.

\subsection{Selection Function}
\label{subsec:selcuts}

\begin{table*}
\begin{minipage}{170mm}
\centering
\caption{Summary of the 6dFGS Fundamental Plane sample selection
  criteria. The criteria apply to central velocity dispersion $s$,
  redshift distance $cz$ (upper and lower limits), morphology, apparent
  magnitude $m$, selection probability $S$, and $\chi^2$. The column
  N$_{exc}$ shows the number of galaxies that would be removed by the
  specified selection cut \emph{alone}. However, the number in brackets
  for each subtotal (or total) is the actual number of galaxies excluded
  when multiple selection limits are \emph{combined} (i.e.\ without
  double-counting the galaxies that are eliminated by more than one
  selection criterion).}
\begin{tabular}{@{}lllll@{}}
\hline
Sample       &  Selection Limit   & N$_g$          & N$_{exc}$   & Comments \\ \hline                        
6dFGSz       &                    & 124646         &             & full redshift sample (with good quality $z$) \\ 
\hline                 
6dFGS$_{\rm FP}$ &                 & 11287          &             & galaxies with derived FP parameters \\ 
\hline
6dFGSv       & $s  \geq 2.05$     &                & 287         & aperture-corrected $s$ cut \\
             & $cz \geq 3000$*    &                &  92         & lower $cz$ limit \\
             & $cz \leq 16120$*   &                & 750         & upper $cz$ limit \\
             & Morphology         &                & 429         & flagged classification (\S\ref{subsec:morphdat}) \\
             & SUBTOTAL:          & 9794           & 1558 (1493) & \\
\hline                 
6dFGSv$_{J}$ & $J \leq 13.65$     &                & 1024        & \\
             & $S \geq 0.05$      &                & 32          & \\
             & $\chi^2 \leq 12$   &                & 48          & \\
             & TOTAL:             & 8901           & 1104 (893)  & $J$ band FP sample \\
\hline          
6dFGSv$_{H}$ & $m \leq 12.85$     &                & 1427        & \\
             & $S \geq 0.05$      &                & 41          & \\
             & $\chi^2 \leq 12$   &                & 45          & \\
             & TOTAL:             & 8568           & 1513 (1226) & $H$ band FP sample \\
\hline                 
6dFGSv$_{K}$ & $m \leq 12.55$     &                & 1398        & \\
             & $S \geq 0.05$      &                & 32          & \\
             & $\chi^2 \leq 12$   &                & 46          & \\
             & TOTAL:             & 8573           & 1476 (1221) & $K$ band FP sample \\
\hline
\multicolumn{5}{l}{*Note: these galaxies are excluded from the fitting
  of the  FP, but are included when deriving FP} \\ 
\multicolumn{5}{l}{distances and peculiar velocities.}
\end{tabular}
\label{tab:selcuts}
\end{minipage}
\end{table*}

In our FP analysis there are selection limits imposed on or inherent in
the sample that the fitting model must incorporate to provide accurate
FP coefficients. In \S\ref{subsec:selcutfit}, we explained how these
limits are included in our model, and now we provide the specific
details of the selection criteria for the 6dFGSv data, as summarised in
Table~\ref{tab:selcuts}.

The 6dFGS FP sample of 11287 galaxies \citep{Campbell:2009} has a
velocity dispersion limit ($s \geq 2.05$) that is set by the
instrumental resolution of the V band 6dFGS spectra. This limit is only
achieved for galaxies with observed redshifts $cz < 16500$\,\kms, since
at higher redshifts crucial spectral features such as Fe 5270\AA, Mg\,b
5174\AA\ and H$\beta$ 4861\AA\ begin to move out of the V band spectra
and into the lower resolution R band spectra. For the 6dFGS peculiar
velocity sample we in fact impose a stricter upper redshift limit of $cz
\leq cz_{max} = 16120$\,\kms\ in the CMB frame in order to avoid an
asymmetry on the sky when redshifts are converted from the heliocentric
frame to the CMB frame (which we use for the peculiar velocities). This
upper redshift limit for the sample excludes 750 galaxies.

We also only include galaxies with CMB-frame redshifts high enough ($cz
\geq cz_{min} = 3000$\,\kms) that their peculiar velocities are not
significant relative to their recession velocities and so do not
appreciably increase the scatter about the FP. This removes a further 92
low-redshift galaxies from the sample. However, unlike other selection
criteria, galaxies excluded from the FP fitting by these upper and lower
redshift limits are re-instated in the sample when deriving distances
and peculiar velocities.

The morphologies and spectra of all the galaxies in the FP sample were
classified by eye, as described in Section \ref{subsec:morphdat}. Based
on this visual inspection, 429 galaxies were removed on the basis of
their morphological type, contamination of their fibre spectrum by a
disk component, the (real or apparent) merger of their image with stars
or other galaxies, or discernible emission line features in their
spectra.

Our sample has slightly brighter flux limits than the original 6dFGS
magnitude limits \citep{Jones:2009}, reflecting the changes in the 2MASS
(and, consequently, 6dFGS) magnitude limits that occurred after the
6dFGS sample was selected. To maintain high completeness in each
passband over the whole sample area, we impose magnitude limits of $J
\leq 13.65$, $H \leq 12.85$ and $K \leq12.55$. At fixed luminosity
distance, the magnitude limit is a strict cut in the $r$--$i$ plane;
given the distance range of the sample, this flux limit translates into
a graduated selection effect in the $r$--$i$ plane. In fitting the FP
distribution we can account for the galaxies excluded by this selection
effect by weighting the likelihood of each galaxy with a selection
probability as described in \S\ref{subsec:selcutfit}.

\begin{figure}
\centering
\includegraphics[width=0.45\textwidth]{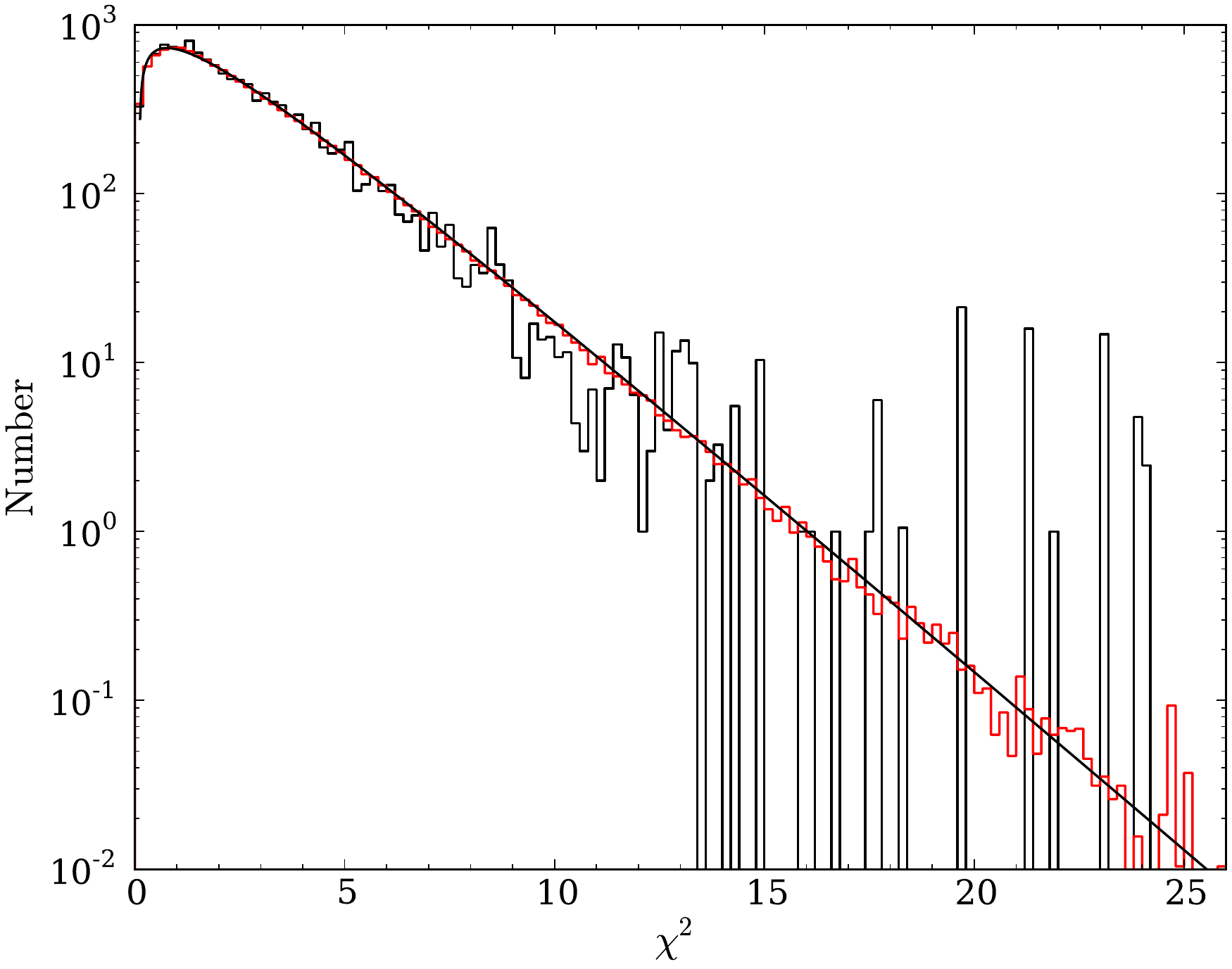} 
\caption{The distribution of $\chi^2$ for the galaxies in the
    observed $J$ band FP sample (black) and for mock galaxies in a
    sample drawn from the best-fitting 3D Gaussian model (red). The
    smooth curve is an analytic $\chi^2$ distribution with 2.65 degrees
    of freedom, derived by fitting to the mock sample (there are fewer
    than 3 degrees of freedom due to the censoring of the 3D Gaussian by
    selection effects).}
\label{fig:chisqdist}
\end{figure}

Finally, in order to reduce the impact on the fit from a small number of
galaxies with extremely low selection probabilities, we impose a minimum
selection probability requirement ($S \geq 0.05$; see
equation~\ref{eq:sprob}). We also remove outliers and blunders by
requiring $\chi^2 \leq 12$. This $\chi^2$ limit was derived
empirically by comparing the $\chi^2$ distributions for the observed
galaxies and for mock galaxies drawn from the best-fitting 3D Gaussian
model, as illustrated in Figure~\ref{fig:chisqdist} for the $J$ band
sample (the $H$ and $K$ band samples are very similar). The number of
observed galaxies at a given $\chi^2$ begins to exceed the number of
mock galaxies for $\chi^2>12$, which we attribute to outliers or
blunders. The $J$, $H$ and $K$ samples have 48, 45 and 46 galaxies
respectively above this limit (see Table~\ref{tab:selcuts}), so we are
typically removing just 0.5\% of each sample.

The selection probability requirement is the only sample selection
criterion that induces a significant residual bias, because it is the
only one not accounted for in the normalisation of the probability
distribution when computing the likelihood. We therefore correct for the
(small) residual biases it produces by calibrating its impact using mock
FP samples, as described in \S\ref{subsec:mockalg}.

After applying all these selection criteria to obtain the samples to
which we fit the FP, the numbers of galaxies remaining in each of the
passbands are 8901 ($J$ band), 8568 ($H$ band) and 8573 ($K$ band). The
numbers of galaxies for which we can derive peculiar velocities are
somewhat larger, since we can reinstate at least the galaxies excluded
by the lower redshift limit.

\subsection{Measurement Uncertainties}
\label{subsec:errors}

Each galaxy in the FP sample has an associated uncertainty from the
measurement errors in each of its FP observables: size, velocity
dispersion and surface brightness. The treatment of these errors is
often simplified or approximated when fitting the FP---e.g.\
\citet{LaBarbera:2010b} use mock galaxy samples to approximate errors
and correlations. However the maximum likelihood method allows us to to
deal with the errors in all the observables (and their correlations) in
a straightforward manner (see \S\ref{subsec:likelihood}). For galaxy
$n$, the measurement uncertainties are included through the error
matrix, $\mathbf{E}_n$, given by
\begin{equation}
\label{eq:errmatrix}
\mathbf{E}_n = \left( \begin{array}{ccc}
\epsilon^2_{r_n} + \epsilon^2_{rp_n} & 0 & \rho_{ri} \epsilon_{r_n} \epsilon_{i_n} \\
0 & \epsilon^2_{s_n} & 0 \\
\rho_{ri} \epsilon_{r_n} \epsilon_{i_n}  & 0 & \epsilon^2_{i_n} ~.
 \end{array} \right)
\end{equation}
The quantities $\epsilon_r$, $\epsilon_s$ and $\epsilon_i$ are the
observational errors on the FP parameters $r$, $s$ and $i$, and their
estimation is discussed in \citet{Campbell:2009}. 

The errors in the velocity dispersions, $\epsilon_s$, are based on the
\citet{Tonry:1979} formula derived for the Fourier cross-correlation
technique, and are dependent on the measured signal-to-noise in the
cross-correlation peak. These error estimates are validated by the large
number of repeat velocity dispersion measurements in the 6dFGS sample.
The typical error on the velocity dispersions in the 6dFGS FP sample is
around 0.054\,dex or 12\%.

The photometric errors, $\epsilon_r$ and $\epsilon_i$, were
estimated by studying the scatter when comparing the sizes and surface
brightnesses obtained from the three independent 2MASS passbands. We
assume that the surface brightness colours (i.e.\ the values of
$i_j-i_h$, $i_j-i_k$, and $i_h-i_k$) are very similar for every galaxy
within each narrow range of apparent magnitude, and that the dominant
cause of variation from one galaxy to the next is the error in the
surface brightness measurements. We then compute the mean square
deviation in surface brightness colour for the $J$ and $H$ bands,
$\delta_{jh}^2$, over the $N$ galaxies within a specified apparent
magnitude bin, given by
\begin{equation}
\delta_{jh}^2 = (\Sigma_{n=1,N} [(i_{j,n} -i_{h,n}) - <i_j - i_h>]^2)/N ~.
\end{equation}
If we assume that $\delta_{jh}^2$ is the sum of the mean square
errors in $i_j$ and $i_h$, and that $\delta_{jk}^2$ and $\delta_{hk}^2$
are likewise the sum of the mean square errors in $i_j$ and $i_h$, and
$i_h$ and $i_k$, respectively, then we can solve for the error in $i_j$
alone, obtaining
\begin{equation}
\epsilon_{i,j} = [0.5(\delta_{jh}^2 + \delta_{jk}^2 - \delta_{hk}^2)]^{1/2} ~.
\end{equation}
This is the error on $i_j$, which we compute separately in apparent
magnitude bins of width 0.2\,mag. We similarly compute $\epsilon_{i,h}$
and $\epsilon_{i,k}$, shifting the magnitude bins by the mean colour of
the galaxies in the sample, to get the surface brightness errors in each
band as a function of apparent magnitude.

\begin{figure}
\centering
\includegraphics[width=0.45\textwidth]{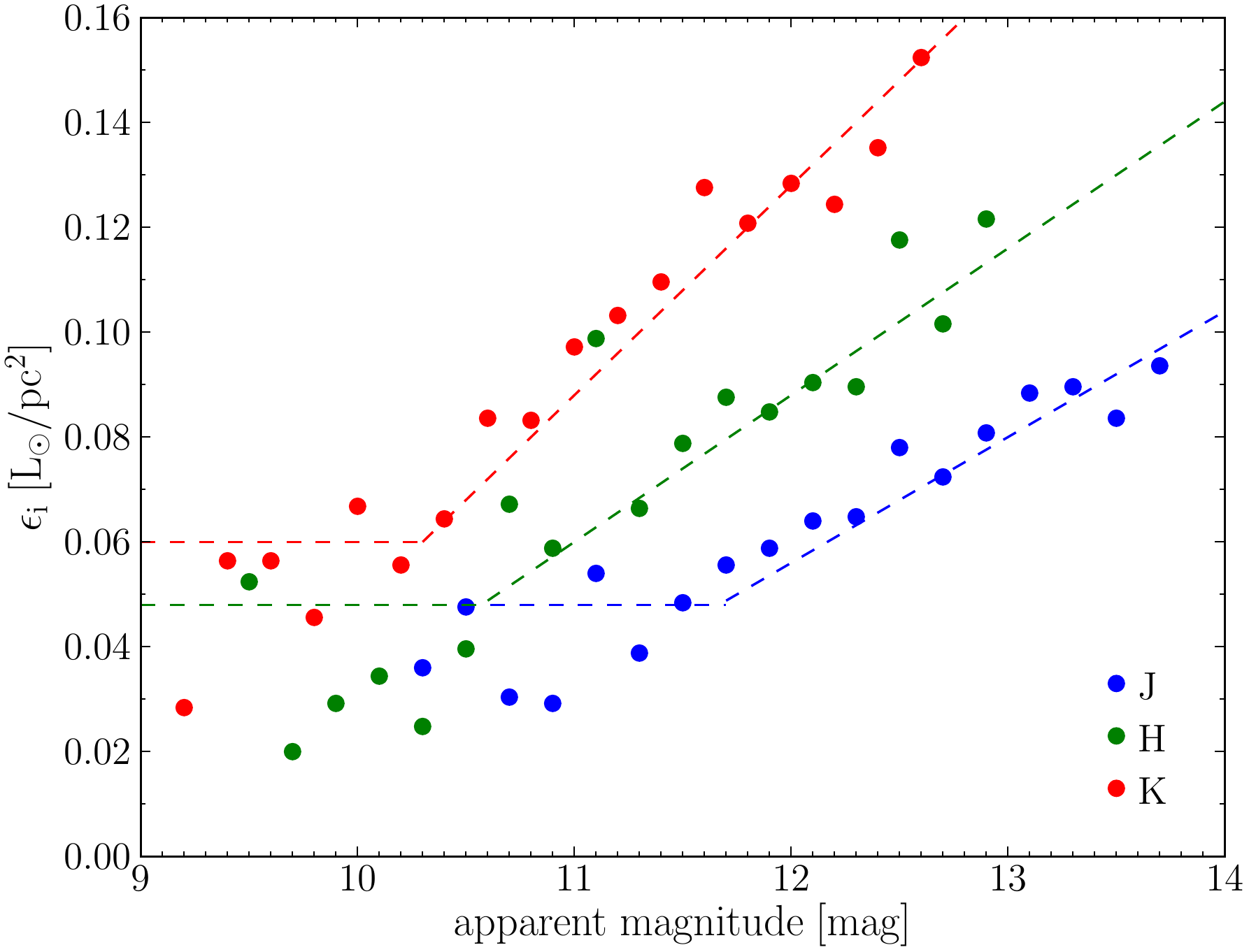} 
\caption{The blue (green, red) points show the derived measurement
    error on $i_j$ ($i_h$, $i_k$) as a function of $m_J$ ($m_H$, $m_K$).
    The measurement errors, $\epsilon_i$, are in units of
    $\log$[L$_{\sun}$\,pc$^{-2}$]. We approximate these measurement error
    relations by the dashed lines of the corresponding colours, which
    are specified by equation~\ref{eq:errests}).}
\label{fig:sberrs}
\end{figure}

Figure~\ref{fig:sberrs} shows the $J$, $H$, and $K$ band surface
brightness errors as a function of $J$, $H$, and $K$ apparent magnitude.
We approximate the errors using the following relations, which are shown
as dashed lines in Figure~\ref{fig:sberrs}:
\begin{align}
\label{eq:errests}
\epsilon_i &= \begin{cases}
0.024~m_{J} - 0.232 &  m_{J} \geq 11.7 \\
0.048 &  m_{J} < 11.7 \\ 
\end{cases} \notag \\
\epsilon_i &= \begin{cases}
0.028~m_{H} - 0.248 &  m_{H} \geq 10.6 \\
0.048 &  m_{H} < 10.6 \\ 
\end{cases}\\
\epsilon_i &= \begin{cases}
0.040~m_{K} - 0.352 &  m_{K} \geq 10.3 \\
0.060 &  m_{K} < 10.3 ~. \notag \\ 
\end{cases}
\end{align}
Note that at bright apparent magnitudes we conservatively truncate the
$J$ and $H$ band errors at 0.048\,mag and the $K$ band error at
0.060\,mag.

There is no correlation between the errors in $s$ and those in $r$ or
$i$, but there is a strong correlation between those in $r$ and $i$.
This is quantified by a correlation coefficient that is determined
empirically by studying the distribution of the differences in $r$
against the differences in $i$ for pairs of independent passbands. The
coefficient is found to be $\rho_{ri} = -0.95$ for all passbands. To
preserve this correlation, the error in $r$ is calculated directly from
the error in $i$ using $\epsilon_r = 0.68\epsilon_i$. For the $J$ band,
the typical error in the effective radius is around 0.049\,dex (11\%)
and in surface brightness is around 0.073\,dex (17\%). However in the
correlated combination in which these quantities appear in the FP,
namely $X_\mathrm{FP}=r-bi$, the typical error in $X_\mathrm{FP}$ is
just 0.016\,dex (4\%).

There is an additional error term for effective radius, $\epsilon_{rp}$,
which allows for the uncertainty in the conversion of angular to
physical units under the assumption that the galaxy is at its redshift
distance (i.e.\ neglecting the unknown peculiar velocity). This error
term is approximated as $\epsilon_{rp_n} =
\log(1+300\mathrm{\,km\,s}^{-1}/cz_n)$, which assumes a typical peculiar
velocity of 300\,\kms\ for the galaxies in the sample
\citep{Strauss:1995}. Because we explicitly exclude from the sample
galaxies at low redshifts, where the peculiar velocities are potentially
large relative to the recession velocities (see \S\ref{subsec:selcuts}),
$\epsilon_{rp}$ is typically $<$3\% and contributes less than 10\% to
the overall error in $r$.

We note that a similar error on surface brightness exists due to the use
of observed redshifts (uncorrected for peculiar velocities) in
computing the cosmological dimming. However, we do not include this in
our measurement error matrix because it is typically less than $0.4\%$,
which is negligible when added in quadrature to the photometric
measurement errors.

\subsection{Group Catalogue}
\label{subsec:groupcat}

Groups and clusters in the the 6dFGS sample were identified using a
friends-of-friends group-finding algorithm (Merson et~al., in prep.).
The algorithm follows a similar procedure to the group-finding method
used to construct the 2dF Percolation-Inferred Galaxy Groups (2PIGG)
Catalogue of the 2dF Galaxy Redshift Survey \citep{Eke:2004}, but it is
re-calibrated to the specifications (redshift depth and sample density)
of the 6dFGS.

This group catalogue is used to test the universality of the Fundamental
Plane (i.e.\ whether the FP coefficients vary with galaxy environment)
and to derive mean redshifts for groups and thus group distances and
peculiar velocities (in addition to distances and peculiar velocities
for single galaxies). Combining galaxies into groups is important to our
future peculiar velocity analysis for two reasons: (i)~it minimises the
`Finger-of-God' distortions of distances and peculiar velocities
produced by virialised structures in redshift space; (ii)~it allows us
to correct any variations in the FP with environment that might bias the
distance and peculiar velocity estimates.

From the initial 11287 galaxies in the 6dFGS FP subsample, there were
3186 galaxies found in groups containing at least four members (and so
deemed to have reliable group membership status). The flux-limited
nature of our survey meant that the faintest members of a group might
not have been observed, so the richness of a group (which we use as
proxy for global environment) is defined as the number, $N_R$, of
observed galaxies in the group brighter than a specified absolute
magnitude, chosen so that galaxies brighter than this would be visible
throughout the sample volume. Any galaxy not in a group was given a
richness $N_R$=0, signifying its status as either a field galaxy or a
bright member of a poor group.

In addition to this group catalogue, we also determine parameters that
define each galaxy's local environment using the method described in 
Wijesinghe et~al. (submitted, 2012).  

In this catalogue, local environment is represented by the projected 
comoving distance, $d_5$ (in Mpc) to the $5^{\rm th}$ nearest neighbour
and the surface density, $\Sigma_5$ (in galaxies\,Mpc$^{-2}$), is therefore 
defined as $\Sigma_5 = 5 / \pi * d_5^2$.  To exclude contamination from 
foreground and background galaxies these density measurements are 
made within a velocity cylinder of $\pm$1000\,\kms. In our final FP sample, 
there are 8258 galaxies for which we can calculate reliable values of these estimators of local environment

\subsection{Morphological Classification}
\label{subsec:morphdat}

All 11287 galaxies in the 6dFGS FP sample were visually inspected to
provide morphological classifications. Each galaxy was examined by up to
four experienced observers, and on average classified twice. This was
done to determine and flag any galaxies without dominant bulges that
might bias, or add scatter to, the FP fits, and also to allow us to test
whether ellipticals, lenticulars and spiral bulges have different FP
distributions.

All of the galaxies were visually inspected using the 2MASS $J$, $H$ and
$K$ band images and also the higher-resolution SuperCOSMOS images in the
$b_J$ and $r_F$ bands. The galaxies were classified into the standard
morphological types: elliptical (E), lenticular (S0), spiral (Sp) and
irregular or amorphous (Irr), plus the transition cases E/S0, S0/Sp and
Sp/Irr.  The presence of dust lanes was also flagged. The galaxy images
had 6.7\,arcsec diameter circles superimposed in order to determine
whether the 6dF fibre enclosed only bulge light or whether there was
significant contamination by light from the disks of S0 and Sp galaxies.
At the same time, the 6dFGS spectra were scrutinised for any discernible
emission features.

From this sample there were 429 galaxies excluded on the basis of one or
more of the criteria defined below. If any one of these criteria was
flagged by two or more classifiers, or flagged by the single classifier
in cases where a galaxy was only classified once, then the galaxy was
excluded as not being bulge-dominated or as problematic in some other
respect. The exclusion criteria were: (i)~galaxy morphology classified
as irregular or amorphous; (ii)~galaxy identified as edge-on with a full
dust lane; (iii)~ significant fraction of light in fibre is from a disk;
and (iv)~light in fibre contaminated by nearby star, galaxy or defect.

\section{Mock Galaxy FP Samples}
\label{sec:genmocks}

We now describe the process of generating mock catalogues from a model
that reproduces all of the main features of the observed data sample as
closely as possible. It is important that the mock samples are robust
and well calibrated, as they serve several functions. We use them: to
perform comparisons of different fitting methods
(\S\ref{subsec:leastsq}); to validate the ML fitting method and the
assumption of a 3D Gaussian model for the data (\S\ref{subsec:mockalg}
and \S\ref{subsec:modelvalid}); to correct for residual bias effects
(\S\ref{subsec:resbiascorr}); and to determine the accuracy and
precision of the fits (\S\ref{subsec:fpparunc}).

\subsection{Mock Sample Algorithm}
\label{subsec:mockalg}

We create mock samples from a given set of FP parameters
$\{a,b,c,\bar{r},\bar{s},\bar{i},\sigma_1,\sigma_2,\sigma_3\}$ using the
following steps to generate each mock galaxy:
\begin{enumerate}
\item Draw values for $v_1$, $v_2$ and $v_3$ at random from a 3D
  Gaussian with corresponding specified variances $\sigma_1$, $\sigma_2$
  and $\sigma_3$.
\item Transform these values from the $\mathbf{v}$-space (principal
  axes) coordinate system to the $\{\mathbf{r,s,i}\}$-space (observed
  parameters) coordinate system using the inverse of the relations in
  equation~\ref{eq:vectors} with the specified FP slopes ($a$ and $b$)
  and FP mean values ($\bar{r}$, $\bar{s}$ and $\bar{i}$).
\item Generate a comoving distance from a random uniform density
  distribution over the volume out to $cz_{max} = 16120$\,\kms\ using
  the assumed cosmology. This comoving distance is converted to an
  angular diameter distance in order to calculate an angular effective
  radius from a physical effective radius.
\item The redshift of each mock galaxy is also derived from this
  comoving distance; it must be greater than the lower limit on $cz$ to
  remain in the mock sample.
\item Derive an apparent magnitude from the surface brightness and
  effective radius (in angular units) of each galaxy, obtained at
  step~(ii), using $m=$\sbmag$-2.5\log[2\pi(R_{e}^{\theta})^2]$.
\item Estimate rms measurement uncertainties from this magnitude via the
  prescription given in \S\ref{subsec:errors}, and use these
  uncertainties to generate Gaussian measurement errors in $\{r,s,i\}$
  from the error matrix in equation~\ref{eq:errmatrix} (including the
  correlation between $\epsilon_r$ and $\epsilon_i$).
\item Add these measurement errors to $\{r,s,i\}$ to obtain the observed
  values of the FP parameters; the velocity dispersion must be greater
  than the lower selection limit to remain in the mock sample.
\item Compute the observed magnitude using the observed values of $r$
  and $i$ (i.e.\ including measurement errors); it must be brighter than
  the limiting magnitude for the galaxy to remain in the sample.
\item Compute the selection probability from the observed magnitude and
  redshift using equations~\ref{eq:dlim} and~\ref{eq:sprob}; it must be
  greater than the minimum selection probability for the galaxy to
  remain in the mock sample.
\end{enumerate}
This process is repeated until the desired number of galaxies is
generated for the mock sample.

\begin{figure*}
\centering
\begin{minipage}{170mm}
\includegraphics[width=0.95\textwidth]{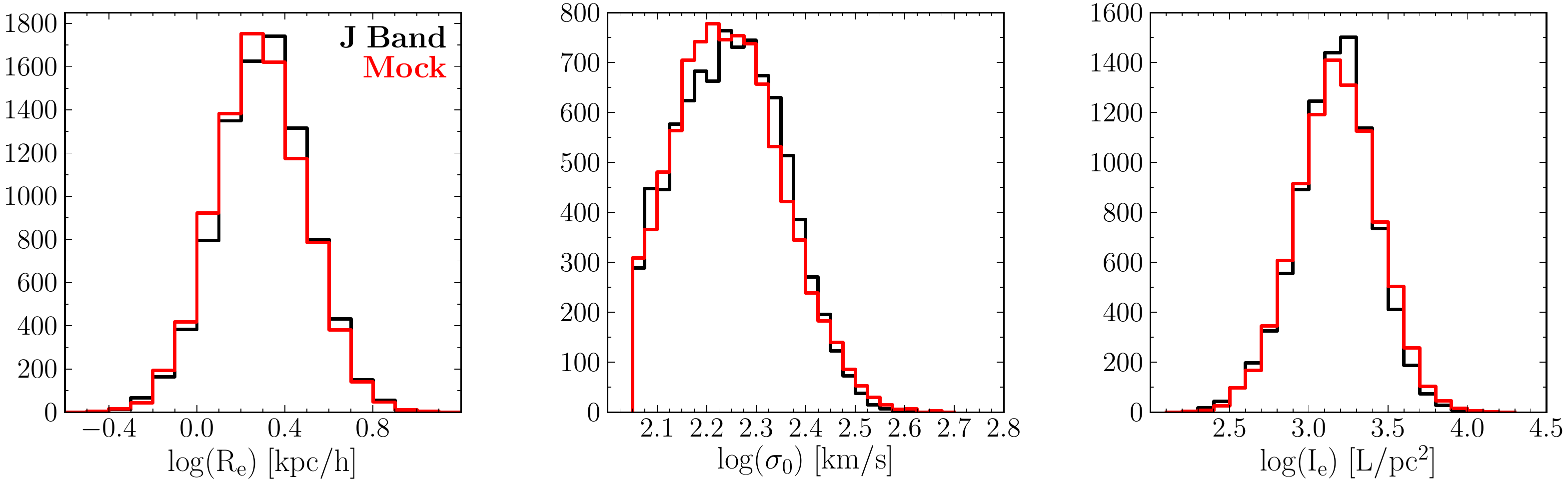} 
\caption{The distribution of the observed Fundamental Plane parameters
  \er, \vd\ and \sblum\ for the 6dFGS $J$ band sample (black) and a mock
  sample (red) of the same size ($N_g=8901$) and the same selection
  criteria, with FP coefficients $a=1.52$ and $b=-0.89$.}
\label{fig:rsicompare}
\end{minipage}
\end{figure*}

\begin{figure*}
\centering
\begin{minipage}{170mm}
\includegraphics[width=0.95\textwidth]{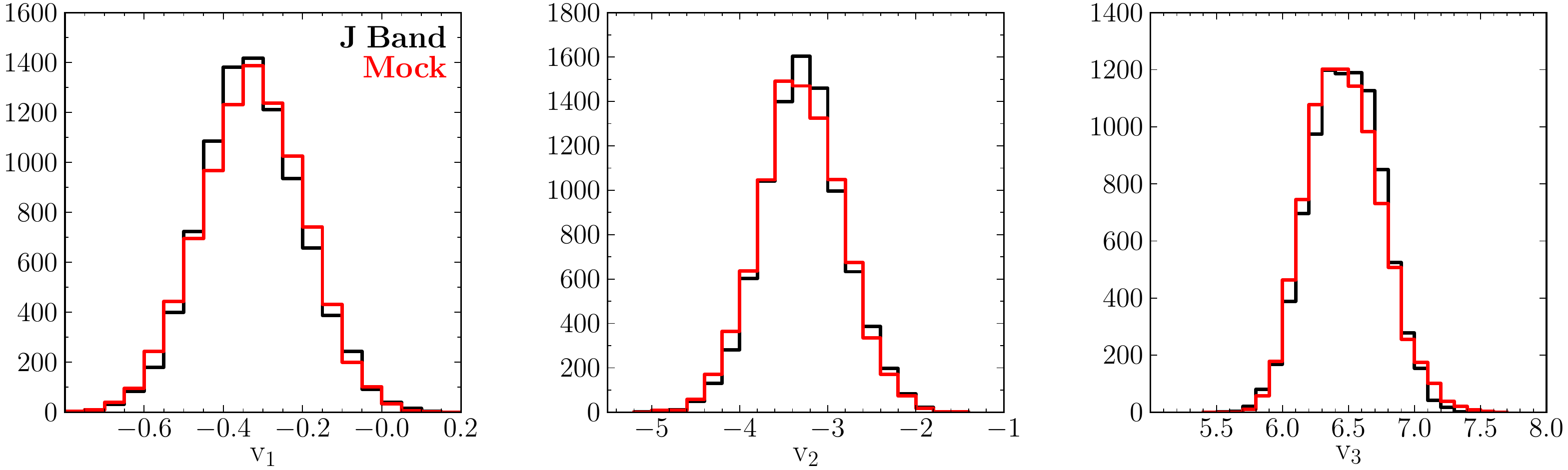} 
\caption{The distribution of the natural Fundamental Plane parameters
  $v_1$, $v_2$ and $v_3$ for the 6dFGS $J$ band sample (black) and a mock
  sample (red) of the same size ($N_g=8901$) and the same selection
  criteria, with FP coefficients $a=1.52$ and $b=-0.89$.}
\label{fig:v1v2v3compare}
\end{minipage}
\end{figure*}

Figure~\ref{fig:rsicompare} compares the distributions of effective
radius, velocity dispersion and surface brightness for the 6dFGS $J$ band
FP sample and a mock sample generated from the best-fitting 3D Gaussian
model (see below) having the same number of galaxies, the same selection
criteria and the same observational errors. The mock sample accurately
replicates the distributions of the galaxies in FP space, both for the
observed parameters ($r$, $s$ and $i$) and the `natural' parameters
($v_1$, $v_2$ and $v_3$), which are shown in
Figure~\ref{fig:v1v2v3compare}. This close match between the model and
the data justifies our use of a 3D Gaussian model for the distribution
of galaxies in FP space.

\subsection{Residual Bias Corrections}
\label{subsec:resbiascorr}

The only effect that is not explicitly corrected for in the maximum
likelihood fitting process, and which introduces a (small) bias, is the
exclusion of low-selection-probability (i.e.\ high-weight) galaxies.
These galaxies are excluded because: (a)~they may be outliers; and
(b)~they enter the likelihood with high weights and may therefore
distort the fits. They cannot be directly accounted for in the ML fit
because we do not have an explicit model for the distribution of
outliers.

In practice this bias is small because only a small number of galaxies
are excluded, and may be quantified under the assumptions of our model
using mock samples. By applying the same selection criteria to the mocks
as we do to the data, we can recover the correction $\Delta y$ for the
residual bias in some parameter $y$ as the difference between the value
$y_{\rm obs}$ obtained from fitting the observed data and the value
$y_{\rm mock}$ recovered as the average from ML fits to many mock
samples:
\begin{equation}
\Delta y = y_{\rm obs} - y_{\rm mock}
\end{equation}
where $y$ can be any of the parameters describing the 3D Gaussian model,
$\{a,b,c,\bar{r},\bar{s},\bar{i},\sigma_1,\sigma_2,\sigma_3\}$. To
correct fits to the observed data for residual bias, these corrections
should be added to the best-fit FP parameter values to recover the
`true' parameters:
\begin{equation}
y_{\rm cor} = y_{\rm obs} + \Delta y  ~.
\end{equation}

These corrections were obtained for mock samples of increasing sample
size, with $N_g$ ranging from $1000$ to $10000$ galaxies. For all
parameters the bias correction was found to be constant for all sample
sizes. We have therefore employed a fixed bias correction for each
parameter regardless of sample size. These corrections are listed for
each fitted FP parameter in Table~\ref{tab:bias}.

\begin{table}
\centering
\caption{Bias corrections for each of the FP parameters. These corrections
  are added to the fitted parameters to remove the residual bias. Note
  that these corrections are small for all parameters.} 
\renewcommand{\tabcolsep}{3.6pt}
\begin{tabular}{@{}ccccccccc@{}}
\hline
$a$ & $b$ & $c$ & $\bar{r}$ & $\bar{s}$ & $\bar{i}$ & $\sigma_1$ & $\sigma_2$ & $\sigma_3$ \\
\hline
0.022 & -0.008 & -0.027 & -0.006 & -0.001 & 0.004 & 0.0002 & 0.0026 & 0.0013  \\
\hline
\end{tabular}
\label{tab:bias}
\end{table}

\section{The Fundamental Plane}
\label{sec:thefp}

\subsection{The 3D Fundamental Plane}
\label{sec:3dfp}

Fundamental Plane studies in optical passbands are relatively abundant,
while studies in near-infrared passbands are less so. It is only
recently that large, homogeneous FP data sets across both optical and
near-infrared wavelengths have become available
\citep{Hyde:2009,LaBarbera:2010a}. Using near-infrared photometry
in FP analyses is advantageous because in these passbands the lower
extinction reduces the variations due to dust and the dominance of
older stellar populations reduces the variations due to recent
star-formation (at least in the absence of a significant population of
intermediate-age AGB stars---cf.\ \citealt{Maraston:2005}). Comparison
of optical and near-infrared observations can reveal the effect of
variations in the mass-to-light ($M/L$) ratios on the Fundamental Plane.

\begin{figure*}
\centering
\begin{minipage}{180mm}
\includegraphics[width=0.95\textwidth]{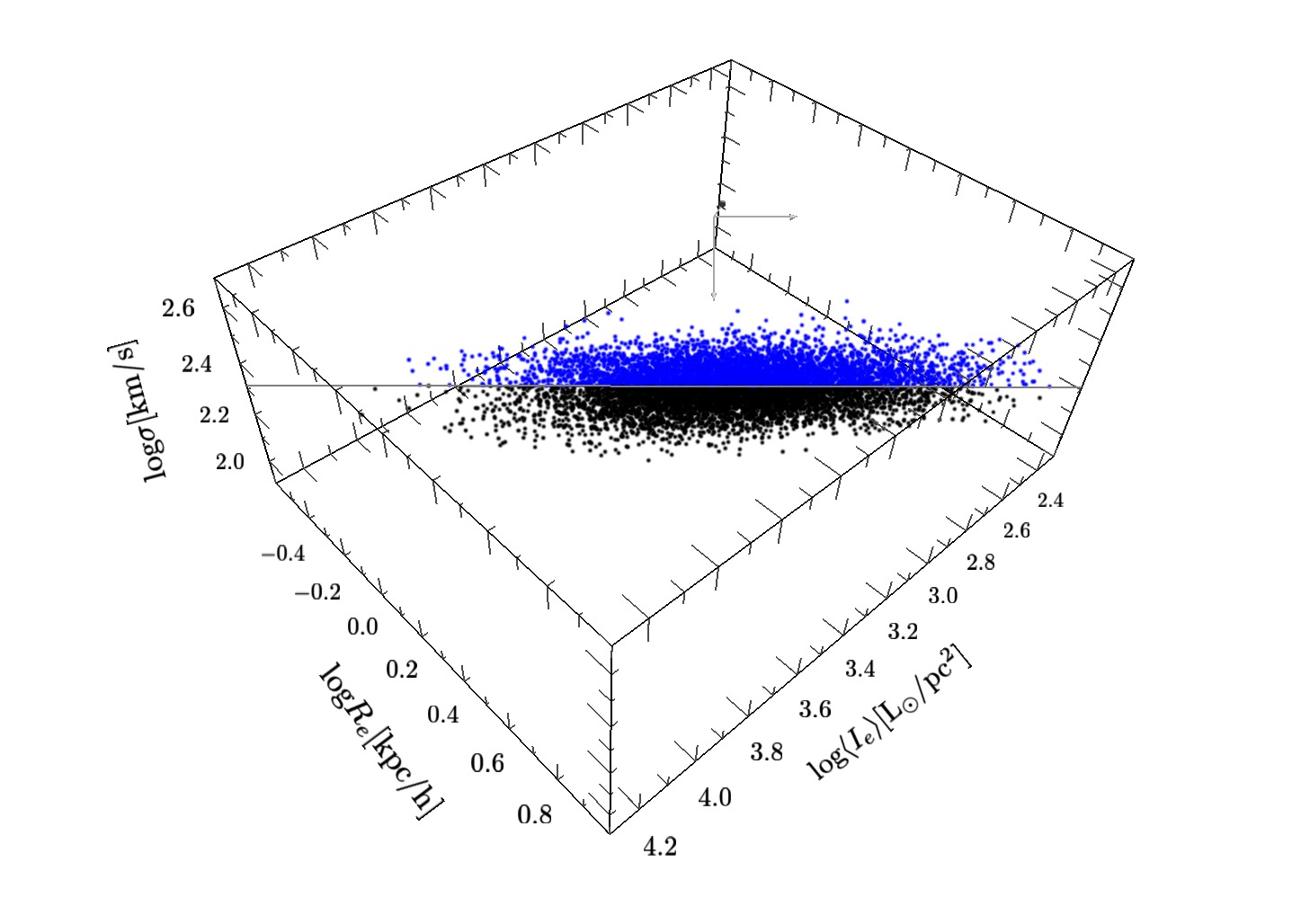}  
\caption{Interactive 3D visualisation of the 6dFGS $J$ band Fundamental
  Plane in $\mathbf{r,s,i}$ space. The best fitting plane (in grey) has
  slopes $a=1.523$ and $b=-0.885$, and an offset $c=-0.330$. The galaxies
  are colour-coded according to whether they are above (blue) or below
  (black) the best-fit plane. The 1$\sigma$, 2$\sigma$, and 3$\sigma$
  contours of the 3D Gaussian distribution (light grey) can be toggled in the 
  interactive plot environment.
  (Readers using Acrobat Reader v8.0 or higher can enable interactive 3D
  viewing of this schematic by mouse clicking on the version of this figure found in the ancillary files; see
  Appendix~\ref{sec:usage3D} for more detailed usage instructions.) }
\label{fig:jbandfp}
\end{minipage}
\end{figure*}

Figure~\ref{fig:jbandfp} is a 3D visualisation of the 6dFGS $J$ band FP
sample that can be interactively viewed in the full 3D space of the
observed parameters $r$, $s$ and $i$. This figure (like
Figure~\ref{fig:3dvectors}) was created with the S2PLOT programming
library. It is important to show the 3D view of the FP, rather than the
2D plots usually found in the literature, because information is lost in
projecting the FP onto two dimensions from its native three dimensions,
and the true properties of the 3D distribution of the FP are disguised.
Figure~\ref{fig:jbandfp} reveals in 3D the well-known features of the
FP, including the small scatter in the edge-on view relative to the
other two dimensions, and the Gaussian nature of the distribution in all
three dimensions; the impact of sampling effects, such as the hard
selection limit in velocity dispersion, are also readily apparent.
 
\subsection{Fundamental Plane parameters and uncertainties}
\label{subsec:fpparunc}

Using our maximum likelihood fitting routine we recover the best-fit FP
in the $J$, $H$ and $K$ passbands for samples containing 8901, 8568 and
8573 galaxies respectively. The full details of the FP fits in these
bands are given in Table~\ref{tab:fpfits}, including all eight fitted
parameters together with the constant of the fit ($c$), the offset of
the plane in the $r$-direction ($r_{0}$; see below), the total rms
scatter about the FP in the $r$-direction ($\sigma_r$), and the total
rms scatter in distance ($\sigma_d$); the difference between these two
scatters is discussed in \S\ref{sec:discuss}.

\begin{table*}
\renewcommand{\tabcolsep}{2.5pt}
\begin{minipage}{180mm}
  \caption{Best-fit 6dFGS FP parameters (including bias corrections) and
    their associated uncertainties for: (i)~the full $J$, $H$ and $K$
    samples; (ii)~the $J$ band $N_R$ richness subsamples (field, low,
    medium and high); (iii)~the $J$ band $\Sigma_5$ local environment
    subsamples (low, medium and high); and (iv)~the $J$ band morphology
    subsamples (early and late types). As well as the nine FP
    parameters, the table also lists: $N_g$, the number of galaxies in
    each sample; $r_0$, the location of the FP at the fiducial point
    $(s_0=2.3,i_0=3.2)$; $\sigma_r$, the scatter about the FP in the
    $r$-direction (see \S\ref{subsec:fpscatter}); and $\sigma_d$, the
    scatter in the distance (see \S\ref{subsec:disterrs}).}
\label{tab:fpfits}
{\scriptsize
\begin{tabular}{lccccccccccccc}
\hline
Sample                      & $N_g$& $a$               & $b$                & $c$                & $\bar{r}$       & $\bar{s}$       & $\bar{i}$       & $r_0$           & $\sigma_1$      & $\sigma_2$      & $\sigma_3$      & $\sigma_r$   & $\sigma_d$    \\ 
\hline
$J$ band                    & 8901 & 1.523$\pm$0.026 & -0.885$\pm$0.008 & -0.330$\pm$0.054 & 0.184$\pm$0.004 & 2.188$\pm$0.004 & 3.188$\pm$0.004 & 0.345$\pm$0.002 & 0.053$\pm$0.001 & 0.318$\pm$0.004 & 0.170$\pm$0.003 & 0.127 (29.7\%) & 0.097 (22.5\%) \\
$H$ band                    & 8568 & 1.473$\pm$0.024 & -0.876$\pm$0.008 & -0.121$\pm$0.051 & 0.175$\pm$0.004 & 2.190$\pm$0.003 & 3.347$\pm$0.004 & 0.465$\pm$0.002 & 0.051$\pm$0.001 & 0.318$\pm$0.004 & 0.167$\pm$0.003 & 0.123 (28.8\%) & 0.096 (22.3\%) \\
$K$ band                    & 8573 & 1.459$\pm$0.024 & -0.858$\pm$0.008 & -0.103$\pm$0.050 & 0.153$\pm$0.005 & 2.189$\pm$0.003 & 3.430$\pm$0.005 & 0.511$\pm$0.003 & 0.050$\pm$0.001 & 0.329$\pm$0.004 & 0.166$\pm$0.003 & 0.123 (28.8\%) & 0.095 (22.1\%) \\
\hline
$N_R$$\leq$1                & 6495 & 1.512$\pm$0.030 & -0.882$\pm$0.010 & -0.307$\pm$0.063 & 0.183$\pm$0.005 & 2.187$\pm$0.004 & 3.197$\pm$0.005 & 0.351$\pm$0.002 & 0.053$\pm$0.001 & 0.315$\pm$0.005 & 0.161$\pm$0.003 & 0.127 (29.7\%) & 0.097 (22.5\%) \\
2$\leq$$N_R$$\leq$5         & 1248 & 1.582$\pm$0.058 & -0.899$\pm$0.021 & -0.436$\pm$0.122 & 0.154$\pm$0.014 & 2.170$\pm$0.012 & 3.168$\pm$0.012 & 0.331$\pm$0.005 & 0.051$\pm$0.002 & 0.324$\pm$0.011 & 0.201$\pm$0.009 & 0.126 (29.3\%) & 0.098 (22.7\%) \\
6$\leq$$N_R$$\leq$9         &  546 & 1.573$\pm$0.088 & -0.862$\pm$0.029 & -0.538$\pm$0.187 & 0.220$\pm$0.017 & 2.208$\pm$0.012 & 3.154$\pm$0.015 & 0.327$\pm$0.006 & 0.044$\pm$0.003 & 0.325$\pm$0.014 & 0.181$\pm$0.011 & 0.120 (28.0\%) & 0.094 (21.8\%) \\
$N_R$$\geq$10               &  612 & 1.504$\pm$0.094 & -0.903$\pm$0.029 & -0.248$\pm$0.195 & 0.228$\pm$0.016 & 2.220$\pm$0.012 & 3.171$\pm$0.014 & 0.324$\pm$0.006 & 0.054$\pm$0.003 & 0.316$\pm$0.013 & 0.173$\pm$0.011 & 0.129 (30.1\%) & 0.095 (22.1\%) \\
\hline
$\Sigma_5$$\leq$0.07        & 2664 & 1.486$\pm$0.051 & -0.848$\pm$0.014 & -0.354$\pm$0.113 & 0.190$\pm$0.008 & 2.192$\pm$0.005 & 3.203$\pm$0.006 & 0.354$\pm$0.004 & 0.053$\pm$0.002 & 0.314$\pm$0.007 & 0.147$\pm$0.004 & 0.126 (29.3\%) & 0.095 (21.9\%) \\
0.07$<$$\Sigma_5$$\leq$0.25 & 2812 & 1.516$\pm$0.043 & -0.915$\pm$0.015 & -0.220$\pm$0.090 & 0.175$\pm$0.008 & 2.183$\pm$0.007 & 3.189$\pm$0.007 & 0.343$\pm$0.003 & 0.053$\pm$0.002 & 0.313$\pm$0.007 & 0.173$\pm$0.006 & 0.126 (29.5\%) & 0.097 (22.5\%) \\
$\Sigma_5$$>$0.25           & 2782 & 1.564$\pm$0.039 & -0.889$\pm$0.013 & -0.418$\pm$0.079 & 0.188$\pm$0.009 & 2.190$\pm$0.007 & 3.170$\pm$0.008 & 0.335$\pm$0.003 & 0.050$\pm$0.001 & 0.326$\pm$0.007 & 0.185$\pm$0.006 & 0.127 (29.6\%) & 0.097 (22.5\%) \\
\hline
E+E/S0+S0                   & 6956 & 1.535$\pm$0.029 & -0.879$\pm$0.010 & -0.384$\pm$0.060 & 0.156$\pm$0.005 & 2.199$\pm$0.004 & 3.230$\pm$0.004 & 0.339$\pm$0.002 & 0.052$\pm$0.001 & 0.296$\pm$0.004 & 0.170$\pm$0.003 & 0.128 (29.8\%) & 0.096 (22.3\%) \\
Sp bulges                   & 1945 & 1.586$\pm$0.067 & -0.861$\pm$0.017 & -0.512$\pm$0.138 & 0.305$\pm$0.009 & 2.151$\pm$0.008 & 3.016$\pm$0.008 & 0.384$\pm$0.006 & 0.052$\pm$0.002 & 0.319$\pm$0.008 & 0.157$\pm$0.005 & 0.127 (29.7\%) & 0.097 (22.6\%) \\
\hline
\end{tabular}
}
\end{minipage}
\end{table*}

\begin{figure*}
\begin{minipage}{180mm}
\centering
\includegraphics[width=0.95\textwidth]{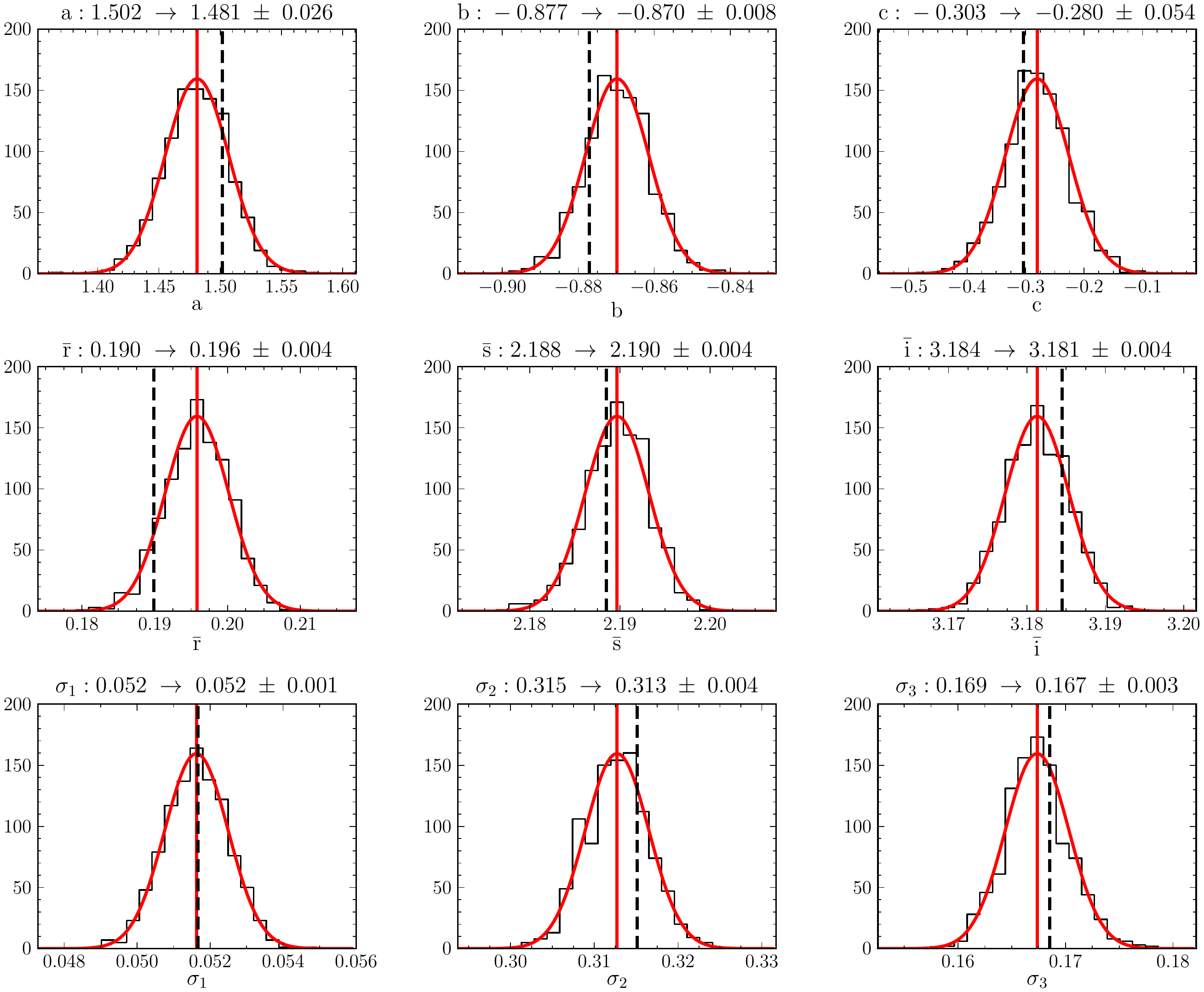} 
\caption{Histograms of the maximum-likelihood best-fit values of the
  $J$ band FP parameters
  $\{$a$,$b$,$c$,\bar{r},\bar{s},\bar{i},\sigma_1,\sigma_2,\sigma_3\}$ from
  1000 simulations. Each panel is labelled at the top with the name of
  the parameter, the input value of the parameter for the 1000 mock
  samples, and the mean and rms of the best-fit parameters obtained from
  ML fits to these mocks; a Gaussian with this mean and rms is
  overplotted on the histograms. The vertical dashed line shows the
  input value of the parameter and the vertical solid line shows the
  mean of the best-fit values. The residual bias correction (see
  \S\ref{subsec:resbiascorr}) is the offset between the dashed line and
  the solid line; in all cases this is comparable to or smaller than the
  modest rms scatter in the fitted parameter.}
\label{fig:histo}
\end{minipage}
\end{figure*}

The errors in the best-fit FP parameters that are given in
Table~\ref{tab:fpfits} are estimated as the rms scatter in fits to
multiple mock samples generated as described in \S\ref{sec:genmocks}
using the parameters of the best-fit FP. The distribution of the
parameters derived from ML fits to 1000 mock samples (each sample
containing 8901 galaxies, as for the 6dFGS $J$ band sample) are shown in
Figure~\ref{fig:histo}. Note that the residual bias corrections (the
differences between the input parameters and the mean of the fitted
parameters) are comparable to or less than the rms scatter in the fits
(i.e.\ comparable to or less than the random errors in the fitted
values). This highlights the accuracy with which the ML method recovers
the FP parameters even in the presence of significant observational
errors and various types of sample censoring.

Both the bias corrections and the random errors are small; the
fractional errors in the FP slopes ($a$ and $b$) and dispersions
($\sigma_1$, $\sigma_2$ and $\sigma_3$) are all less than 2\%. For the
offset of the FP, $c\equiv\bar{r}-a\bar{s}-b\bar{i}$, the uncertainty is
0.054\,dex or 12\%. However as a measure of the uncertainty in the
relative sizes and distances of galaxies due to the fit this `intercept'
offset is misleading. A better measure is the uncertainty in $\bar{r}$,
which is 0.9\%; but even this is an over-estimate of the practical
impact of the uncertainty in the fit, as the point
$(\bar{r},\bar{s},\bar{i})$ is at the edge of the observed distribution
(i.e.\ the observed distribution is well-fitted by a Gaussian centred
close to the velocity dispersion limit). The most realistic estimate of
the uncertainty in the $r$-axis offset of the fitted FP, as it affects
size and distance estimates for 6dFGS galaxies, is given by the
uncertainty in $r_0$, the $r$-value of the fitted FP at a fiducial point
in the middle of the observed sample: $s_0\equiv2.3$ and $i_0\equiv3.2$.
The rms scatter in $r_0 \equiv a s_0 + b i_0 + c$ is just 0.5\%.

\subsection{Model validation}
\label{subsec:modelvalid}

\begin{figure}
  \includegraphics[width=0.45\textwidth]{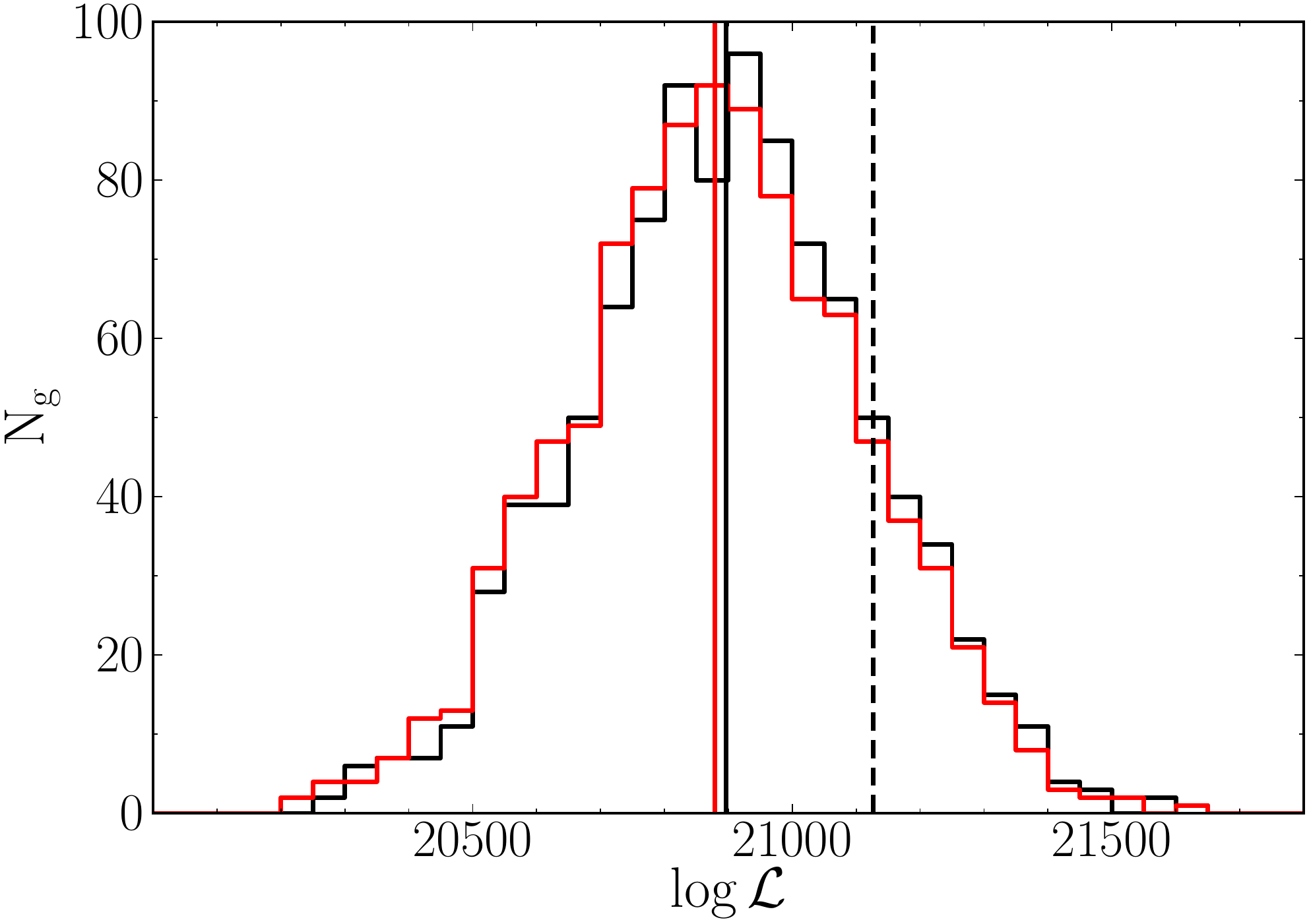}
  \caption{Distribution of likelihood values from 1000 mock samples. The
    FP coefficients used to generate these mock simulations are the same
    values used to generate the mocks in Figure~\ref{fig:histo}. The
    likelihood values in the red histogram were calculated for each mock
    sample using these identical input FP values, whereas the likelihood
    values in the black histogram were calculated using the individual
    best fit for each mock. The mean likelihoods from these mocks (red:
    $\ln \mathcal{L} = 20878\pm225$; black: $\ln \mathcal{L} =
    20897\pm224$) are indicated by the solid lines, and are comparable
    to but lower than the best-fit likelihood obtained for the actual
    data ($\ln\mathcal{L} = 21126$), shown by the dashed black line. }
  \label{fig:likehisto} 
\end{figure}

That our 3D Gaussian model is a good representation of the observed
distribution of galaxies in FP space is verified by the remarkable similarities
between the mock and data likelihoods.  The histogram
of log-likelihood values in Figure~\ref{fig:likehisto} gives the
distribution from the same 1000 mock simulations as
Figure~\ref{fig:histo}, derived in two ways: first by calculating the
likelihoods for all the mocks using the best-fit FP of the data (red
histogram), and second, by calculating the likelihoods using the best-fit
FP values from each individual mock (black histogram). It makes little
difference which method is used, as the distribution of likelihoods for
these two situations are very similar.

The mean of each histogram (red: $\ln\mathcal{L} = 20878\pm225$;
black: $\ln\mathcal{L} = 20897\pm224$) is plotted as a solid line. The
likelihood of the best fit to the actual data ($\ln\mathcal{L} = 21126$)
is shown by the dashed vertical line, and is larger than these means but
still well within the range of likelihoods spanned by the mock samples.
The fact that the likelihood recovered from the data is higher than that
from the mocks (i.e.\ $\ln\mathcal{L}$ is more positive) is a result of
excluding the $\chi^2$ outliers from the data, which may also remove the
extreme tail of the Gaussian distribution. Genuine outliers do not exist
in the mock samples and so no $\chi^2$ clipping is applied, and the
lower likelihoods of the mock samples in Figure~\ref{fig:likehisto}
reflect this difference. 

In summary, the similarity in likelihood values indicates that the
fitting algorithm has accurately recovered the input FP and also that
the 3D Gaussian model is a suitable representation of the observed FP
distribution.

\subsection{Additional $\sigma$-component of 3D Gaussian vectors}
\label{subsec:scomp}

Our 3D Gaussian model of the FP assumes that the $s$-component of the
$\mathbf{v_2}$ vector is zero; i.e.\ that the vector representing the longest
axis of the 3D Gaussian lies wholly in the $r$--$i$ plane. This is based
in part on previous studies \citep{Saglia:2001,Colless:2001}, and in
part assumed for convenience and simplicity.

We can test how accurate this assumption is by extending the vector
definitions of equation~\ref{eq:vectors} to include this component, with
coefficient $k$, defining the set of orthogonal axes
\begin{align}
\label{eq:ninevectors}
\mathbf{v_1} &= \mathbf{\hat{r}} - a \mathbf{\hat{s}} - b \mathbf{\hat{i}},  \notag \\
\mathbf{v_2} &= \mathbf{\hat{r}} - k \mathbf{\hat{s}} + (1 - k a) \mathbf{\hat{i}}/b, \\
\mathbf{v_3} &= (k a^2 - a + k b^2) \mathbf{\hat{r}} + (k a - 1 - b^2)\mathbf{\hat{s}} + (k b+ a b)\mathbf{\hat{i}} \notag
\end{align}
and then including this extra parameter in our fitting algorithm. We
then perform a nine-parameter maximum likelihood fit with the same $J$
band FP sample of galaxies and find a best-fit value
$k=0.09\pm0.01$, and a $J$ band FP given by
\begin{equation}
r = (1.51\pm0.03)s - (0.86\pm0.01)i - (0.39\pm0.06) ~. 
\end{equation}

Therefore, when there are no constraints placed on the components of
$\mathbf{v_2}$, the $s$-component is close to---but slightly larger
than---zero. The coefficient of $s$ is much smaller than the
coefficients of any of the other vector components, the intrinsic
scatter about the plane ($\sigma_1=0.052$) is the same to within 0.5\%,
and the error in distances is 24.3\% (i.e.\ slightly larger than for the
standard 8-parameter model). Hence the addition of this ninth parameter
provides no practical advantages, and we retain the simplifying
approximation of fixing $k\equiv0$.

\subsection{Adding age to the Fundamental Plane model}
\label{subsec:agemodel}

\begin{figure*}
\centering
\begin{minipage}{180mm}
\includegraphics[width=0.96\textwidth]{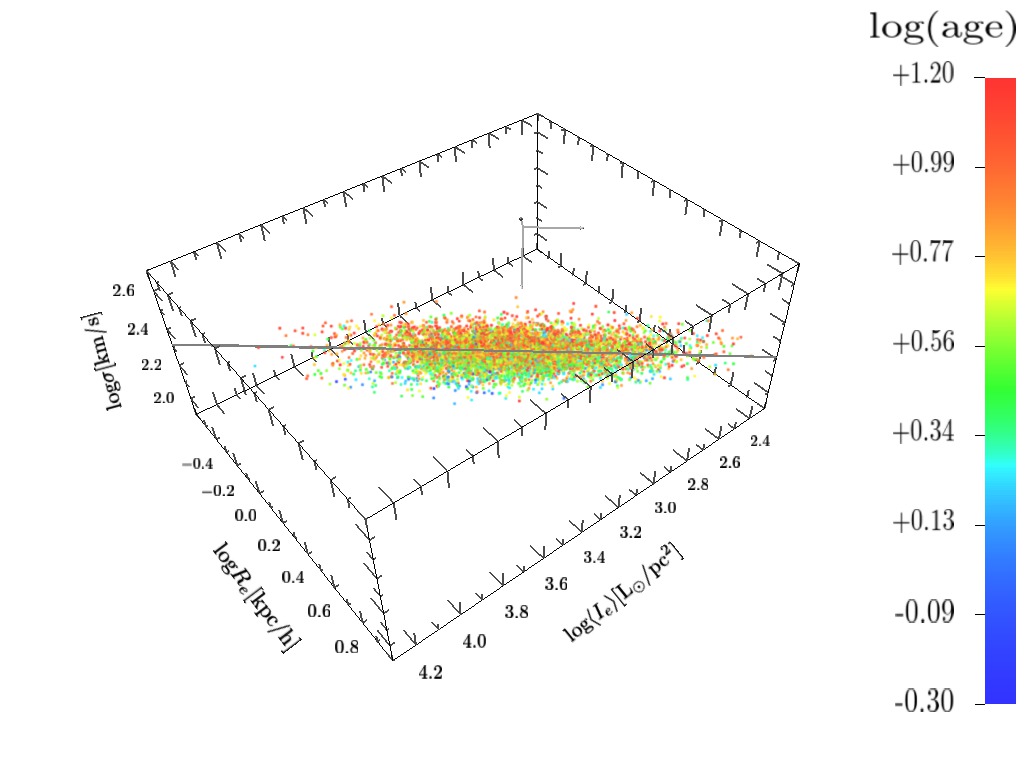} 
\caption{Interactive 3D visualisation of the 6dFGS $J$ band Fundamental
  Plane with individual galaxies colour-coded by $\log(\mathrm{age})$.
  (Readers using Acrobat Reader v8.0 or higher can enable interactive 3D
  viewing of this schematic by mouse clicking on the version of this figure found in the ancillary files; see
  Appendix~\ref{sec:usage3D} for more detailed usage instructions.)}
\label{fig:agefp}
\end{minipage}
\end{figure*}

\citet{Springob:2012} found that there is a clear trend of
galaxy age through the FP (i.e.\ along the $\mathbf{v_1}$ direction), as expected
from models of the effect of stellar populations on mass-to-light ratios
\citep[e.g.][]{Bruzual:2003, Korn:2005}. The variation of age through
the FP is shown in Figure~\ref{fig:agefp}, a 3D plot of the FP-space
distribution of the sub-sample of 6579 galaxies with stellar population
parameters measured by \citet{Springob:2012}, with colour encoding
$\log(\mathrm{age})$. Here we investigate whether this trend in age can be
incorporated into the FP model and used to reduce the overall scatter of
the FP by exploring a very simple extension of the model that allows for
a linear trend of age through the FP.

We include an age component in our existing FP model by adding
$\log(\mathrm{age})$ as a fourth dimension in FP space along with $r$,
$s$, and $i$. We assume that age varies almost entirely in the
$\mathbf{v_1}$ direction (normal to the plane), as suggested by the
results of \citet{Springob:2012}. We therefore assume the $\mathbf{v_2}$
and $\mathbf{v_3}$ vectors have no age component, and derive a fourth
$v-$space vector that is orthogonal to the other three vectors. The
resulting vector definition of this new 4D Gaussian model is
\begin{align}
\label{eq:agevectors}
\mathbf{v_1} &= \mathbf{\hat{r}} - a \mathbf{\hat{s}} - b \mathbf{\hat{i}} - k_A\mathbf{\hat{A}},  \notag \\
\mathbf{v_2} &= \mathbf{\hat{r}} + \mathbf{\hat{i}}/b, \notag \\
\mathbf{v_3} &= -\mathbf{\hat{r}}/b - (1 + b^2)\mathbf{\hat{s}}/(ab) + \mathbf{\hat{i}} \notag \\
\mathbf{v_4} &= \mathbf{\hat{r}} - a \mathbf{\hat{s}} - b \mathbf{\hat{i}} + (1 + a^2 + b^2) \mathbf{\hat{A}}/k_A
\end{align}
where $k_A$ is the component of $A = \log({\rm age})$ in the
$\mathbf{v_1}$ direction. Additional parameters that need to be fitted
along with $k_A$ in this model are the mean of the 4D Gaussian in
$\log({\rm age})$ ($\bar{A}$) and the intrinsic scatter in the
$\mathbf{v_4}$ vector ($\sigma_4$); this gives a total of 11 free
parameters to be fitted. Both the intrinsic variance matrix,
$\mathbf{\Sigma}$, and the observed measurement error matrix,
$\mathbf{E}$, are also extended to four dimensions to include $\sigma_4$
and age measurement errors, respectively.

The 4D Gaussian model including age is then fit to this subsample
resulting in an FP given by
\begin{equation}
r = (1.56\pm0.03)s - (0.89\pm0.01)i - (0.13\pm0.01)A - 0.43\pm0.06
\end{equation}
with $\sigma_1 = 0.048\pm0.001$ and $\sigma_4 = 0.40\pm0.01$. Although
the intrinsic scatter through the FP ($\sigma_1$) is reduced from its
value in the standard 3D Gaussian model (where $\sigma_1 = 0.053$), the
large scatter in $\sigma_4$ and steeper slope in $s$ suggest that the
scatter in distance has not been reduced by including an age component.
In fact, the scatter in distance (see \S\ref{subsec:disterrs}) is
slightly larger, at $\sigma_d = 0.010\,\mathrm{dex}$ (23.3\%), than for
the standard 3D Gaussian model, where $\sigma_d = 0.097\,\mathrm{dex}$
(22.5\%).

We conclude that: (i)~there is a statistically significant contribution
from age variations to the scatter through the FP, which is slightly
reduced by including age in the FP model; and (ii)~the combination of
large measurement errors on individual galaxy ages, intrinsic scatter in
age about the FP, and the tilt of the FP (specifically, the angle
between $\mathbf{v_1}$ and $\mathbf{r}$), means that---for the 6dFGS sample---including
age does not improve the distance estimates obtained from the FP. This
might change, however, if substantially more precise age measurements
were available.

\subsection{Bayesian model selection}

To justify our choice of the standard 3D Gaussian model, as defined in
\S\ref{sec:mlfit}, over the alternative models we have considered in
\S\ref{subsec:scomp} and \S\ref{subsec:agemodel}, we compare these
models using the Bayes information criterion \citep{Schwarz:1978}.

The Bayes information criterion, or $\mathrm{BIC}$, can be used to
choose between different models and determine whether increasing the
number of free parameters in the model will result in over-fitting. It
has the advantages of being easy to compute and independent of the
assumed priors for the models, and in the limit of large sample size it
approaches $-2\ln(B)$, where $B$ is the the Bayes factor that gives the
relative posterior odds of the models under comparison. The BIC depends
on the size of the sample ($N$), the log-likelihood of the best fit
($\ln\mathcal{L}$), and the number of free parameters in the model
($k$), and is given by
\begin{equation}
\mathrm{BIC} = -2 \ln(\mathcal{L}) + k \ln(N) ~.
\end{equation}
The model with the lowest BIC value is preferred.

For the standard 8-parameter model of \S\ref{sec:mlfit}, the BIC value
is $-$42075, as compared to $-$42287 for the 9-parameter model including
an additional $\sigma$ component in the $\mathbf{v_2}$ vector
(\S\ref{subsec:scomp}) and $-$31833 for the 11-parameter model including
age as an additional parameter (\S\ref{subsec:agemodel}). Therefore the
BIC indicates that the 11-parameter model including age is not an
improvement on the standard model, as was previously concluded in
\S\ref{subsec:agemodel}. However the 9-parameter model that includes a
$\sigma$-component in the $\mathbf{v_2}$ vector does have a lower BIC value than
the standard 8-parameter model, and so is the objectively preferred
model. We nonetheless choose to employ the standard 8-parameter 3D
Gaussian model because of its simpler physical interpretation, reduced
computational burden, and marginally better precision in estimating
distances.

\subsection{Fundamental Plane differences between passbands}
\label{subsec:fpbandvar}

\begin{figure*}
\centering
\begin{minipage}{170mm}
\includegraphics[width=0.95\textwidth]{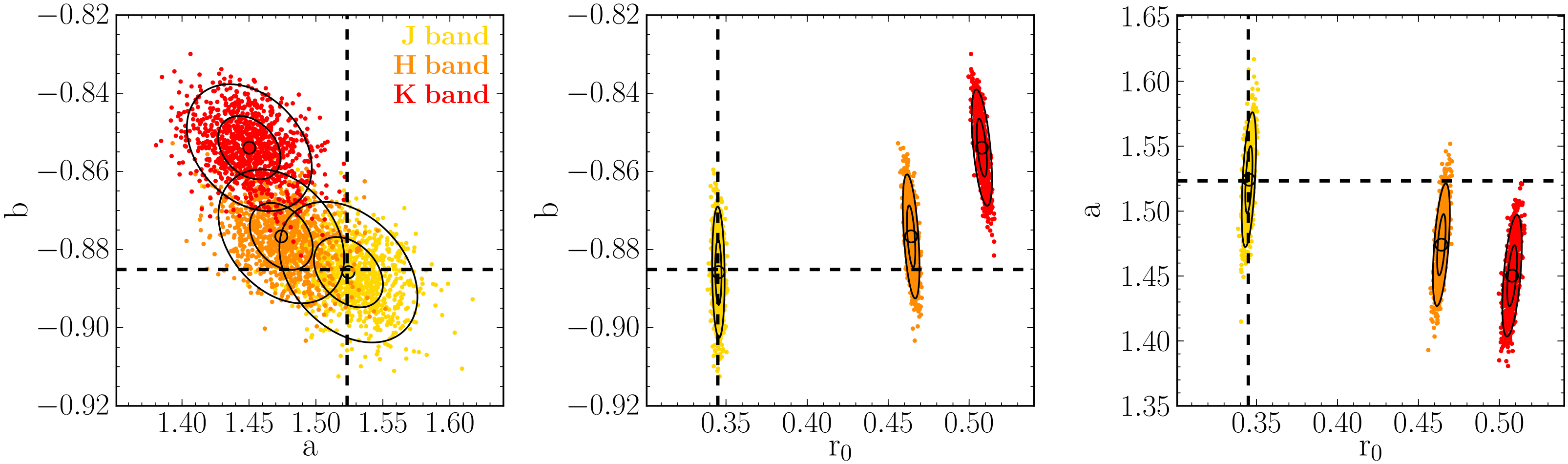}
\caption{Uncertainties on the FP parameters for the 6dFGS $J$ (yellow;
  $N_g$ = 8901), $H$ (orange; $N_g$ = 8568) and $K$ (red; $N_g$ = 8573)
  samples. The points show the best-fit FP parameters for each of 1000
  mock samples that take as input the best-fit FP parameters for the
  observed sample in each band. The mean values of the fitted FP
  parameters from the mocks, and their 1$\sigma$ and 2$\sigma$ contours,
  are also plotted. For reference, the input FP parameters used to
  generate the samples for the $J$ band are indicated as dotted lines.
  {\em Left:} $b$ versus $a$, showing similar FP coefficients although
  with a very weak trend of decreasing $a$ and increasing $b$ with
  increasing wavelength. {\em Centre:} $b$ versus $r_0$, showing
  significant offsets between the FPs in the three passbands.
  {\em Right:} $a$ versus $r_0$, again showing the FP offsets.}
\label{fig:abpass}
\end{minipage}
\end{figure*}

Table~\ref{tab:fpfits} gives the best-fit FP parameters for each of the
$J$, $H$ and $K$ bands. The FP slopes $a$ and $b$ are consistent between
these passbands at about the joint 1$\sigma$ and 2$\sigma$ levels
respectively. All three samples also have the same (small) intrinsic
scatter orthogonal to the FP, $\sigma_1 = 0.05$\,dex (12\%).
Figure~\ref{fig:abpass} illustrates the variation with wavelength of the
fitted FP slopes $a$ and $b$, and also the offset of the FP in the $r$
direction (the latter quantified by $r_0$, defined above in
\S\ref{subsec:fpparunc}). The figure shows the results of fitting FPs to
1000 mock samples in each passband with input parameters given by the
best-fit FP for the corresponding observed sample (as per
Table~\ref{tab:fpfits}). It also shows the mean values of the fitted
parameters for the mock samples, and the 1$\sigma$ and 2$\sigma$
contours of their distributions. As expected, the bias-corrected mean
coefficients accurately recover the input values; for reference, the
coefficients of the best-fit FP for the observed $J$ band sample are
marked in each plot as a pair of dashed black lines. 

The marginally significant (2$\sigma$) difference in the slopes
between the $J$ and $K$ bands may be due to the fact that $J$ band
mass-to-light ratios are almost independent of metallicity, whereas this
is not the case in the $K$ band \citep{Worthey:1994}. In this regard it
is worth noting that the $J$ band FP is (marginally) closer to the
virial plane than the $K$ band FP.

In the central and right panels, there is a clear offset in $r_0$
between passbands, with $r_0$ increasing at longer wavelengths. We
expect the differences in $r_0$ between passbands should be consistent
with the mean colours. To quantify the mean difference in $r_0$ (i.e.\
$\Delta r_0$) as a function of mean colour and surface brightness, we
assume that the FP slopes are consistent in each band (a good
approximation given the similarity of the coefficients in
Table~\ref{tab:fpfits}) and that the galaxies are homologous. These
approximations lead to the following relation:
\begin{equation}
\label{eq:meanr0}
\Delta r_0 = b (\Delta i_0 + 0.4 \langle J-H \rangle)~,
\end{equation}
where $\langle J-H \rangle$ is the mean colour in the $J$ and $H$ bands
(or similarly $\langle J-K \rangle$ for the $J$ and $K$ bands) and
$\Delta i_0$ is the mean difference in $i_0$, the surface brightness
offset of the FP at a fiducial point (here taken to be $s_0=2.3$ and
$r_0=0.35$). For $b=-0.88$, the mean offset in $r_{0}$ between $J$ and
$H$ bands (as calculated from equation~\ref{eq:meanr0}) is $-0.14$ as
compared to the offset of $-0.12$ observed directly from the fits (see
the $r_{0}$ values in Table~\ref{tab:fpfits}). Similarly, for the $J$
and $K$ bands, the predicted $\Delta r_0$ is $-0.19$, as compared to the
observed offset of $-0.17$ from the fits. 

The predicted values are very close to the offsets observed, so we
conclude that the offsets in $r_0$ between passbands are a consequence
of the mean colours, as expected. Equivalently, allowing for the mean
colours the FP is consistent between the $J$, $H$ and $K$ bands.

\subsection{Comparison to literature}
\label{subsec:litcomp}

\begin{table*}
\centering
\begin{minipage}{150mm}
  \caption{Best-fitting FP slopes $a$ and $b$ as reported by previous
    studies in the literature. Also listed are the passband, sample size
    and fitting method used in each study. FP fits in optical and
    near-infrared passbands are shown respectively in the upper and
    lower halves of the table. Where available, the observed scatter
    orthogonal to the FP ($\sigma_{\perp}$) and the scatter about the FP
    in $\log R_{e}$ ($\sigma_r$) are given.\label{tab:litcomp}}
\begin{tabular}{@{}lclllccl@{}}
\hline
Survey & Band & $N_{g}$ & \multicolumn{1}{c}{$a$} & \multicolumn{1}{c}{$b$} & $\sigma_{\perp}$ & $\sigma_r$ & Type of fit \\
\hline
\citet{Dressler:1987}      & $B$     &       97  & 1.33$\pm$0.05 & $-$0.83$\pm$0.03 & -        & 20\% & inverse regression \\
\citet{Djorgovski:1987}  & r$_G$ &      106 & 1.39$\pm$0.14 & $-$0.90$\pm$0.09 & -        & 20\% & 2-step inverse regression \\ 
\citet{Lucey:1991b}        & $V$     &        66 & 1.26                    & $-$0.82                     & -        & 17\% & forward regression \\
\citet{Guzman:1993}      & $V$     &        37 & 1.14                    & $-$0.79                     & -        & 17\% & forward regression \\
\citet{Jorgensen:1996}  & $r$      &      226 & 1.24$\pm$0.07 & $-$0.82$\pm$0.02 & -        & 17\% & orthogonal regression \\
\citet{Hudson:1997}       & $R$     &     352 & 1.38$\pm$0.04 & $-$0.82$\pm$0.03 & -         & 21\% & inverse regression \\
\citet{Mueller:1998}        & $R$     &       40 & 1.25                    & $-$0.87                     & -         & 19\% & orthogonal regression \\
\citet{Gibbons:2001}       & $R$    &     428 & 1.37$\pm$0.05 & $-$0.84$\pm$0.03 & -         & 21\% & inverse regression \\
\citet{Colless:2001}        & $R$     &     255 & 1.22$\pm$0.09 & $-$0.84$\pm$0.03 & 11\% & 20\% & ML Gaussian \\
\citet{Bernardi:2003b}    & $g$     &   5825 & 1.45$\pm$0.06 & $-$0.74$\pm$0.01 & 13\% & 25\% & ML Gaussian \\
\citet{Bernardi:2003b}    & $r$      &   8228 & 1.49$\pm$0.05 & $-$0.75$\pm$0.01 & 12\% & 23\% & ML Gaussian \\
\citet{Hudson:2004}        & \VR     &      694 & 1.43$\pm$0.03 & $-$0.84$\pm$0.02 & -         & 21\% & inverse regression \\
\citet{DOnofrio:2008}      & $V$     &           - & 1.21$\pm$0.05 & $-$0.80$\pm$0.01 & -         & -    & orthogonal regression \\
\citet{LaBarbera:2008}   & $r$      &   1430 & 1.42$\pm$0.05 & $-$0.76$\pm$0.008 & 15\% & 28\% & orthogonal regression \\ 
\citet{Gargiulo:2009}       & $R$    &        91 & 1.35$\pm$0.11 & $-$0.81$\pm$0.03 & -         & 21\% & orthogonal regression \\ 
\citet{Hyde:2009}             & $g$     & 46410 & 1.40$\pm$0.05 & $-$0.76$\pm$0.02 & 16\% & 31\% & orthogonal regression \\ 
\citet{Hyde:2009}             & $r$      & 46410 & 1.43$\pm$0.05 & $-$0.79$\pm$0.02 & 15\% & 30\% & orthogonal regression \\ 
\citet{LaBarbera:2010b} & $g$     &   4589 & 1.38$\pm$0.02 & $-$0.79$\pm$0.003 & -       & 29\% & orthogonal regression \\
\citet{LaBarbera:2010b} & $r$      &   4589 & 1.39$\pm$0.02 & $-$0.79$\pm$0.003 & -       & 26\% & orthogonal regression \\
\hline
\citet{Scodeggio:1997}   & $I$     &     109 & 1.25$\pm$0.02 & $-$0.79$\pm$0.03   & -         & 20\% & mean regression \\
\citet{Pahre:1998a}         &$K$    &      251 & 1.53$\pm$0.08 & $-$0.79$\pm$0.03   & -         & 21\% & orthogonal regression \\
\citet{Bernardi:2003b}    & $i$     &    8022 & 1.52$\pm$0.05 & $-$0.78$\pm$0.01   & 11\% & 23\% & ML Gaussian \\
\citet{Bernardi:2003b}    & $z$    &    7914 & 1.51$\pm$0.05 & $-$0.77$\pm$0.01   & 11\% & 22\% & ML Gaussian \\
\citet{LaBarbera:2008}   & $K$   &    1430 & 1.53$\pm$0.04 & $-$0.77$\pm$0.008 & 14\% & 29\% & orthogonal regression \\
\citet{Hyde:2009}             & $i$     & 46410 & 1.46$\pm$0.05 & $-$0.80$\pm$0.02   & 15\% & 29\% & orthogonal regression \\ 
\citet{Hyde:2009}             & $z$    & 46410 & 1.47$\pm$0.05 & $-$0.83$\pm$0.02   & 15\% & 29\% & orthogonal regression \\ 
\citet{LaBarbera:2010b} & $J$    &    4589 & 1.53$\pm$0.02 & $-$0.80$\pm$0.003 & -        & 26\% & orthogonal regression \\
\citet{LaBarbera:2010b} & $H$   &    4589 & 1.56$\pm$0.02 & $-$0.80$\pm$0.005 & -        & 27\% & orthogonal regression \\
\citet{LaBarbera:2010b} & $K$   &    4589 & 1.55$\pm$0.02 & $-$0.79$\pm$0.005  & -        & 28\% & orthogonal regression \\
6dFGS (this paper, Table~\ref{tab:fpfits}) & $J$ & 8901 & 1.52$\pm$0.03 & $-$0.89$\pm$0.008 & 15\% & 30\% & ML Gaussian \\
6dFGS (this paper, Table~\ref{tab:fpfits}) & $H$ & 8568 & 1.47$\pm$0.02 & $-$0.88$\pm$0.008 & 15\% & 29\% & ML Gaussian \\
6dFGS (this paper, Table~\ref{tab:fpfits}) & $K$ & 8573 & 1.46$\pm$0.02 & $-$0.86$\pm$0.008 & 15\% & 29\% & ML Gaussian \\
\hline
\end{tabular}
\end{minipage}
\end{table*}

A summary of previous FP slope determinations from the literature is
given in Table~\ref{tab:litcomp}, along with the passband, sample size
and fitting method of each study. Where more than one regression method
was employed, the slopes from the orthogonal regression fit are given.
The coefficients of surface brightness, $b$, were converted to  the units 
used in this work (i.e.\ as the coefficient of $i\equiv$~\sblum\ rather than
 \sbmag, where the conversion is $b_i=-2.5 b_\mu$). In those studies
where an orthogonal rms scatter about the plane was quoted (based on 
an orthogonal regression or ML fit), we have listed this value in the 
$\sigma_{\perp}$ column and converted it to an rms scatter in the 
$r\equiv$~\er\ direction using $\sigma_r=\sigma_{\perp}(1+a^2+b^2)^{1/2}$ 
(for reference, this scaling factor is 2.0 for $a=1.5$ and $b=0.88$).
Note that the rms scatter in $r\equiv$~\er\ in dex, $\delta_r$, is 
conventionally converted to a fractional scatter in $R_e$ in percent, 
$\sigma_r$, using $\sigma_r\equiv(10^{+\delta_r}-10^{-\delta_r})/2$.

Table~\ref{tab:litcomp} shows the increase over time in the size of the samples being
studied and also the variety of fitting techniques employed, with the
more recent studies generally preferring orthogonal regression or ML
fits. The fitted value of the FP coefficient of velocity dispersion,
$a$, is typically found to be 1.2--1.4 at optical wavelengths and
1.4--1.5 in the near-infrared. {\em Within} individual studies in the
optical, $a$ tends to be larger in redder passbands; {\em between}
studies the differences are at least as large as this trend. By
contrast, $b$ is generally consistent with being constant across
passbands within any individual study, although it varies over the range
$-$0.74 to $-$0.90 when comparing different studies.

A direct comparison of the 6dFGS FP to the results of other studies is
constrained by the fact that only one study uses $J$ and $H$ band samples 
\citep{LaBarbera:2010b}, and only two studies use $K$ band samples
\citep{Pahre:1998a,LaBarbera:2010b}. Moreover,
neither of these studies use a ML fitting technique, so we have chosen
to compare with orthogonal regressions, where available, as the
next-best fitting method. Our $s\equiv$~\vd\ slope ($a=1.52$) is
consistent with the other near-infrared FP fits in being steeper than is
generally found in optical passbands. Our $i\equiv$~\sblum\ slope
($b=-0.89$) is at the upper end of the range of previous results. Due to
the large sample and homogeneous data afforded by the 6dFGS, the
fractional errors on both slopes (for $a$ less than 2\% and for $b$ less
than 1\%) are significantly smaller than is the case for older FP
samples, and comparable to those obtained for the similarly large and
homogeneous SDSS and UKIDSS samples \citep{Hyde:2009,LaBarbera:2010b}.

The most recent FP studies analysing large data sets across
multiple passbands have found a steepening of the FP slope $a$ going
from shorter to longer wavelengths, while in general the slope $b$ 
remains constant \citep{Hyde:2009,LaBarbera:2010b}. This trend, however, is
observed across optical to near-infrared wavelengths, but (as here) not
over the $JHK$ passbands (see Table~\ref{tab:litcomp}). This implies, as
expected, that there is relatively little variation with mass or size in
the dominant stellar populations (and hence the stellar $M/L$) across
these near-infrared passbands. 

The recent SPIDER FP study by \citet{LaBarbera:2010b} provides the
closest match to 6dFGS in both sample size and passbands: we can compare
the $J$, $H$ and $K$ ML Gaussian FP fits for more than 8500 6dGFS
galaxies with orthogonal regression FP fits in the same bands for 4589
SPIDER galaxies. The two studies obtain almost identical values of $a$
in the $J$ band (1.52 and 1.53), but 6dFGS finds $a$ to be significantly
smaller in the $H$ and $K$ bands (1.47 and 1.46), while SPIDER finds
slightly larger values in these bands (1.56 and 1.55). The differences
between the two studies in the $H$ and $K$ band values of $a$ are
significant relative to the estimated uncertainties (3.2$\sigma$).
Within each of the 6dFGS and SPIDER studies the values of $b$ are
consistent across the three bands; however 6dFGS finds $b$ in the range
$-0.89$ to $-0.86$, while SPIDER obtains a more positive value,
$b=-0.79$. This difference in $b$ is highly significant relative to the
estimated uncertainties ($>$8$\sigma$), but may be at least partly
attributed to the fact that orthogonal regression tends to find
systematically higher values of $b$, as shown in Figure
\ref{fig:fiterror}.

As well as comparing the {\em slopes} of the FP fits, it is interesting
to consider the {\em scatter} about the FP found in different studies.
The rms scatter about the FP relation projected in the \er\ direction
($\sigma_r$ in Table~\ref{tab:litcomp}) is usually taken as an estimate
of the rms uncertainty in distances and peculiar velocities when the FP
is used as a distance estimator. This uncertainty is widely quoted as
being 20\% or even smaller, a figure reflected in
Table~\ref{tab:litcomp} for the older FP samples. However the scatter in
\er\ calculated in this way for the most recent studies
\citep{LaBarbera:2008,Hyde:2009,LaBarbera:2010b}, and for the 6dFGS
sample, is in fact almost 30\%. This is somewhat surprising, given that
these recent samples are large and generally contain good-quality
homogeneous measurements of the FP parameters. In part the difference
may be due to the fact that these larger samples may contain a more
heterogeneous mix of galaxy types than the older `hand-picked' samples
(see \S\ref{sec:fpmorph} below). However a major source of this
discrepancy is that it is {\em not} correct to interpret the rms scatter
about an orthogonal regression or ML fit, projected in \er, as the
uncertainty in distance. As discussed in detail in
\S\ref{subsec:disterrs}, if one correctly accounts for the distribution
of galaxies in the FP, then the true distance error, $\sigma_d$, is
significantly smaller than $\sigma_r$. For the 6dFGS sample, while the
rms scatter about the FP in the \er\ direction is $\sigma_r=29\%$, the
rms scatter in the distance estimates is in fact $\sigma_d=23\%$.
 
\section{Environment and the Fundamental Plane}
\label{sec:fpenviro}

We investigate possible variations in the FP with {\em group}
environment, characterised by richness, and with {\em local}
environment, characterised by a nearest-neighbour density measure.

First, we consider potential environmental effects that correlate with
the scale of the dark matter halos that galaxies inhabit, using the
richness estimates from the group catalogue described in
\S\ref{subsec:groupcat} as a proxy for halo mass. We define four
subsamples according to richness $N_R$: galaxies in the field or very
low richness groups ($N_R \leq 1$), galaxies in low-richness groups ($2
\leq N_R \leq 5$), galaxies in medium-richness groups ($6 \leq N_R \leq
9$), and those galaxies in high-richness groups and clusters ($N_R \geq
10$). There are 6495 field galaxies, 1248 in low-richness groups, 546 in
medium-richness groups, and 612 in high-richness groups and clusters.

\begin{figure*}
\centering
\begin{minipage}{180mm}
\includegraphics[width=0.96\textwidth]{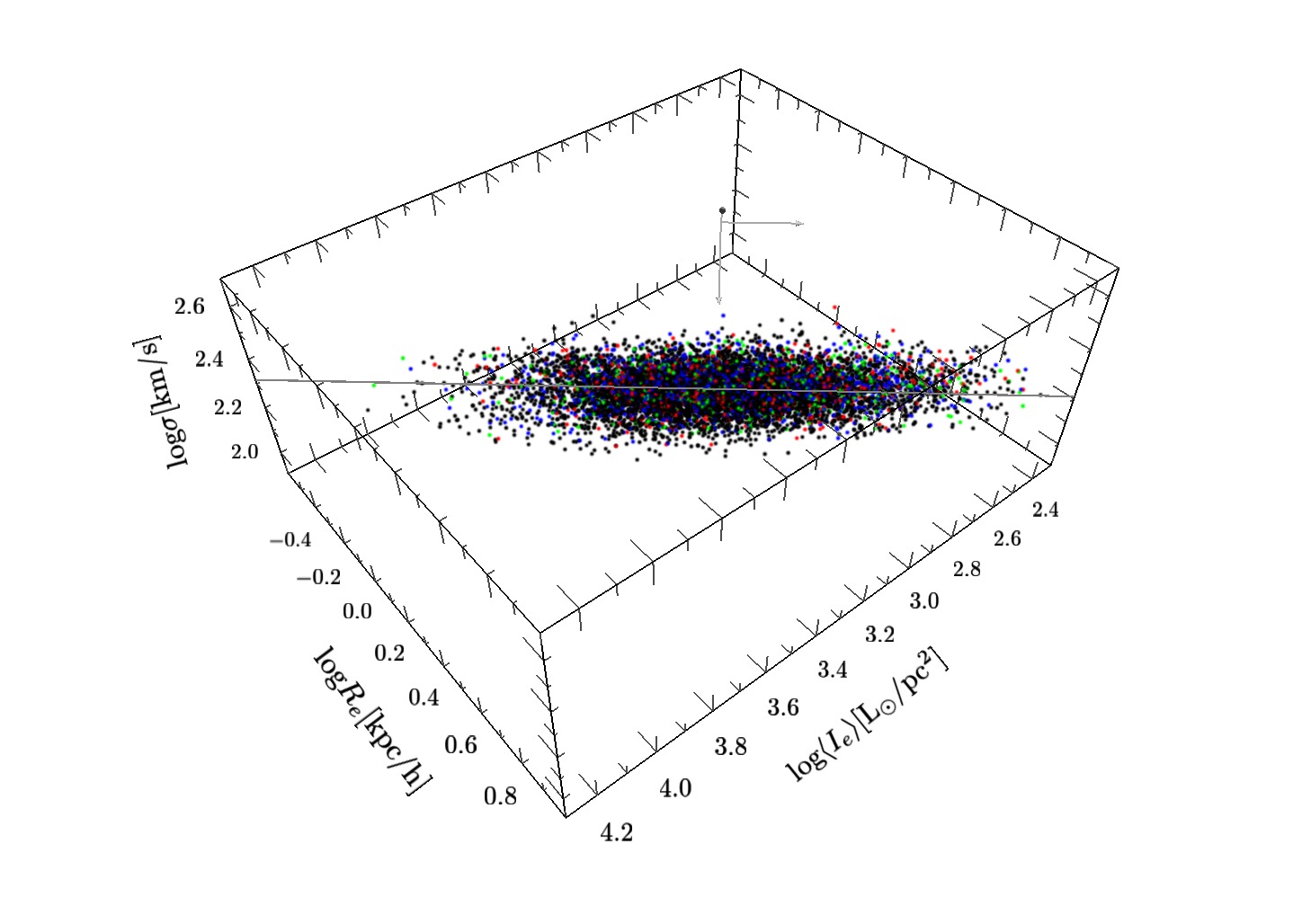} 
\caption{Interactive 3D visualisation of the 6dFGS $J$ band Fundamental
  Plane with individual galaxies colour-coded by the richness of the
  group environment they inhabit: 6495 field galaxies in black; 1248
  galaxies in low-richness groups in blue; 546 galaxies in
  medium-richness groups in green; and 612 galaxies in high-richness
  groups in red (these richness classes are defined in the text). The
  best-fitting plane (in grey) for the entire sample (with $a=1.523$, $b=-0.885$
  and  $c=-0.330$) is shown for reference. (Readers using Acrobat
  Reader v8.0 or higher can enable interactive 3D viewing of this
  schematic by mouse clicking on the version of this figure found in the ancillary files; see
  Appendix~\ref{sec:usage3D} for more detailed usage instructions.)}
\label{fig:richfp}
\end{minipage}
\end{figure*}

The distribution of these richness subsamples in FP space can be viewed
in the interactive 3D visualisation of Figure~\ref{fig:richfp}, where
the galaxies in the 6dFGS $J$ band FP sample are colour-coded by the
richness of the group environment they inhabit. From examination of
these distributions it is apparent that these subsamples tend to
populate similar FPs. This is broadly confirmed by the
best-fit FP parameters for each of these richness subsamples given in
Table~\ref{tab:fpfits}. The FP slopes $a$ and $b$ are similar within
1$\sigma$ for all four richness subsamples and the full $J$ band sample,
and the offset of the FP, given by $r_0$, is similar for the three
subsamples of galaxies in groups. The one significant difference is
between the offset for the field galaxy subsample and the group
subsamples.

\begin{figure*}
\centering
\begin{minipage}{170mm}
\includegraphics[width=0.95\textwidth]{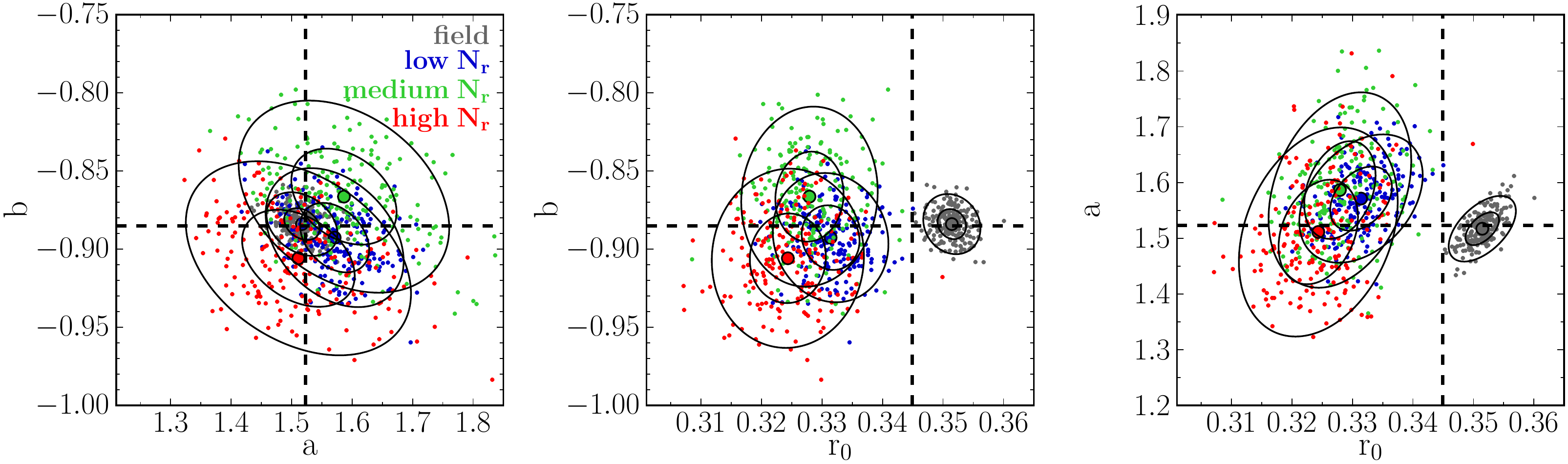} 
\caption{As for Figure~\ref{fig:abpass}, but comparing the FP fits to
  four richness samples spanning field (grey; $N_g = 6495$), low
  richness (blue; $N_g = 1248$), medium richness (green; $N_g = 546$)
  and high richness (red; $N_g = 612$) galaxy samples. The points in
  each panel are the fits to 200 mocks of each of these four subsamples;
  the large black circles show the means and the ellipses the 1$\sigma$
  and 2$\sigma$ contours of the distribution of fitted parameters. The
  dashed lines show, for reference, the best-fit parameters for the full
  $J$ band sample.}
\label{fig:abrich}
\end{minipage}
\end{figure*}

These similarities and differences are clarified in
Figure~\ref{fig:abrich}, which shows the best-fitting parameters of each
richness subsample, along with the 1$\sigma$ and 2$\sigma$ error
contours determined from 200 mock samples. The consistency of the FP
slopes is shown in the left panel of this figure, while the difference
in FP offsets between the field and group subsamples is shown in the
centre and right panels. This offset is $\Delta r_0 \approx 0.02$\,dex,
which is relatively small compared to the total scatter in $r$ of the
full sample ($\sigma_r = 0.127$\,dex). Nonetheless, it corresponds to a
systematic size or distance offset of about 4.5\%, and is statistically
significant at $>$3.7$\sigma$. Such an offset would have an appreciable
impact on estimates of the relative distances of field and group
galaxies if it were not accounted for.

\begin{figure*}
\centering
\begin{minipage}{170mm}
\includegraphics[width=0.95\textwidth]{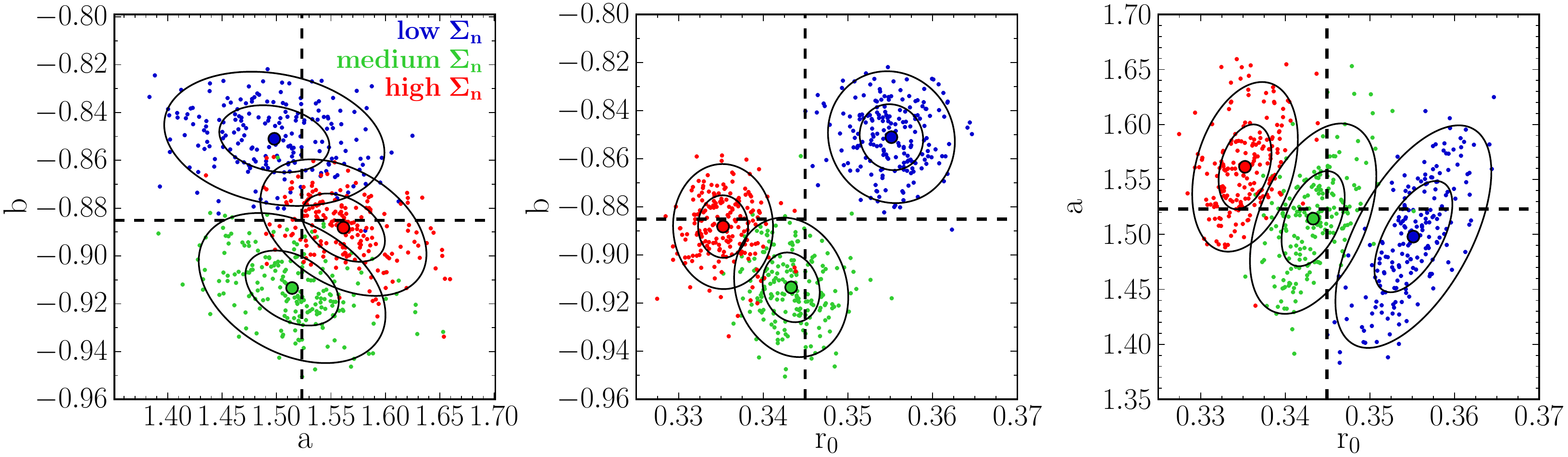} 
\caption{As for Figure~\ref{fig:abpass}, but comparing the FP fits to
  three local surface density, $\Sigma_5$, samples spanning low
  $\Sigma_5$ (blue; $N_g = 2664$), medium $\Sigma_5$ (green; $N_g =
  2812$) and high $\Sigma_5$ (red; $N_g = 2782$) galaxy samples. The
  points in each panel are the fits to 200 mocks of each of these three
  subsamples; the large black circles show the means and the ellipses
  the 1$\sigma$ and 2$\sigma$ contours of the distribution of fitted
  parameters. The dashed lines show, for reference, the best-fit
  parameters for the full $J$ band sample.}
\label{fig:absurfdens}
\end{minipage}
\end{figure*}

We repeat this analysis for the sample of 8258 galaxies for which we
have local environment estimates, as described in
\S\ref{subsec:groupcat}. This sample is divided by local surface
density ($\Sigma_5$) into three approximately equal-sized subsamples:
2664 galaxies in low-density environments ($\Sigma_5 \leq 0.07$), 2812
galaxies in medium-density environments ($0.07 < \Sigma_5 \leq 0.25$)
and 2782 galaxies in high-density environments ($\Sigma_5 > 0.25$). We
fit FPs to each of these subsamples individually, deriving the best-fit
parameters given in Table~\ref{tab:fpfits}. The coefficient of velocity
dispersion, $a$, is similar across the three subsamples and also with
respect to the global sample. There is weak variation (at the $2\sigma$
level), in the surface brightness coefficient, $b$, with galaxies in
denser environments tending to have an FP with a shallower $b$ slope;
galaxies in the low surface density sample exhibit the largest variation
in $b$ from the global FP. However, the strongest trend with local
environment is in the offset of the FP, where $r_0$ is systematically
smaller for galaxies with higher surface density. The significance of
this trend is clearly shown in the centre and right panels of
Figure~\ref{fig:absurfdens}, where we plot the best-fit FP slopes, $a$
and $b$, and the $r_0$ offset from 200 mock simulations of each local
surface density subsample.

Comparing the local density FP fits illustrated in
Figure~\ref{fig:absurfdens} to those for richness shown in
Figure~\ref{fig:abrich}, we find the same consistency in $a$ and the
same trend with environment in $r_0$. The trend in $b$ as a function of
local surface density is not seen for global environment, although this
may possibly be because our higher richness subsamples have too few
galaxies to recover such a weak trend.

Suggestions of environmental dependence in the FP (or the
$D_{n}$--$\sigma$ relation) first emerged in studies where a weak offset
between galaxies in clusters (such as Coma and Virgo) and the field was
detected \citep{Lucey:1991a,deCarvalho:1992}. However it was later
suggested that these differences could be attributed to errors in
measurement, as no such offset in the FP was subsequently found between
field and cluster galaxies in other similar studies
\citep{Burstein:1990,Lucey:1991b,Jorgensen:1996}. As samples of
early-type galaxies increased, and the range covered in environment and
mass was extended, trends with environment were found for {\em local}
density indicators such as cluster-centric distance
\citep{Bernardi:2003b} and local galaxy density \citep{DOnofrio:2008}.
The latter study also found a strong trend in the FP slopes $a$ and $b$
with {\em local} galaxy density, but no trend with {\em global}
environment parameters such as richness, R$_{200}$ and velocity
dispersion. More recently, \citet{LaBarbera:2010c} explored the role of
environment in the FP and found a strong trend with local galaxy density
(and a weaker trend with normalised cluster-centric distance),
independent of passband. Evidence of this trend is indicated by a lower
offset of the FP for galaxies in high-density regions compared to
low-density regions, consistent with previous results
\citep{Bernardi:2003b,DOnofrio:2008}. The slope $a$ was found to
decrease in high-density regions (in all passbands), while $b$ tended to
weakly increase with local galaxy density (a trend that disappears in
the near-infrared). Similar trends in the FP parameters were found for
galaxies in groups and the field.

The results obtained for the 6dFGS sample are consistent with other
recent studies, in that the variation of the FP is more pronounced for
parameters that reflect \emph{local} density or environment than for
those that are proxies for \emph{global} environment. Even though we
compare the offset between FPs using $r_0$ rather than $c$ (as
\citealt{LaBarbera:2010c} do), the trend we find with surface density
(i.e.\ lower $r_0$ for galaxies in higher-density environments), is at
least qualitatively consistent with that of the SPIDER study. However,
to anticipate the discussion in \S\ref{subsec:residtrends}, these
variations in the FP with environment are smaller than the variation
found with age; if the age of the stellar population were the main
driver of FP variations, then the environmental variations might be
primarily the result of correlations between environment and stellar
population.

\section{Morphology and the Fundamental Plane}
\label{sec:fpmorph}

We examine the morphological variation of the Fundamental Plane using a
visual classification of each galaxy's morphology from multiple people,
as described in \S\ref{subsec:morphdat}. The $J$ band FP sample was
divided into two morphological subsamples: 6956 elliptical and
lenticular galaxies (those classified as E, E/S0 or S0) and 1945
early-type spiral bulges (those classified as S0/Sp or Sp and having
bulges filling the 6dF fibre aperture). Note that the initial NIR
selection criteria mean there are relatively few of the latter class,
and that these may have some degree of bias towards larger \er. We do
not separate the E and S0 galaxies into separate subsamples since there
is significant overlap in our morphological classifications for these
two classes. We note that the FP is, in general, found to be consistent
between samples of E and S0 galaxies
\citep{Jorgensen:1996,Colless:2001}, and that, in fitting the E and S0
galaxies as one morphological subsample, we find the same scatter about
the FP as the for full sample.

\begin{figure*}
\centering
\begin{minipage}{180mm}
\includegraphics[width=0.96\textwidth]{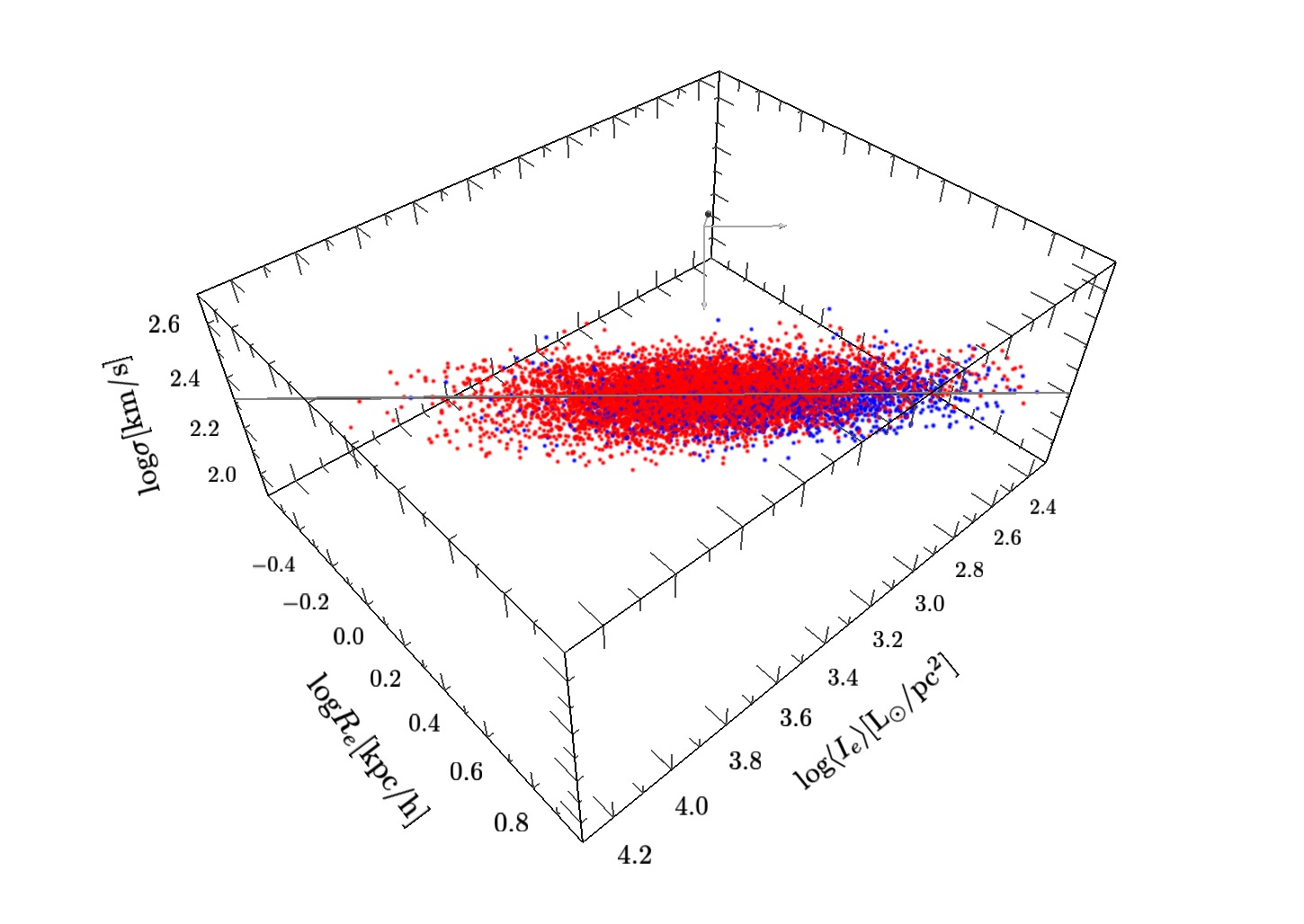} 
\caption{Interactive 3D visualisation of the 6dFGS $J$ band Fundamental
  Plane in $(\mathbf{r,s,i})$-space. The best-fitting plane (in grey) for the
  $J$ band (with $a=1.523$, $b=-0.885$ and $c=-0.330$) is plotted
  for reference. The galaxies are colour-coded according to morphology:
  6956 early types in red and 1945 late types in blue. (Readers using
  Acrobat Reader v8.0 or higher can enable interactive 3D viewing of
  this schematic by mouse clicking on the version of this figure found in the ancillary files; see
  Appendix~\ref{sec:usage3D} for more detailed usage instructions.)}
\label{fig:morphfp}
\end{minipage}
\end{figure*}

Figure~\ref{fig:morphfp} is an interactive 3D visualisation of the
$J$ band FP sample colour-coded by morphology, with the ellipticals and
lenticulars in red and the early-type spiral bulges in blue. This figure
shows that the two morphological subsamples populate slightly different
locations {\em within} the FP, with the early-type spiral bulges more
common at larger \er.

\begin{figure*}
\centering
\begin{minipage}{170mm}
\includegraphics[width=0.95\textwidth]{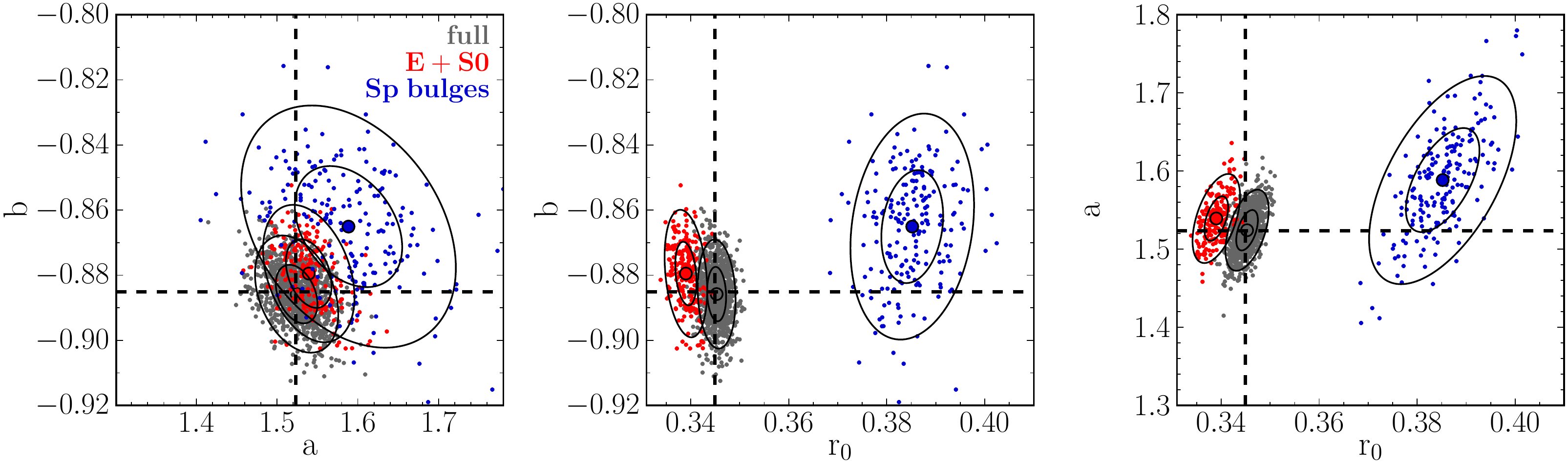} 
\caption{As for Figure~\ref{fig:abpass} but comparing the FP fits to the
  two morphological subsamples: 6956 elliptical and lenticular galaxies
  (E/S0) in red and 1945 early-type spiral bulges (Sp bulges) in blue;
  the full $J$ band sample of 8901 galaxies is shown in grey. The points
  in each panel are the fits to 200 mocks of the two morphological
  subsamples and to 1000 mocks of the full sample; the large black
  circles show the means of the fitted parameters and the ellipses show
  the 1$\sigma$ and 2$\sigma$ contours of the distribution. The dashed
  lines show, for reference, the best-fit parameters for the full
  observed $J$ band sample.}
\label{fig:abmorph}
\end{minipage}
\end{figure*}

The best-fit FP parameters for these two subsamples are given in
Table~\ref{tab:fpfits}, and their relative values and errors are
illustrated using mock samples in Figure~\ref{fig:abmorph}. The figure
shows that the FP slopes, $a$ and $b$, are consistent for the different
morphological classes but that the offset in \er, while small ($\Delta
r_0 = 0.045$\,dex) relative to the overall scatter in \er, is highly
significant (7$\sigma$) and corresponds to a systematic error of 10\% in
sizes and distances. An offset of this amplitude would have a
substantial impact on estimates of the relative distances of E/S0
galaxies and Sp bulges if it were not accounted for.

In addition to the difference in FP offset, there is a large shift in
the centroid of the distribution {\em within} the FP, with the
early-type spiral bulges having $\bar{r}=0.304$ while the ellipticals
and lenticulars have $\bar{r}=0.155$; i.e.\ the spiral bulges are
typically 35\% larger. We speculated that this may be due to the
selection criteria imposed, namely that the spiral bulges had to fill
the 6dF fibre apertures. We therefore re-sampled the
elliptical/lenticular sample to have the same apparent size distribution
as the spiral bulges, and re-fit the FP to this subsample; this did not
induce an offset in $r_0$ as observed in the spiral bulges. We conclude
that this offset is not primarily a selection effect, but rather a real
difference between the FPs of the ellipticals/lenticulars and the early
spiral bulges.

\section{Discussion}
\label{sec:discuss}

\subsection{The Fundamental Plane as a 3D Gaussian}
\label{subsec:FP3DGaussian}

Although throughout this paper we emphasise the value of fitting a 3D
Gaussian model to the FP, this is {\em not} saying that the intrinsic FP
is necessarily Gaussian. That may be the case in some axes, but in
others (e.g.\ in luminosity or velocity dispersion) the intrinsic
distribution very likely takes some other form (such as a Schechter
function)---a form that is only approximated by a Gaussian over the
range of values in our sample (i.e.\ the bright/large/massive end of the
distribution).

We have chosen to use a Gaussian model because it is computationally
easy and because empirically it fits the data in our sample (as
evidenced by Figure~\ref{fig:v1v2v3compare}). In practice the observed
FP is consistent with (well modelled by) a Gaussian partly due to either
(or both) the sample selection criteria and the observational errors.
The errors are approximately Gaussian and are relatively large in the
raw quantities $r$, $s$ and $i$ (although not in some combined
quantities like $r-bi$). Convolving these errors with the intrinsic FP
results in a more Gaussian distribution.

This effect is compounded by the selection criteria. For example, the
velocity dispersion cutoff truncates the probable Schechter function of
the intrinsic distribution in such a way that the truncated distribution
can be fit by a truncated Gaussian (the exponential part of a Schechter
function is similar to a Gaussian that is truncated near its peak). This
truncated distribution is then blurred and made more Gaussian by the
observational errors.

In sum, although a Gaussian intrinsic distribution is statistically a
sufficiently good model for the data in the 6dFGS sample (as well as
being computationally convenient), the substantial effects due to the
sample selection criteria and observational errors mean that we cannot
conclude that the underlying physical distribution is Gaussian. While
the ML method successfully fits a Gaussian to the intrinsic FP
distribution, a more realistic distribution might fit as well or better.

\subsection{Fundamental Plane scatter}
\label{subsec:fpscatter}

In general, the total scatter in $r$ that we recover for the 6dFGS FP
($\sigma_r$\,$\approx$\,29\%) is comparable to that found in other
recent studies \citep{Gargiulo:2009,Hyde:2009,LaBarbera:2010b}, but
larger than the value typically quoted as the FP distance error
($\sigma_r$\,$\sim$\,20\%) found in earlier studies (see
Table~\ref{tab:litcomp}). However, it is important to note that the
larger value of $\sigma_r$ found in recent studies (and here) is the rms
scatter, projected along the $r$-direction, about the best-fitting
orthogonal or maximum-likelihood FP. In \S\ref{subsec:disterrs}, we show
that this over-estimates the actual FP distance errors.

Here we examine the individual components contributing to the overall
scatter about the FP. This scatter results from a combination of
intrinsic scatter in the FP relation (the physical origins of which are
subject to investigation), observational errors and contamination from
outliers (such as non-early-type galaxies or merging objects). To
understand how each of these contribute to the total rms scatter in $r$,
we split $\sigma_r$ into the quadrature sum of these components:
\begin{equation}
\label{eqn:sigmar}
\sigma_r^2 = (a \epsilon_s)^2 + \epsilon_X^2 + \sigma_{r,int}^2 ~.
\end{equation}

The first term represents the effect of the rms observational scatter in
velocity dispersion, $\epsilon_s$, on the overall scatter in $r$.
Because $\epsilon_s$ is scaled by $a$, the FP coefficient of $s$, this
term is larger for samples with larger FP slopes. Since $a$ tends to
increase with wavelength ($a$\,$\approx$\,1.2--1.4 in optical passbands
and $a$\,$\approx$\,1.4--1.5 in near-infrared passbands), this term is
generally larger for near-infrared selected samples (such as 6dFGS) than
for optically selected samples (such as SDSS). The rms velocity
dispersion error of the 6dFGS sample is $\epsilon_s = 0.054$\,dex
\citep[i.e.\ 12\%, comparable to other large survey samples; see][]
{Campbell:2009}. So, given our $J$ band slope of $a = 1.52$, this term
amounts to a contribution to the overall scatter of about 18\%. To more
directly determine the effect of the errors in $s$ on the FP fits, we
have fitted subsamples restricted to smaller $\epsilon_s$ values (see
Table~\ref{tab:fpds}). We find no change in the FP slopes (at the
1$\sigma$ level), a small but significant change in the offset, and a
modest reduction (at most 5\%) in the overall scatter in \er, consistent
with that expected from the smaller value of $\epsilon_s$ and the above
formula for the total scatter.

\begin{table}
\centering
\caption{Best-fit FP dependence on velocity dispersion error. 
  \label{tab:fpds}} 
\begin{tabular}{@{}lccccc@{}}
\hline
$\epsilon_s$ & N$_g$ & a               & b                & $r_0$           & $\sigma_r$ \\
\hline
no limit     &  8901 & 1.523$\pm$0.026 & -0.885$\pm$0.008 & 0.345$\pm$0.002 & 0.127 \\
$\leq$0.07   &  7913 & 1.523$\pm$0.026 & -0.896$\pm$0.009 & 0.346$\pm$0.002 & 0.124 \\
$\leq$0.06   &  6694 & 1.529$\pm$0.029 & -0.903$\pm$0.010 & 0.349$\pm$0.002 & 0.122 \\
$\leq$0.05   &  4692 & 1.528$\pm$0.032 & -0.909$\pm$0.011 & 0.356$\pm$0.003 & 0.118 \\
$\leq$0.03   &  1855 & 1.558$\pm$0.053 & -0.894$\pm$0.018 & 0.376$\pm$0.005 & 0.108 \\
\hline
\end{tabular}
\end{table}

The second term in equation~\ref{eqn:sigmar} is the rms observational
scatter in the combined photometric quantity $X_\mathrm{FP} \equiv
r-bi$, which accounts for the high degree of correlation between the
measurement errors in $r$ and $i$ (see \S\ref{subsec:errors}). This
correlation conspires to make the value of this term negligible in
comparison to the other terms; for all the 6dFGS passbands,
$\epsilon_X$\,$\leq$\,4\%.

The final term represents the intrinsic scatter of the FP relation in
the $r$ direction. For a pure 3D Gaussian distribution the intrinsic
scatter in $r$ would be given by $\sigma_r = \sigma_1(1+a^2+b^2)^{1/2}$,
which, for our typical values of $a=1.5$ and $b=-0.88$, yields $\sigma_r
\approx 2.0\,\sigma_1$. However, because our observed distribution is
heavily censored by our selection criteria, the actual distribution of
galaxies in FP space is a truncated 3D Gaussian, and so we cannot apply
this formula. Instead we must calculate $\sigma_r$ either from
equation~\ref{eqn:sigmar}, taking the difference between the total
scatter and the rms measurement errors, or as the rms scatter in
$r-as-bi$ for mock samples drawn from the same intrinsic 3D Gaussian and
the same selection criteria, but with no measurement errors. Both these
approaches yield the same estimate for the intrinsic scatter in $r$ for
our $J$ band sample: $\sigma_{r,int}$\,$\approx$\,23\%. The intrinsic
scatter is therefore the single largest contributor to the overall
scatter about the 6dFGS FP.

Thus we have our total scatter in $r$ of 29\% being the quadrature sum
of 18\% scatter from the measurement errors in velocity dispersion, 4\%
scatter from the measurement errors in the photometric quantities, and
23\% scatter from the intrinsic dispersion of the FP distribution.

\subsection{Distance errors}
\label{subsec:disterrs}

We have found that the scatter about the 6dFGS FP in $r$ is 29\%.
However, this does {\it not} mean that, when we use this FP fit to
measure distances, we will only measure them to this precision. To
understand why this is the case, we must consider the procedure used to
measure distances and peculiar velocities from the FP.

In the most naive approach, one would convert the observed angular
radius of a galaxy to a physical radius assuming that the distance to
the galaxy is given by its redshift distance. The peculiar velocity of
the galaxy would then be approximated by the offset of this galaxy from
the FP in $r$. Since the peculiar velocity is measured from the offset
along the $r$-direction, the average scatter from the FP in $r$ then
represents the total error in galaxy distances and peculiar velocities
(from the combination of measurement errors and intrinsic scatter).

However there is a more general (and precise) way to estimate the
peculiar velocity. The peculiar velocity of a galaxy $n$ is given by its
offset along the $r$-direction from a particular value, $r^*_n$. This
$r^*_n$ is the most likely radius for galaxy $n$, given a particular set
of observed values of the velocity dispersion and surface brightness,
$s_n$ and $i_n$. In the preceding paragraph, we assumed that $r^*_n$ is
a point on the FP, given by $r^*_n = as_n + bi_n +c$. This assumption is
valid if the FP is best modelled as an infinite plane with uniform
scatter. However, the assumption is {\em not} valid if the distribution
of galaxies in FP space is best modelled by a 3D Gaussian and the minor
axis of this Gaussian is not aligned with the $r$-axis.

In equation~\ref{eq:trigauss}, we show the expression for the
probability density distribution of a single galaxy $n$. In
equation~\ref{eq:likelihood}, we give the sum of the log of such
probability densities for all galaxies in our sample. For a single
galaxy $n$, however, the likelihood is
\begin{equation}
\label{eq:gallike}
\begin{split}
\ln (P({\bf x_{n}})) = & -[\frac{3}{2} \ln(2\pi) + \ln(f_{n})  + 
                        \frac{1}{2}\ln(|\mathbf{\Sigma}+\mathbf{E_{n}}|) \\
  &+ \frac{1}{2}\mathbf{x_{n}^{T}}(\mathbf{\Sigma}+\mathbf{E_{n}})^{-1}\mathbf{x_{n}}]~.
\end{split}
\end{equation}
For a particular galaxy with known observational errors, each of these
terms is fixed except the final $\chi^2$ term, which is a quadratic
function of the physical parameters $r$, $s$ and $i$.

Since we directly observe $s$ and $i$, we can fix them at the observed
values $s_n$ and $i_n$. We can then use this equation to give us the
probability density distribution of $r$ for fixed $s=s_n$ and $i=i_n$
(i.e.\ $P(r|s,i)$). This is a quadratic function of the form
\begin{equation}
\ln(P(r|s,i)) = k_0 + k_1 (r-\bar{r}) + k_2 (r-\bar{r})^2 
\end{equation}
where $k_0$, $k_1$ and $k_2$ are functions of $s_n$, $i_n$, the
observational errors for the galaxy, and the FP fit parameters ($a$,
$b$, $\bar{r}$, $\bar{s}$, $\bar{i}$, $\sigma_1$, $\sigma_2$, and
$\sigma_3$). They can thus be obtained by expanding the matrix
multiplication terms in the preceding equations. The effective expectation value
for galaxy distances and peculiar velocities occurs at the maximum
likelihood---i.e.\ the maximum of this quadratic function,
\begin{equation}
r^*-\bar{r} = -k_1 / (2k_2) ~.
\end{equation}
This value varies from galaxy to galaxy, depending both on the
galaxy's position in FP space and its observational errors. If we
evaluate this in the case of no errors, and insert the values of the FP
fit parameters given in Table~\ref{tab:fpfits} for the $J$ band sample,
we find that the effective expectation value for distances is given by the plane
$r^*=1.18s-0.80i+0.152$; this relation differs quite markedly from the 
underlying Fundamental Plane. However, since we do in fact have observational
errors, and they vary from galaxy to galaxy, the peculiar velocity
expectation values for individual galaxies will {\it not} be confined to a
plane. 

We have evaluated this $J$ band zeropoint (i.e.\ the maximum
likelihood distance) for every galaxy in our sample, and find that the
scatter about the zeropoint is 23\%. This, then, is the distance error
in the $J$ band assuming no Malmquist bias corrections; we therefore
anticipate that 23\% does not necessarily represent our final distance
error, which will be explored in a future paper.

This 23\% scatter in distance is significantly smaller than the 29\%
that is naively obtained by calculating the scatter in $r$ about the
best-fit FP. The difference is purely a consequence of the fact that, in
our empirically well-justified 3D Gaussian model for the distribution in
FP space, galaxies are not symmetrically distributed about the FP in the
$r$ direction. Thus for fixed $s$ and $i$ the probability density of
galaxies in $r$ is not maximised on the FP, the expectation value for
the observed distance is not the redshift distance, the expectation
value of the peculiar velocity is not zero, and the scatter in distance
and peculiar velocity relative to this expectation value is less than
the scatter relative to the FP.

\subsection{The Fundamental Plane in $\kappa$-space}
\label{subsec:kspace}

\begin{figure*}
\centering
\begin{minipage}{170mm}
\includegraphics[width=0.95\textwidth]{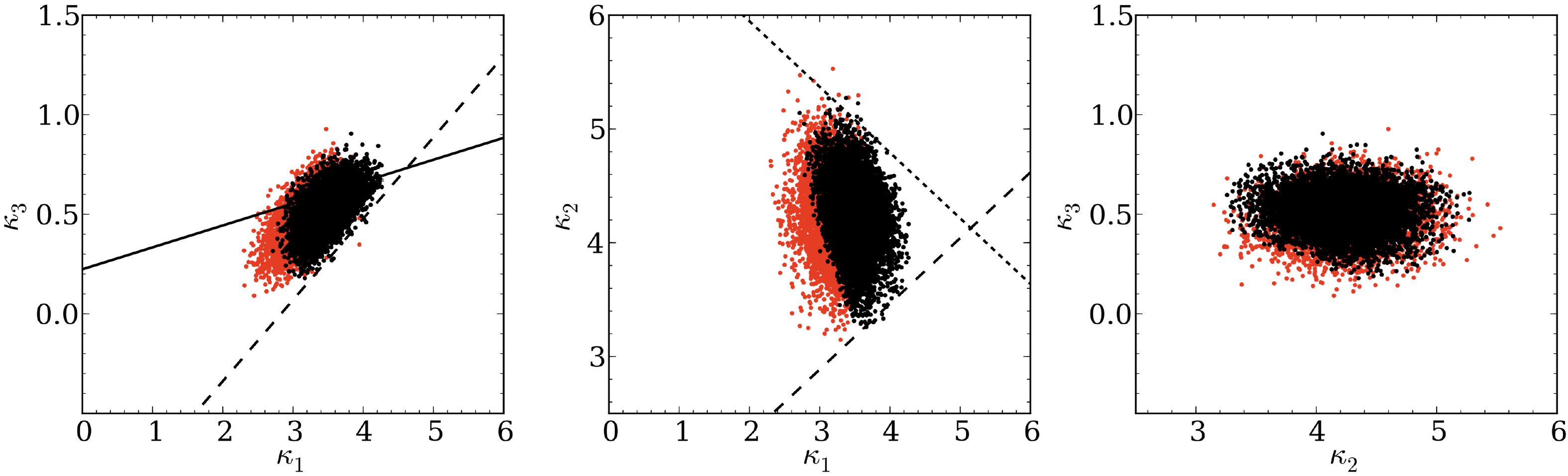} 
\caption{The $\kappa$-space distribution of the 6dFGS $J$ band FP sample
  (black) and the galaxies excluded by our selection criteria from a
  corresponding mock sample (red). {\em Left:} the
  $\kappa_3$--$\kappa_1$ projection of the FP showing the best-fit
  relation ($\kappa_3 \propto 0.110\kappa_1$, solid line) and the lower
  limit on $M/L$ as a function of mass ($2\sqrt{3}\kappa_3 -
  \sqrt{2}\kappa_1 > -4.0$; long-dashed line). {\em Centre:} the
  $\kappa_2$--$\kappa_1$ projection showing the upper limit defining the
  `zone of exclusion' for dissipation ($\kappa_1 + \sqrt{3}\kappa_2 <
  12.3$; short-dashed line), similar to that proposed by
  \citet{Bender:1992}; also the apparent lower limit on luminosity
  density ($\sqrt{3}\kappa_2 - \kappa_1 > 2.0$; long-dashed line). 
  {\em Right:} the $\kappa_3$--$\kappa_2$ projection.}
\label{fig:6dfkspace}
\end{minipage}
\end{figure*}

\citet{Bender:1992} proposed studying the FP using $\kappa$-space, a
coordinate system related to key physical parameters such as galaxy mass
($M$) and luminosity ($L$). Bender et~al.\ took as their observed
parameters $\log\sigma_0^2$, $\log I_e$ and $\log R_e$ (with $\sigma_0$
in units of \kms, $R_e$ in units of $\mathrm{kpc}$ and $I_e$ in units of
L$_{\sun}$\,pc$^{-2}$) and defined $\kappa$-space in terms of the
orthogonal set of basis vectors given by
\begin{align}
\label{eq:kappaspace}
\kappa_1 &\equiv (\log\sigma_0^2+\log R_e)/\sqrt{2} = (2s+r)/\sqrt{2},\notag \\
\kappa_2 &\equiv (\log\sigma_0^2+2\log I_e-\log R_e)/\sqrt{6} = (2s+2i-r)/\sqrt{6},\notag \\
\kappa_3 &\equiv (\log\sigma_0^2-\log I_e-\log R_e)/\sqrt{3} = (2s-i-r)/\sqrt{3}.
\end{align}
In this coordinate system, $\kappa_1$ is proportional to $\log M$,
$\kappa_2$ is proportional to $\log(I_e^3\,M/L)$ and $\kappa_3$ is
proportional to $\log(M/L)$.

FP samples in $\kappa$-space
\citep{Bernardi:2003b,Burstein:1997,Kourkchi:2012} are often plotted in
the $\kappa_3$--$\kappa_1$ projection (to show an almost edge-on view of
the FP) and the $\kappa_2$--$\kappa_1$ projection (to show an almost
face-on view of the FP). Figure~\ref{fig:6dfkspace} shows the
$\kappa$-space distribution for the $J$ band 6dFGS FP sample (black
points) in all three 2D projections of $\kappa$-space. The galaxies
rejected from a mock set of galaxies by the 6dFGS sample selection
criteria are also shown (in red) to illustrate the effects of censoring
on the observed $\kappa$-space distribution.

We can compute the principal axes of the FP distribution in
$(r,s,i)$-space, $(v_1,v_2,v_3)$, in terms of
$(\kappa_1,\kappa_2,\kappa_3)$ using the inverse of the transform
defined by equation~\ref{eq:kappaspace} to map from $\kappa$-space to
$(r,s,i)$-space followed by the transform defined by equations
\ref{eq:vectors}~\&~\ref{eq:vectors2} to then map to $(v_1,v_2,v_3)$.
Inserting the values of $a$ and $b$ for the best-fit $J$ band FP given
in Table~\ref{tab:fpfits}, we obtain
\begin{align}
\label{eq:vkappa}
\ v_1 &= +0.083\kappa_1+0.002\kappa_2-0.754\kappa_3 ~, \notag \\
\ v_2 &= -0.469\kappa_1+0.882\kappa_2-0.050\kappa_3 ~, \notag \\
\ v_3 &= -0.631\kappa_1-0.312\kappa_2+0.422\kappa_3 ~.
\end{align}
As expected, $v_1$ (the direction normal to the FP) is very close to
$\kappa_3$, which is proportional to $\log M/L$. However, because the
transformation from $(r,s,i)$-space to $\kappa$-space is non-orthogonal,
there is significant mixing in $\kappa$-space between $v_1$ and $v_3$,
with $v_1 \cdot v_3 = -0.6$.

In $\kappa$-space the best-fit $J$ band FP derived in $(r,s,i)$-space is
given by
\begin{equation}
\label{eq:kappafp}
\kappa_3 = 0.110\kappa_1 + 0.002\kappa_2 + 0.216 ~. 
\end{equation}
This is significantly shallower than the relation found by
\citet{Bender:1992}, which was $\kappa_3 \propto 0.15\kappa_1$ (although
the difference is in part due to the fact that Bender et al.\ were
working in the $B$ band and the 6dFGS result is for the $J$ band).
Because the coefficient of $\kappa_2$ is so small,
equation~\ref{eq:kappafp} is essentially a relation between
$\kappa_3\propto\log M/L$ and $\kappa_1\propto\log M$. Neglecting the
$\kappa_2$ term and using the definitions of $\kappa_1$ and $\kappa_3$
given in equation~\ref{eq:kappaspace} yields
\begin{equation}
\label{eq:mlvsm}
\frac{\log M/L}{\sqrt{3}} = 0.110\frac{\log M}{\sqrt{2}} + {\mathrm constant} ~,
\end{equation}
which corresponds to $M/L \propto M^{0.135}$.

It is illuminating to derive this same relationship starting from the
{\em assumption} that mass-to-light ratio has a simple power-law
dependency on mass. Letting $m=\log\,M$ and $l=\log\,L$, and assuming
that (ignoring constants) $m=2s+r$ and $l=2r+i$, if the mass-to-light
ratio is a power of mass, $m-l = \alpha m$, then we can write the FP as
\begin{equation}
r = 2 \left(\frac{1-\alpha}{1+\alpha}\right) s 
    - \left(\frac{1}{1+\alpha}\right) i + {\mathrm constant} ~.
\end{equation}
By equating FP coefficients with equation~\ref{eq:fpdef} we get two
relations for $\alpha$, namely $\alpha=(2-a)/(2+a)$ and
$\alpha=-(1+b)/b$. For an arbitrary FP relation there is no requirement
that these two relations give consistent values for $\alpha$. However,
as it happens, for the particular values $a\approx1.52$ and
$b\approx-0.88$ that are very close to the best-fit $J$ band FP for the
6dFGS sample, these relations give consistent values of
$\alpha\approx0.136$. Hence our best-fit FP is consistent with (but does
not require) a simple scenario in which mass-to-light ratio is a power
of mass, namely $M/L \propto M^{0.136}$ (or, equivalently, $M/L \propto
L^{0.157}$).

This relation (strictly, the relation given by equation~\ref{eq:kappafp}
with $\kappa_2$ fixed at its mean value of 4.2) is shown as the solid
line in Figure~\ref{fig:6dfkspace}. Because the transformation from
$(\kappa_1,\kappa_2,\kappa_3)$ is, by definition, orthogonal to
$(r,2s,i)$ but {\em not} orthogonal to $(r,s,i)$, the transformation
from $(r,s,i)$-space to $\kappa$-space does not preserve the shape of
the 3D Gaussian. Consequently this linear relation is not a particularly
compelling description of the $\kappa$-space distribution, even though
the transformed 3D Gaussian fit is still a good match to the data (as
shown by the mock galaxy sample).

The 6dFGS galaxies respect the zone of exclusion in the
$\kappa_1$--$\kappa_2$ plane suggested by \citet{Bender:1992},
corresponding to an upper limit on the amount of dissipation that a hot
stellar system of a given mass undergoes. This limit is indicated by the
short-dashed line in the centre panel of Figure~\ref{fig:6dfkspace},
given by $\kappa_1 + \sqrt{3} \kappa_2 < 12.3$. The long-dashed line in
the same panel provides another limit, $\sqrt{3}\kappa_2 - \kappa_1 >
2.0$, corresponding to a lower bound on the luminosity density, $L/R^3$
of an early-type galaxy of a given mass. However this requires further
investigation, as more compact galaxies may be catalogued in the 2MASS
database as stars and consequently would be excluded from our study. The
sharpest and most striking limit is that indicated by the long-dashed
line in the left panel of Figure~\ref{fig:6dfkspace}, $2\sqrt{3}\kappa_3
- \sqrt{2}\kappa_1 > -4.0$. This implies that for these early-type
galaxies there is a minimum mass-to-light ratio that increases with
increasing mass as $(M/L)_{\mathrm min} \propto M^{1/2}$. Since these
galaxies all have similar stellar populations, this suggests that more
massive galaxies have a maximum stellar-to-total mass ratio that
decreases as $M^{-1/2}$. 

\subsection{Fundamental Plane residual trends}
\label{subsec:residtrends}

In \S\ref{sec:fpenviro} and \S\ref{sec:fpmorph} we examined the
dependence of the 6dFGS FP on environment and morphology by comparing
the FP fits for appropriate subsamples of galaxies. Here we take an
alternative approach by looking at the trends of the orthogonal
residuals from the FP (defined as $[r - (as+bi+c)]/\sqrt{1 + a^2 +b^2}$)
with various galaxy properties. As well as morphology, group richness
($\log N_R$) and local density ($\log \Sigma_5$), we also consider three
stellar population parameters discussed in \citet{Springob:2012}:
$\log$\,age, metallicity ([Z/H]) and alpha-enhancement ([$\alpha$/Fe]).
For this particular purpose we convert our morphological classification
scheme to a discrete scale where 0\,=\,elliptical, 2\,=\,lenticular,
4\,=\,spiral and 1, 3 and 5 are the respective transition classes.

\begin{figure*}
\centering
\begin{minipage}{170mm}
\includegraphics[width=0.95\textwidth]{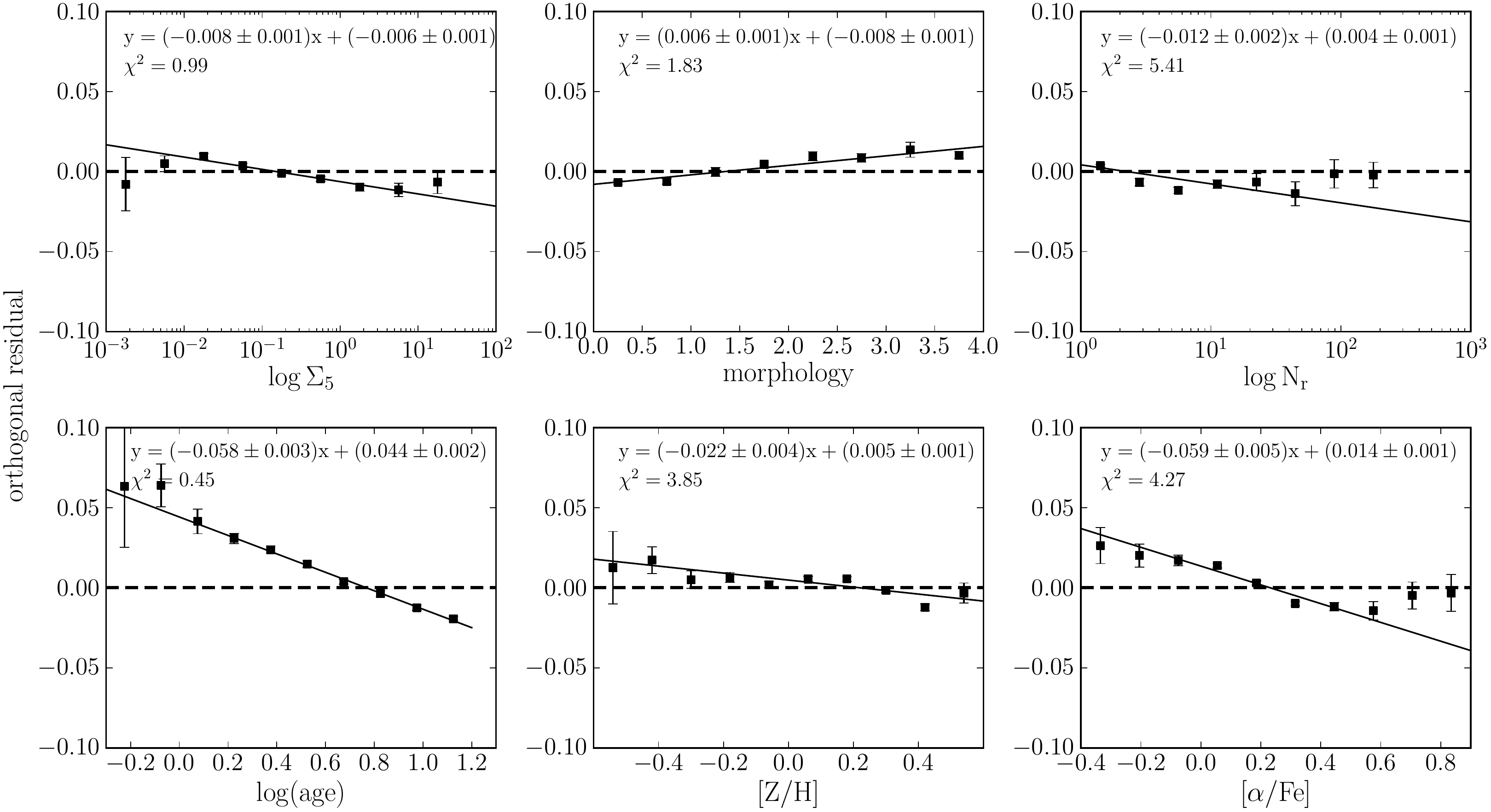} 
\caption{Correlation of orthogonal residuals relative to the best-fit FP
  ($a = 1.52$ and $b = -0.89$) with various galaxy properties: local
  density ($\log \Sigma_5$), morphological type and group
  richness ($\log N_R$) (all as defined in this paper), and $\log$\,age,
  [Z/H] and [$\alpha$/Fe] \citep[as defined in][]{Springob:2012}. In
  each panel the best-fitting line for the binned residuals is given
  along with the corresponding reduced $\chi^2$ value.}
\label{fig:orthresidbin}
\end{minipage}
\end{figure*}

Figure~\ref{fig:orthresidbin} shows the mean residuals orthogonal to the
best-fit global $J$ band FP (with $a = 1.52$ and $b = -0.89$) as a
function of these properties. The mean orthogonal residuals are computed
in bins of $\log \Sigma_5$ (for 8258 galaxies), morphological type and
$\log N_R$ (for 8901 galaxies), and $\log$\,age, [Z/H] and [$\alpha$/Fe]
(for 6679 galaxies). A weighted least-squares regression is performed to
quantify the significance of a linear trend in the binned data. The
slope and offset of the linear fit for each galaxy property (and their
errors) are given at the top of each panel, along with the reduced
$\chi^2$ of the fit.

The strongest trend of the FP residuals is clearly with the age of the
stellar population, and amounts to $\sim$0.08\,dex over the full range
in age; the next strongest trend is with [$\alpha$/Fe], amounting to
$\sim$0.05\,dex over the observed range. Both these trends are highly
statistically significant, although a line is not a good fit to the
relation in the case of [$\alpha$/Fe]. The residuals from the FP show
relatively weaker (although still statistically significant) trends with
morphological type, local density, group richness and metallicity. These
results are consistent with our fits to subsamples defined on the basis
of these properties, and confirm the equivalent analysis by
\citet{Springob:2012}. We refer the interested reader to that
paper for a more extensive investigation of the variations of stellar
populations in FP space, including a detailed comparison to the similar
study by \citet{Graves:2009}, \citet{Graves:2010a} and
\citet{Graves:2010b}.

If galaxy ages could be precisely determined, then these results imply
that it would be possible to reduce the intrinsic scatter about the FP
by a few percent. However the substantial uncertainties in estimating
the ages of stellar populations mean that even this modest gain cannot
be realised with current observational data and existing stellar
population models.

\section{Conclusion}

The 6dFGS Fundamental Plane sample comprises $\sim$10$^4$ early-type
galaxies from the 6dF Galaxy Survey. We provide the first comprehensive
visualisation for the entire Fundamental Plane (FP) parameter space
(without projection) by displaying this large and homogeneous dataset in
fully-interactive 3D plots.

We demonstrate that significant biasing can occur when deriving a
best-fit FP using least-squares regression (the predominant fitting
method used in previous studies). Standard regression techniques
implicitly assume models that fail to accurately represent the
underlying distribution of galaxies in FP space, and moreover do not
fully account for observational errors and selection effects that tend
to bias the best-fit plane. We show that a 3D Gaussian provides an
excellent empirical match to the distribution of galaxies in FP space
for the 6dFGS sample, and we use a maximum likelihood fitting technique
to properly account for all the observational errors and selection
effects in our well-characterised sample.

With this approach we obtain a best-fit FP in the 2MASS $J$ band of
$R_{e}$\,$\propto$\,$\sigma_0^{1.52\pm0.03}I_{e}^{-0.89\pm0.01}$. Fits in
the $H$ and $K$ bands are consistent with this at the 1$\sigma$ level
once allowance is made for differences in mean colour, implying that
$M/L$ variations along the FP are consistent among these near-infrared
passbands.

We deconstruct the scatter in $r$ about the FP, $\sigma_r$, into
contributions from observational errors and intrinsic scatter, and find
that the overall scatter of 29\% is the quadrature combination of an
18\% observational contribution and a 23\% intrinsic contribution. The
observational contribution is strongly dominated by the velocity
dispersion errors, and compounded by the fact that the FP slope is
steeper in near-infrared passbands than in optical passbands---the FP
coefficient of $\sigma_0$ is $a$\,$\approx$\,1.5 for $J$, $H$ and $K$
and $a$\,$\approx$\,1.2--1.4 for $B$, $V$ and $R$, so the same error on
$\sigma_0$ contributes 15--50\% more scatter to $\sigma_r$ for the
near-infrared FP than the optical FP.

The overall scatter in $R_e$ about the 6dFGS FP is larger than the
widely-quoted value of 20\%, but in fact is consistent with virtually
all recent studies of large samples of galaxies (see
Table~\ref{tab:litcomp}). Moreover, the actual scatter in distance
estimates is {\em not} the same as the scatter in $R_e$ about the
best-fit maximum-likelihood FP. We show that the true scatter in
distance (and peculiar velocity) must be calculated relative to the
expectation value of the distance (and peculiar velocity), which does
{\em not} lie in the FP. This is because our empirically-validated 3D
Gaussian model of galaxies in FP space has an {\em asymmetric}
distribution about the FP in the $r$-direction. Consequently, the
expectation value of the distance (and peculiar velocity) lies in a
plane with a shallower slope than the actual FP. When the scatter is
properly computed relative to this expectation value, we find that the
rms scatter in distance (or peculiar velocity) is in fact 23\%
(neglecting any corrections for Malmquist bias).

We investigate possible changes in the FP with environment, looking for
variations with both global environment (quantified by group or cluster
richness) and local environment (quantified by the surface density to
the fifth-nearest neighbour). We find little variation of the 6dFGS FP
slopes (i.e.\ the coefficients of velocity dispersion and surface
brightness) with either of these measures of environment. However there
is a statistically and physically significant offset of the FP with
environment in the sense that, at fixed velocity dispersion and surface
brightness, galaxies in the field and low-density regions are on average
about 5\% larger than those in groups and higher-density regions.

Morphological classification of our FP sample allows us to separate the
galaxies into two broad types: elliptical (E) and lenticular (S0)
galaxies are combined into one subsample and early-type spiral (Sp)
galaxies define the other type. For the latter, the construction of our
sample means that we are effectively determining the FP parameters for
the bulges of these galaxies. We find that this sample of early-type Sp
bulges has FP slopes and scatter consistent with the E/S0 galaxy sample,
although the FPs are offset in the sense that, at fixed velocity
dispersion and surface brightness, early-type Sp bulges are on average
about 10\% larger than E/S0 galaxies. Contrary to our expectations, this
does not appear to be a selection effect. Since the 6dFGS FP sample is
dominated by E/S0 galaxies (6956 E/S0's and 1945 Sp bulges), the
additional scatter in the overall FP from the offset in the FPs of the
two types of galaxies is negligible.

Complementing the analysis of \citet{Springob:2012}, we determine the
trends in the residuals of the FP as functions of group richness, local
density, morphology, and the age, metallicity and $\alpha$-enhancement
of the stellar population. We find that the strongest trend is
with age, and we speculate that, of the galaxy properties considered
here, age is the most important systematic source of offsets from the
FP, and may drive (through the correlations of age with environment,
morphology and metallicity) most of the variations with the other galaxy
properties. Demonstrating that this is the case, however, requires
detailed analysis of the covariances between all these quantities, which
we defer to a future paper.

The contributions to the intrinsic scatter about the FP from the mix of
morphologies, environments and stellar populations present in the 6dFGS
sample are at most (in the case of the ages of the stellar populations)
a few percent. Although it is in principle possible to compensate for
these effects, any corrections based on the mean relations between FP
residuals and the properties of individual galaxies would in practice
introduce more scatter than they would remove, due to the substantial
uncertainties in determining these properties. In any case, the bulk of
the intrinsic scatter would appear to be due either to physical
parameters not considered here or to genuinely stochastic variations in
the structure of galaxies.

Nonetheless, the systematic offsets of the FP for galaxies with
different morphologies, environments and stellar populations are
significant, and will need to be accounted for when, in future papers,
we use these FP determinations to derive distances and peculiar
velocities for this sample of $\sim$10$^4$ early-type galaxies covering
most of the southern hemisphere and reaching out to 16500\,\kms.

\section*{Acknowledgements}

We thank the AAO staff who supported the observations for the 6dF Galaxy
Survey on the UK Schmidt Telescope; without their professionalism and
dedication this ambitious survey would not have been possible.
Three-dimensional visualisation was achieved with the S2PLOT programming
library \citep{Barnes:2006}. We particularly thank Chris Fluke for
showing us how to construct the interactive 3D figures that make such a
difference to understanding intrinsically multi-dimensional datasets
like the Fundamental Plane.

This publication makes use of data products from the Two Micron All Sky 
Survey, which is a joint project of the University of Massachusetts and the
Infrared Processing and Analysis Center/California Institute of Technology, 
funded by the National Aeronautics and Space Administration and the 
National Science Foundation.

We acknowledge support from Australian Research Council (ARC)
Discovery Projects Grant (DP-0208876), administered by the Australian 
National University.  CM and JM acknowledge support from ARC
Discovery Projects Grant (DP-1092666). CM is also supported by 
a scholarship from the AAO.

\bibliographystyle{mn2e}
\bibliography{references}

\appendix
\section{Likelihood Normalisation}
\label{sec:likenorm}

For a trivariate Gaussian with lower selection limits of $r_{cut},
s_{cut}$ and $i_{cut}$, the likelihood normalisation integral is
\begin{equation}
\label{eq: norm}
f_{n}=\int_{r_{cut}}^{\infty}\int_{s_{cut}}^{\infty}\int_{u_{cut}}^{\infty} 
\frac{\exp[\frac{1}{2}(\mathbf{x_{n}^{T}}(\mathbf{\Sigma}+ 
                       \mathbf{E_{n}})^{-1}\mathbf{x_{n}})]}
{\sqrt{(2\pi)^3|\mathbf{\Sigma}+\mathbf{E_{n}}}|}\,dx
\end{equation}
where $\mathbf{x_{n}} = (r_{n},s_{n},i_{n})$. To determine $f_{n}$
numerically, we transform the integral using the Cholesky decomposition
of the matrix sum $\mathbf{\Sigma + E_{n}} = \mathrm{C}$ and then again
using the standard normal distribution function, $\Phi(y)$, given by
\begin{equation}
\Phi(y)=\frac{1}{2\pi}\int^{y}_{-\infty}\exp^{-\frac{1}{2}\theta^2}\,d\theta
\end{equation}
A final substitution is made to perform the integration over a unit
cube, resulting in the integral
\begin{equation}
\begin{split}
f_n&=(1-\Phi(\frac{r_{cut}}{C_{00}})) \\
   &\times\int_0^1(1-\Phi[\frac{s_{cut}}{C_{11}}-\frac{C_{10}}{C_{11}}
    \Phi^{-1}((1-w_0)\Phi(\frac{r_{cut}}{C_{00}})+w_0)]) \\
   &\times\int_0^1(1-\Phi[\frac{u_{cut}}{C_{22}}-\frac{C_{20}}{C_{22}}
    \Phi^{-1}((1-w_0)\Phi(\frac{r_{cut}}{C_{00}})+w_0) \\
   &-\frac{C_{21}}{C_{22}}\Phi^{-1}((1-w_1)
     \Phi[\frac{s_{cut}}{C_{11}} \\
   &-\frac{C_{10}}{C_{11}}\Phi^{-1}((1-w_0)
     \Phi(\frac{r_{cut}}{C_{00}})+ w_0 )]+w_1)]) \int_0^1\,d\mathbf{w} 
\end{split}
\end{equation}
In practice our model only includes an explicit selection cut in
velocity dispersion ($\sigma \geq \sigma_{cut}$). The above equation
then reduces to
\begin{equation}
f_{n}=\int_0^1 1-\Phi[\frac{s_{cut}}{C_{11}}-
     \frac{C_{10}}{C_{11}}\Phi^{-1}(w_0)]~dw_0
\end{equation}

\section{Interactive 3D Figures}
\label{sec:usage3D}

Several of the figures presented here (namely Figures~\ref{fig:3dvectors},
\ref{fig:jbandfp}, \ref{fig:agefp}, \ref{fig:richfp} and \ref{fig:morphfp}) can be
accessed as 3D interactive visualisations when viewing the 3D versions of these figures found in the ancillary files with
Acrobat Reader v8.0 or higher. Once 3D viewing is enabled by clicking on
the figure, the 3D mode allows the reader to rotate, pan and zoom the
view using the mouse.

The toolbar on each 3D figure contains a whole host of interactive
elements which can help in exploring the 3D visualisation. We
particularly direct the reader's attention to the following toolbar
features: (i)~you can restore the initial default view at any time using
the home button; (ii)~you can rotate to any orientation you prefer and,
where relevant, to special, author-selected 3D views (e.g.\ the edge-on
view of the FP); these can be selected from the Views drop-down menu;
(iii)~you can toggle the model tree, which allows individual plot
features (e.g.\ scatter points, planes, vectors) of the 3D figure to be
turned on and off, giving the viewer greater control of the interactive
figure. Suggested interactions with particular 3D figures include:

(a)~In Figure~\ref{fig:3dvectors}, use the model tree to toggle the
$\mathbf{v}$-space vectors and mass/luminosity vectors one at a time to
see how they compare in our 3D Gaussian model. Also, rotate the figure
to view the small angle between $\mathbf{v_1}$ and $\mathbf{m}
-\mathbf{l}$ and also $\mathbf{v_2}$ and $\mathbf{l} - 3\mathbf{r}$.

(b)~Figure~\ref{fig:jbandfp} not only contains the $J$ band FP sample of
galaxies, but also the $H$ band and $K$ band samples. They can be 
enabled in the model tree by selecting `H Band' or `K Band' respectively.
For an unimpeded view of the individual galaxies, toggle the best-fit 
plane (called `Fundamental Plane' in the model tree); this also applies to Figures~\ref{fig:richfp} and \ref{fig:morphfp}. In the Views drop-down menu, 
select `Edge-on' to view the Fundamental Plane in the projection with the smallest scatter.

(c)~In Figure~\ref{fig:richfp}, rotate and pan across the FP galaxies to
explore where the richness subsamples lie on the Fundamental Plane.

(d)~In Figure~\ref{fig:morphfp}, toggle the individual points of each
morphology subsample to see the differences in the way their
distributions populate FP space.
  
\end{document}